\documentclass[a4paper,11pt]{article}
\usepackage{graphicx}
\pdfoutput=1 

\usepackage{jheppub} 

\usepackage{preprintcover}  
\PreprintCoverPaperTitle{Search for supersymmetry at \rts=8 TeV in final states with jets and
 two same-sign leptons or three leptons with the ATLAS detector}  
\PreprintIdNumber{CERN-PH-EP-2014-044}  
\PreprintCoverAbstract{A search for strongly produced supersymmetric particles is
  conducted using signatures involving multiple energetic jets and either two
  isolated leptons ($e$ or $\mu$) with the same electric charge, or at least
  three isolated leptons. The search also utilises jets originating
  from $b$-quarks, missing
  transverse momentum and other observables to extend its sensitivity.
 The analysis uses a data
  sample corresponding to a total
  integrated luminosity of 20.3 \ifb\ of $\rts = 8$ TeV proton--proton
  collisions recorded with the ATLAS detector at the Large Hadron
  Collider in 2012. No deviation from the Standard Model expectation is observed. 
  New or significantly improved exclusion limits are set on a wide
  variety of supersymmetric models in which the lightest squark can be of the
  first, second or third generations, and in which R-parity can be
  conserved or violated.}  
\PreprintJournalName{JHEP} 

\usepackage[T1]{fontenc} 
\usepackage{atlasphysics}
\usepackage{subfigure}
\usepackage{verbatim}

\usepackage{float}
\usepackage{bm}            
\usepackage{hyperref}   
\usepackage{multirow}
\usepackage{placeins}

\usepackage{datetime}
\date{\currenttime}

\title{Search for supersymmetry at \rts=8 TeV in final states with jets and
 two same-sign leptons or three leptons with the ATLAS detector}

\author{The ATLAS Collaboration}

\abstract{
 A search for strongly produced supersymmetric particles is
  conducted using signatures involving multiple energetic jets and either two
  isolated leptons ($e$ or $\mu$) with the same electric charge, or at least
  three isolated leptons. The search also utilises jets originating
  from $b$-quarks, missing
  transverse momentum and other observables to extend its sensitivity.
 The analysis uses a data
  sample corresponding to a total
  integrated luminosity of 20.3 \ifb\ of $\rts = 8$ TeV proton--proton
  collisions recorded with the ATLAS detector at the Large Hadron
  Collider in 2012. No deviation from the Standard Model expectation is observed. 
  New or significantly improved exclusion limits are set on a wide
  variety of supersymmetric models in which the lightest squark can be of the
  first, second or third generations, and in which R-parity can be
  conserved or violated.}

\usepackage{pdfpages}

\begin{document} 
\setlength{\parskip}{0pt}

\widowpenalty10000
\clubpenalty10000

\newcommand{\figref}[1]{Figure~\ref{#1}}
\newcommand{\figsref}[1]{Figures.~\ref{#1}}
\newcommand{\Secref}[1]{Section~\ref{#1}}
\newcommand{\tabref}[1]{Table~\ref{#1}}

\newcommand{\Nfive}{\ensuremath{N_{5}}}
\newcommand{\Mmess}{\ensuremath{M_{\mathrm{mess}}}}
\newcommand{\Cgrav}{\ensuremath{C_{\mathrm{grav}}}}
\newcommand{\meff}{\ensuremath{m_{\mathrm{eff}}}} 
\newcommand{\mt}{\ensuremath{m_\mathrm{T}}}
\newcommand{\mttwo}{\ensuremath{m_\mathrm{T2}}}
\newcommand{\mc}{\ensuremath{m_\mathrm{C}}}
\newcommand{\mct}{\ensuremath{m_\mathrm{CT}}}
%
\newcommand{\gravitino}{\ensuremath{\tilde{G}}}
\newcommand{\sneutrino}{\ensuremath{\tilde{\nu}}}
\def\chinoonemp{\ensuremath{\mathchoice%
      {\displaystyle\raise.4ex\hbox{$\displaystyle\tilde\chi^\mp_1$}}%
         {\textstyle\raise.4ex\hbox{$\textstyle\tilde\chi^\mp_1$}}%
       {\scriptstyle\raise.3ex\hbox{$\scriptstyle\tilde\chi^\mp_1$}}%
 {\scriptscriptstyle\raise.3ex\hbox{$\scriptscriptstyle\tilde\chi^\mp_1$}}}}

\def\chinojmp{\ensuremath{\mathchoice%
      {\displaystyle\raise.4ex\hbox{$\displaystyle\tilde\chi^\mp_j$}}%
         {\textstyle\raise.4ex\hbox{$\textstyle\tilde\chi^\mp_j$}}%
       {\scriptstyle\raise.3ex\hbox{$\scriptstyle\tilde\chi^\mp_j$}}%
 {\scriptscriptstyle\raise.3ex\hbox{$\scriptscriptstyle\tilde\chi^\mp_j$}}}}

\def\chinoipm{\ensuremath{\mathchoice%
      {\displaystyle\raise.4ex\hbox{$\displaystyle\tilde\chi^\pm_i$}}%
         {\textstyle\raise.4ex\hbox{$\textstyle\tilde\chi^\pm_i$}}%
       {\scriptstyle\raise.3ex\hbox{$\scriptstyle\tilde\chi^\pm_i$}}%
 {\scriptscriptstyle\raise.3ex\hbox{$\scriptscriptstyle\tilde\chi^\pm_i$}}}}

\def\chinojpm{\ensuremath{\mathchoice%
      {\displaystyle\raise.4ex\hbox{$\displaystyle\tilde\chi^\pm_j$}}%
         {\textstyle\raise.4ex\hbox{$\textstyle\tilde\chi^\pm_j$}}%
       {\scriptstyle\raise.3ex\hbox{$\scriptstyle\tilde\chi^\pm_j$}}%
 {\scriptscriptstyle\raise.3ex\hbox{$\scriptscriptstyle\tilde\chi^\pm_j$}}}}
 
\def\ninoi{\ensuremath{\mathchoice%
      {\displaystyle\raise.4ex\hbox{$\displaystyle\tilde\chi^0_i$}}%
         {\textstyle\raise.4ex\hbox{$\textstyle\tilde\chi^0_i$}}%
       {\scriptstyle\raise.3ex\hbox{$\scriptstyle\tilde\chi^0_i$}}%
 {\scriptscriptstyle\raise.3ex\hbox{$\scriptscriptstyle\tilde\chi^0_i$}}}}

\def\ninoj{\ensuremath{\mathchoice%
      {\displaystyle\raise.4ex\hbox{$\displaystyle\tilde\chi^0_j$}}%
         {\textstyle\raise.4ex\hbox{$\textstyle\tilde\chi^0_j$}}%
       {\scriptstyle\raise.3ex\hbox{$\scriptstyle\tilde\chi^0_j$}}%
 {\scriptscriptstyle\raise.3ex\hbox{$\scriptscriptstyle\tilde\chi^0_j$}}}}
 
\def\Ptmiss{\ensuremath{\vec{E}_\mathrm{T}^{\mathrm{miss}}}}

\def\lsim{\mathrel{\rlap{\lower4pt\hbox{\hskip1pt$\sim$}}
    \raise1pt\hbox{$<$}}}                
\def\gsim{\mathrel{\rlap{\lower4pt\hbox{\hskip1pt$\sim$}}
    \raise1pt\hbox{$>$}}}                

\makeatletter
\renewcommand{\p@subfigure}{\arabic{figure}}
\renewcommand{\thesubfigure}{(\alph{subfigure})}
\makeatother

\maketitle
\flushbottom

\section{Introduction}
\label{sec:intro}

Supersymmetry (SUSY)
\cite{Miyazawa:1966,Ramond:1971gb,Golfand:1971iw,Neveu:1971rx,Neveu:1971iv,Gervais:1971ji,Volkov:1973ix,Wess:1973kz,Wess:1974tw} 
is a generalisation of space-time symmetries that 
predicts new bosonic partners for the fermions and new fermionic partners for the bosons
of the Standard Model (SM). If $R$-parity is conserved~\cite{Fayet:1977yc,Farrar:1978xj},
SUSY particles are produced in pairs and the lightest supersymmetric particle (LSP) is stable. 
In a large variety of models, the LSP is the lightest neutralino ($\ninoone$) and provides
a possible candidate for dark matter. 
The coloured superpartners of
quarks and gluons, the squarks ($\tilde q$) and gluinos ($\gluino$), could be produced in strong interaction processes at the Large Hadron Collider (LHC) and
decay via cascades ending with a stable $\ninoone$. The undetected
$\ninoone$ would result in substantial missing transverse momentum
($\bm{p}\mathrm{^{miss}_T}$ and its magnitude $\met$).
The rest of the cascade would yield final states with multiple jets and
possibly leptons arising from the decay of sleptons ($\tilde\ell$),
the superpartners of leptons,  or
$W$, $Z$ and Higgs ($h$) bosons. 
If $R$-parity is violated (RPV), the LSP is not stable, which would lead to
similar signatures but with lower, or no, \met.

In the Minimal Supersymmetric Standard
Model~\cite{Fayet:1976et,Fayet:1979sa,Dimopoulos:1981zb} (MSSM), 
the scalar partners of right-handed and left-handed quarks,
$\tilde q_R$ and $\tilde q_L$, can mix to form 
two mass eigenstates, $\tilde q_1$ and $\tilde q_2$, where 
$\tilde q_1$ denotes the lighter particle. This
mixing effect is proportional to the corresponding SM
fermion masses and therefore is more important for the third generation.
Furthermore, SUSY can solve the hierarchy problem of the SM 
(also referred to as the naturalness problem)
\cite{Witten:1981nf,Dine:1981za,Dimopoulos:1981au,Sakai:1981gr,Kaul:1981hi}
if the masses of the gluinos, higgsinos\footnote{The charginos $\tilde\chi^{\pm}_{1,2}$
and neutralinos $\tilde\chi^0_{1,2,3,4}$ are the 
mass eigenstates formed from the linear superposition of the SUSY partners of the Higgs
and electroweak gauge bosons (higgsinos, winos and binos).} (the superpartners of Higgs bosons) and top
squarks ($\stop$) are not heavier than the $\mathcal{O}$(TeV) scale. 
A light left-handed top squark also implies
that the left-handed bottom squark ($\sbottom_L$) may be
relatively light because of the SM weak-isospin symmetry. As a
consequence, the lightest bottom squark ($\sbottomone$) and top squark ($\stopone$)
could be produced with relatively large cross sections at the LHC,
either directly in pairs or through $\gluino\gluino$ production
followed by $\gluino \rightarrow \sbottomone b$ or $\gluino
\rightarrow \stopone t$ decays (gluino-mediated production).

In this paper, events containing multiple jets and either two leptons of the same electric charge
(same-sign leptons, SS) or at least three leptons (3L) are used to search for strongly produced 
supersymmetric particles.  Throughout this paper, the term leptons
($\ell$) refers to electrons and/or muons only.
Signatures with SS or 3L are predicted in many SUSY scenarios.
Gluinos produced in pairs or in association with a squark  
can lead to SS signatures when decaying to any final state that includes leptons because gluinos are Majorana fermions.
Squark production, directly in pairs
or through $\gluino \gluino$ or $\gluino \tilde q$  production with 
subsequent $\gluino \to q\tilde q$ decay,
can also lead to
SS or 3L signatures when the squarks decay in cascades involving top quarks ($t$), charginos, 
neutralinos or sleptons, which subsequently decay as $t\to bW$,
$\tilde\chi^{\pm}_i\to W^{\pm (*)}\tilde\chi^0_j$,
$\tilde\chi^0_i\to h/Z^{(*)} \tilde\chi^0_j$, or
$\tilde\ell\to \ell\ninoone$, respectively.
Similar signatures are also
predicted by non-SUSY models such as minimal Universal Extra
Dimensions (mUED) ~\cite{Cheng:2002ab}.
Since this search benefits from low 
SM backgrounds, it allows the use of relatively loose kinematic
requirements on
$\met$, increasing the sensitivity to scenarios with
small mass differences between SUSY particles (compressed scenarios)
or where $R$-parity is violated. 
This search is thus sensitive to a wide variety of models based on
very different assumptions.

\begin{figure}[t]
\centering
\includegraphics{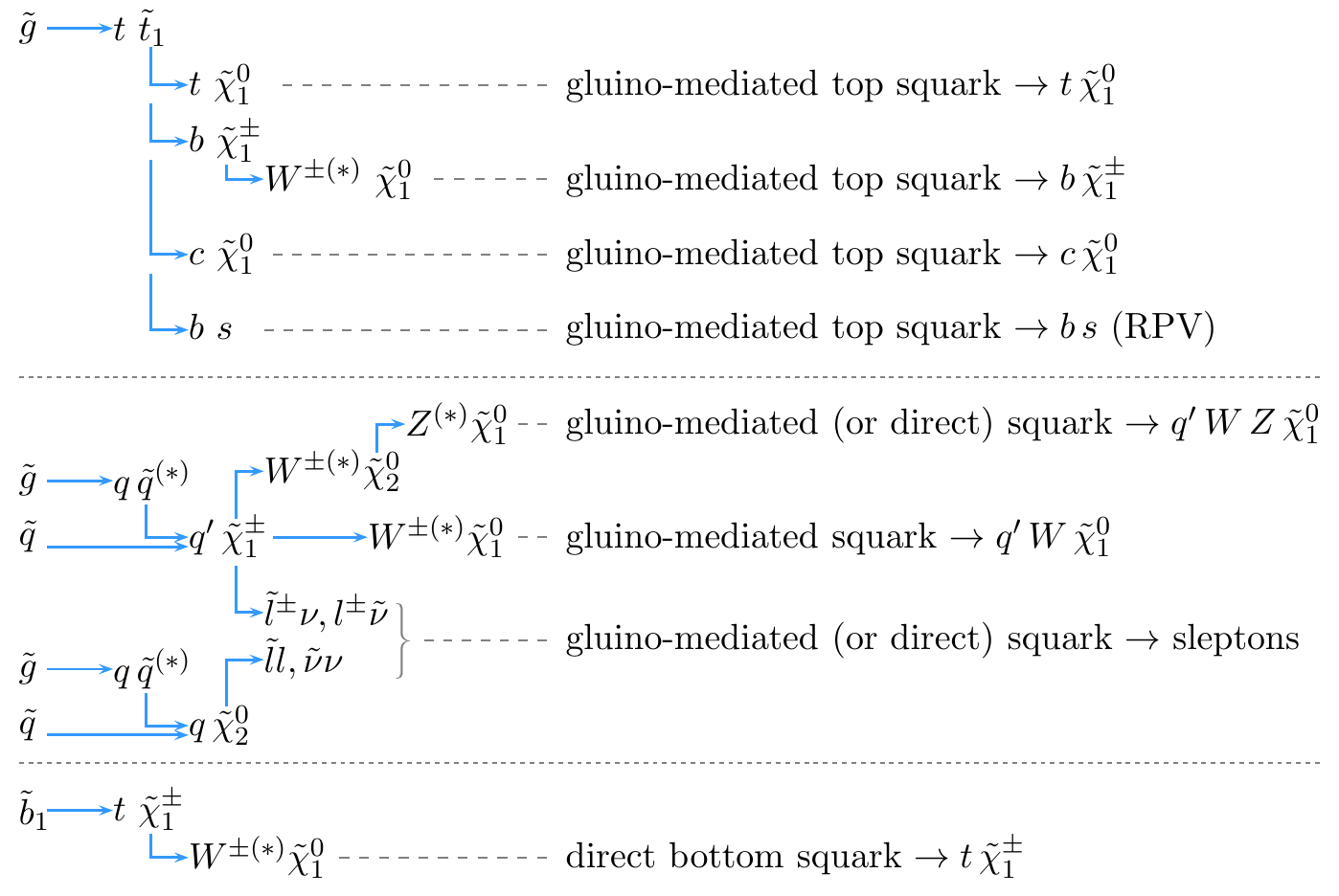}
\caption{Overview of the SUSY processes considered in the analysis.
The initial supersymmetric particles are always produced in pairs: 
$pp\rightarrow \gluino\gluino,\ \sbottom \bar\sbottom$ or $\squark \bar\squark$.
The notation $q$ ($\squark$) refers to quark (squark) of the first or second generation. 
The slepton and sneutrino decay as $\tilde\ell\to \ell\ninoone$ and
$\tilde\nu\to \nu\ninoone$, respectively. Leptons in the final state
can arise from the decay of any $W$ or $Z$ bosons or
sleptons that are produced. The charge-conjugate processes are also considered.}
\label{fig:SUSYmodels}
\end{figure}

The analysis uses $pp$ collision data from the full 2012 data-taking period, corresponding to 
an integrated luminosity of 20.3 \ifb\ collected at $\rts$=8 TeV, and
significantly extends the reach of previous searches 
performed by the ATLAS \cite{ATLAS:2012ai} and CMS
\cite{Chatrchyan:2012paa,Chatrchyan:2012ira,Chatrchyan:2012sa,Chatrchyan:2014}
Collaborations.
Five statistically independent signal regions (SR) are designed to cover the SUSY processes
illustrated in figure~\ref{fig:SUSYmodels}.
Two signal regions requiring SS and jets identified to originate from $b$-quarks ($b$-jets) are optimised for
gluino-mediated top squark and direct bottom squark production.
These are complemented with a signal region requiring a $b$-jet veto, optimised for the gluino-mediated production of 
first- and second-generation squarks.
Two signal regions requiring 3L are designed for scenarios characterised by multi-step
decays. 

Backgrounds with prompt SS or 3L events arising from rare SM
processes, such as
$t\bar{t} W$, $\ttbar Z$, $W^\pm W^\pm$ and $WZ$, are estimated with Monte Carlo simulations.
Backgrounds from hadrons mis-identified as leptons, 
leptons originating from heavy-flavour decays, 
electrons from photon conversions, 
and electrons with mis-measured charge are estimated with data-driven methods.
The background predictions are cross-checked with alternative methods and 
tested with data in validation regions chosen to be close in phase
space to the signal regions.
The probability ($p$-value) of the background-only hypothesis is then estimated independently in each signal region.
To maximise the sensitivity of the analysis across the entire phase space,
a simultaneous fit is performed in all signal regions to place model-dependent
exclusion limits on several SUSY benchmark scenarios.

\section{ATLAS detector and data sample}
\label{sec:detector}

ATLAS is a multi-purpose detector \cite{Aad:2008zzm} 
designed for the study of $pp$ and heavy-ion collisions at the LHC. It provides nearly
full solid angle\footnote {ATLAS uses a
  right-handed coordinate system with its origin at the nominal
  interaction point (IP) in the centre of the detector and the $z$-axis
  along the beam pipe. The $x$-axis points from the IP to the centre of
  the LHC ring, and the $y$-axis points upward. Cylindrical coordinates
  ($r$,$\phi$) are used in the transverse plane, $\phi$ being the azimuthal angle
  around the beam pipe. The pseudorapidity is defined in terms of the
  polar angle $\theta$ as $\eta = - \ln \tan(\theta/2)$.} coverage around the interaction point.
Charged particles are tracked
by the inner detector, which covers the pseudorapidity region $|\eta| <
2.5.$ In order to
measure their momenta, the inner detector is embedded in the 2 T magnetic field
of a thin superconducting solenoid.
Sampling calorimeters span the pseudorapidity range up to $|\eta| = 4.9$. High-granularity liquid-argon (LAr) electromagnetic calorimeters are present up
to $|\eta| = 3.2$. Hadronic calorimeters with scintillating tiles as
active material cover $|\eta| < 1.7$ while LAr technology is used
for hadronic calorimetry from $|\eta| = 1.5$ to $|\eta| = 4.9$.
The calorimeters are surrounded by a muon spectrometer. The magnetic field
is provided by air-core toroid magnets. Three layers of precision
gas chambers track muons up to $|\eta|$ = 2.7 and muon trigger chambers
cover  the range $|\eta|$ < 2.4. A three-level trigger system is used to select
interesting events for storage and subsequent analysis.

The data set, after the application of beam, detector and
data quality requirements, has an integrated luminosity of $20.3 \pm 0.6$
\ifb. The luminosity is measured using
techniques similar to those described in ref.~\cite{Aad:2013ucp} with a preliminary
calibration of the luminosity scale derived from beam-overlap scans
performed in November 2012. The number of $pp$ interactions occurring in the
same bunch crossing varies between approximately 10 and 30 with an
average of 20.7 for this data set.

\section{Simulated event samples}
\label{sec:mc}

Simulated events are used to model the SUSY signal, optimise the event selection requirements, compute
systematic uncertainties and estimate some of the SM
backgrounds with prompt same-sign lepton pairs or three leptons. These include
top quark(s) plus bosons ($W/Z/H$), diboson ($W^\pm W^\pm$, $WZ$, $ZZ$, $WH$, $ZH$),
triboson ($WWW$, $WZZ$, $ZZZ$) and $\ttbar\ttbar$ production. 
Other sources of background such as $\ttbar$, $W/Z$+jets, $W
\gamma$, $W^+ W^-$, $\ttbar \gamma$ and
single-top production are estimated with data-driven methods
described in section~\ref{sec:bkgd}. 

Samples of 
$\ttbar V$+jets ($V= W,Z$), $\ttbar WW$, single top quark plus a $Z$ boson,
$VVV$+jets and $\ttbar\ttbar$ 
are generated with {\sc MadGraph-5.1.4.8}~\cite{Alwall:2007st} interfaced to 
{\sc Pythia-6.426}~\cite{Sjostrand:2006za}.  Alternative $\ttbar V$+jets samples 
generated with  {\sc 
  Alpgen-2.14}~\cite{SAMPLES-ALPGEN} interfaced with  {\sc
  Herwig-6.520}~\cite{SAMPLES-HERWIG} and {\sc
  Jimmy-4.31}~\cite{Butterworth:1996zw} are employed to estimate 
the sensitivity of the analysis to Monte Carlo modelling. 
The {\sc Pythia-8.165}~\cite{Sjostrand:2007gs} generator is used to model $\ttbar H$
production, for which the Higgs boson
  mass is set to 125 GeV.
The $WZ$ and $W^\pm W^\pm$ processes are modelled
using {\sc Sherpa-1.4.1}~\cite{Gleisberg:2008ta} with matrix elements producing
up to three final-state partons.  
The $ZZ$
process is generated with 
{\sc Powheg-1.0}~\cite{Frixione:2007vw} interfaced to {\sc
  Pythia-8.165}. Monte Carlo modelling systematic uncertainties for the $ZZ$
process are estimated using two 
sets of a{\sc Mc@nlo}~\cite{Frederix:2011ss} samples where next-to-leading-order (NLO) matrix elements
are matched to either {\sc Pythia-6.426} or {\sc Herwig-6.520} with
{\sc Jimmy-4.31} parton showers according to
the {\sc Mc@nlo} formalism~\cite{Frixione:2002ik}.
Monte Carlo samples of $\ttbar$ events are used to provide corrections to the
data-driven background estimates, described in section~\ref{sec:bkgmet}, for
kinematic regions where the sample size is not sufficient to measure the
$\ttbar$ contribution directly in data. Four different samples are used: {\sc
  Powheg-1.0} interfaced with {\sc Pythia-6.426}, {\sc Powheg-1.0} interfaced with
{\sc Herwig-6.520} and {\sc Jimmy-4.31}, 
{\sc Mc@nlo-4.06} interfaced with  {\sc Herwig-6.520}
and {\sc Jimmy-4.31}
and {\sc Alpgen-2.14}
interfaced with  {\sc Herwig-6.520} and {\sc Jimmy-4.31}. 

 The NLO CT10~\cite{Lai:2010vv} parton distribution function (PDF) set is used with {\sc
Sherpa}, {\sc Powheg} and {\sc Mc@nlo} while the CTEQ6L1
\cite{Pumplin:2002vw} PDF set is used with {\sc MadGraph}, {\sc
  Pythia} and  {\sc Alpgen}. 
The predicted background yields are obtained by normalising the simulated samples to theoretical cross sections
from the most precise available calculations~\cite{Campbell:2012dh, Garzelli:2012bn, Campbell:2011bn}.

The SUSY signal samples are generated with 
{\sc Herwig++2.5.2}~\cite{Bahr:2008pv} or {\sc MadGraph-5.1.4.8}
interfaced with {\sc Pythia-6.426}, in both cases using the PDF set
CTEQ6L1. 
Signal cross sections are calculated to next-to-leading order in the
strong coupling constant, adding the resummation of soft gluon
emission at next-to-leading-logarithmic accuracy
(NLO+NLL)~\cite{nlo1,nlo2,nlo3,nlo4,nlo5}. 
The cross section and its uncertainty are taken from an
envelope of cross-section predictions using different PDF sets and
factorisation and renormalisation scales, as described in
ref.~\cite{Kramer:2012bx}.
The mUED samples are
generated with {\sc Herwig++2.5.2} using the
CTEQ6L1 PDF set and the leading-order cross
section from {\sc Herwig++}.

The parton shower parameters of the simulated samples were tuned to
match \mbox{ATLAS} data observables sensitive to initial- and final-state QCD
radiation, colour reconnection, hadronisation,
and multiple parton interactions. The tuned parameter set {\sc AUET2}~\cite{auet2} is used with {\sc Pythia 6}, {\sc Herwig
  6} and {\sc Pythia 8} (except that the tune {\sc P2011C}
\cite{Skands:2010ak} is used for the   {\sc
  Powheg + Pythia} $\ttbar$ sample), and the set {\sc
  UEEE3}~\cite{Gieseke:2012ft} is used with {\sc
  Herwig++}.
The effect of additional proton-proton collisions in the same or
neighbouring bunch crossings, called ``pile-up'', is modelled by overlaying 
minimum-bias events, simulated with {\sc Pythia-8.160} using the  {\sc
  AUET2} tune, onto the original hard-scattering event. 
Simulated events are weighted to reproduce the observed distribution of the average
number of collisions per bunch crossing in data. 
Monte Carlo samples are passed through a detector simulation \cite{Aad:2010ah} based on
{\sc Geant4}~\cite{Agostinelli:2002hh} or on a fast simulation using a
parametric response to the showers in the electromagnetic
and hadronic calorimeters \cite{ATL-PHYS-PUB-2010-013} and {\sc
  Geant4}-based simulation elsewhere.

Simulated events are reconstructed with the same algorithms as data.
Corrections derived from data control samples are applied to account for
differences between data and simulation for the lepton trigger and
reconstruction efficiencies, momentum scale and resolution, and for
the efficiency and mis-tag rate for tagging jets originating from $b$-quarks.

\section{Physics object reconstruction}
\label{sec:objectdefn}

Jets are reconstructed from topological clusters \cite{topocluster,Aad:2011he} formed from calorimeter cells by using the anti-$k_t$ algorithm
\cite{Cacciari:2008gp,Cacciari:2005hq}  with a cone size parameter of
0.4 implemented in the {\sc FastJet} package
\cite{Cacciari:2011ma}.
Jet energies are corrected \cite{Aad:2011he} for detector
inhomogeneities and the non-compensating response of the calorimeter using
factors derived from test beam, cosmic ray and {\it pp} collision
data, as well as from the detailed {\sc Geant4} detector simulation. 
The impact of multiple overlapping {\it pp} interactions is accounted
for using a technique, based on jet areas, that provides an
event-by-event and jet-by-jet pile-up correction~\cite{JetArea}.
Selected jets are required to have transverse momentum
$\pt>40~\GeV$\  and $|\eta|<2.8$. The identification of
$b$-jets is performed using a neural-network-based
$b$-tagging algorithm \cite{ATLAS-CONF-2014-004} with 
an efficiency of 70\% in simulated $\ttbar$ events.
The probabilities for mistakenly $b$-tagging a
jet originating from a $c$-quark or a light-flavour parton are
approximately 20\% and 1\%~\cite{ATLAS-CONF-2013-109,CONF-2012-040}, respectively. The kinematic
requirements on $b$-jets are
$\pt>20$~\GeV\ and $|\eta|<2.5$. 
Signal jets and $b$-jets are selected independently, hence $b$-jets with $\pt > 40$~\GeV\ are included in both jet and $b$-jet multiplicities.

Electron candidates are reconstructed using a cluster in the
electromagnetic calorimeter matched to a track in the inner
detector. Preselected electrons must satisfy the ``medium'' selection criteria
described in ref.~\cite{Aad:2011mk}, re-optimised for 2012 data, and fulfil $\pt > 10$~\GeV, $\vert\eta\vert <
2.47$ and requirements on the impact parameter of the track.
Muon candidates are identified by matching an extrapolated inner detector track to one or more track segments in the muon spectrometer 
\cite{Aad:1299479}. Preselected muons must fulfil $\pt > 10$~\GeV\ and
$\vert\eta\vert < 2.5$.

Signal leptons are defined by requiring tighter quality criteria  and
increasing the \pt\ threshold to 15 \GeV. 
Signal electrons must satisfy the ``tight'' selection
criteria~\cite{Aad:2011mk}. In addition, for both the signal electrons and muons, isolation requirements based on tracking and calorimeter
information and impact parameter requirements are
applied. 
The electron track isolation discriminant is computed as the summed scalar \pt\
of additional tracks inside a cone of radius
$\Delta R = \sqrt{ (\Delta \eta)^2 + (\Delta \phi)^2}= 0.2$ around the electron. The tracks considered must
originate from the same vertex associated with the electron and have $\pt > 0.4$~\GeV. The
electron calorimeter isolation discriminant is defined as the scalar sum of the
transverse energy,
\et, of topological clusters 
within a cone of radius $\Delta R = 0.2$ around 
the electron cluster and is corrected for any contribution from
the electron energy and pile-up.
 The muon track and calorimeter isolation discriminants are the same
 as the ones used for electrons, except for the isolation cone radius being $\Delta
 R =0.3$ and calorimeter cells around the muon
 extrapolated track being used for the calorimeter isolation discriminant.
For leptons with $\pt <$ 60 \GeV, both track and
calorimeter isolation are required to be smaller than 6\% and 12\% of the
electron's and muon's \pt, respectively.
For leptons with $\pt >$ 60 \GeV, an upper limit of
3.6 \GeV\ and 7.2 \GeV\ is imposed on both the calorimeter and track isolation
requirements for electrons and muons, respectively.
The track associated with the electron or muon candidate must have a longitudinal impact 
parameter $z_0$ satisfying $\vert z_0 \sin\theta\vert<0.4$~mm and
fulfil the requirement for the significance of the transverse impact
parameter, $d_0$, of $|d_0/\sigma(d_0)| <3$. 
The track parameters $z_0$ and $d_0$ are defined with respect to the
reconstructed primary vertex. For events
with multiple vertices along the beam axis, the vertex with the
largest $\sum \pt^2$ of associated tracks is taken as the
primary vertex. Furthermore, the primary vertex must be 
made of at least five tracks
with $\pt >$ 0.4 \GeV\ and its position
must be consistent with the beam spot
envelope.

Ambiguities between the reconstructed jets and leptons are resolved by applying the following criteria sequentially. 
Jets with a separation $\Delta R < 0.2$ from an electron candidate are
rejected. Any lepton candidate with a
distance $\Delta R < 0.4$ to the closest remaining jet is discarded.
If an electron and a muon have a separation
$\Delta R < 0.1$, the electron is discarded. For these
requirements, jets with $\pt > 20$ 
\GeV\ and preselected leptons are considered.

The missing transverse momentum vector, $\bm{p}\mathrm{_T^{miss}}$
with magnitude \met, is constructed as the negative
of the vector sum of the calibrated transverse momenta of all muons and electrons
with $\pt >10$ \GeV, jets with $\pt >20$ \GeV\ and calorimeter
energy clusters with $|\eta| < 4.9$ not assigned to these objects \cite{Aad:2012re}. 

\section{Event selection}
\label{sec:eventSel}

\par Events are selected
using a combination (logical OR) of \met\ and non-isolated single-lepton
and dilepton triggers. 
The thresholds applied to \met\ and the leading and subleading
lepton \pt\ are lower than those applied offline to ensure that
trigger efficiencies are constant
in the phase space of interest. 
The trigger threshold for \met\ is 80 \GeV.
The \pt\ thresholds for single-lepton triggers are 60 \GeV\ and 36 \GeV\
for electrons and muons, respectively. The dilepton triggers feature
lower thresholds in \pt, down to 12 \GeV\ for electrons and 8 \GeV\
for muons,  allowing events with multiple
soft leptons to be kept.
The efficiencies of \mbox{\met-only} triggers in the phase space of interest are close to 100\%.
The electron triggers reach efficiencies
above 95\% and muon triggers have efficiencies between 75\% and 100\%,
being lowest in the region $|\eta| < 1.05$.

\par Events from non-collision backgrounds
are rejected using dedicated quality criteria
\cite{Aad:2011he}. 
Events of interest are selected if they contain at least two leptons passing the requirements
described in section~\ref{sec:objectdefn} and if the highest-\pt\ lepton satisfies $\pt >$ 20
\GeV. 
Events with a leading pair of leptons having an
 invariant mass $m_{\ell\ell}<$ 12 \GeV\ are
removed. This requirement rejects events with pairs of energetic leptons from
decays of heavy hadrons and has
negligible impact on the signal acceptance.

\subsection{Signal regions}

The signal regions are determined with an optimisation procedure
using simulated events from the simplified models illustrated in figure~\ref{fig:SUSYmodels}.
The data are divided into two mutually exclusive SS and 3L samples.
In the SS sample, the
two highest-\pt\ leptons
must have the same electric charge and fulfil $\pt > $ [20,15] \GeV,
and there must be no other signal lepton with $\pt >$ 15~\GeV. 
In the 3L sample, the three highest-\pt\ leptons must
fulfil $\pt > $ [20,15,15] \GeV, respectively. No requirements
on the total electric charge are applied to this sample.
Good sensitivity to the signatures in all signal models is obtained
by defining five non-overlapping signal regions with
selection requirements based on the following kinematic
variables: \met; jet and $b$-jet multiplicities ($N_{\rm jets}$ and
$N_{b\mathrm{-jets}}$);
effective mass \meff\ computed from all signal leptons 
and selected jets as 
$m_{\mathrm{eff}}=\met~+~\sum\limits p^{\ell}_\mathrm{T}~+~\sum p^{\rm
  jet}_\mathrm{T}$;
transverse mass computed from the highest-\pt\ lepton ($\ell_1$) and \met\
as $\mt= \sqrt{2 \pt^{\ell_1} \met(1-\cos[\Delta\phi(\ell_1, p\mathrm{_T^{miss}})])}$;
and invariant mass $m_{\ell\ell}$ computed with
opposite-charge same-flavour leptons.

\begin{table}
\begin{center}
\resizebox{\textwidth}{!}{
\begin{tabular}{lcccc}
\hline\noalign{\smallskip}
\hline\noalign{\smallskip}
 SR &  Leptons & $N_{b\mathrm{-jets}}$ & Other variables &
   Additional requirement \\
 & & & &
 on \meff \\
\hline\noalign{\smallskip}
 SR3b & SS or 3L & $\ge $3 & $N_{\rm jets} \ge$ 5& \meff $>$350~\GeV \\
\noalign{\smallskip}
SR0b & SS & = 0     & $N_{\rm jets} \ge$ 3, \met $>$
 150~\GeV,  & \meff $>$400~\GeV \\
& & & \mt$>$ 100~\GeV & \\
\noalign{\smallskip}
 SR1b & SS & $\ge $1 & $N_{\rm jets} \ge$ 3, \met $>$ 150~\GeV,  & \meff $>$700~\GeV \\
& & &  \mt
 $>$100~\GeV,  SR3b veto & \\
\noalign{\smallskip}
 SR3Llow & 3L &  -    & $N_{\rm jets} \ge$ 4, 50~$<$ \met $<$ 150~\GeV,  & \meff $>$400~\GeV\\
& & & $Z$ boson
 veto, SR3b veto & \\
\noalign{\smallskip}
 SR3Lhigh & 3L &  -    & $N_{\rm jets} \ge$ 4, \met $>$ 150~\GeV, SR3b veto & \meff $>$400~\GeV\\
\hline\noalign{\smallskip}
\end{tabular} 
}
\end{center}
\caption{Definition of the signal regions (see text for details).}
\label{tab:SRDefn}
\end{table}
 
As detailed in table~\ref{tab:SRDefn}, the selection requirements of the five signal regions are:
\begin{itemize}
\item \textbf{SR3b}: SS or 3L events with at least five jets and at
  least three $b$-jets;
\item \textbf{SR0b}: SS events with at least three jets, zero
  $b$-jets, 
  large \met\ and large \mt; 
\item \textbf{SR1b}: similar to SR0b, but with at least one $b$-jet;
\item \textbf{SR3Llow}: 3L events with at least four jets, small \met\
  and $Z$ boson veto; 
\item \textbf{SR3Lhigh}: 3L events with at least four jets and large \met.
\end{itemize}
The $Z$ boson veto in SR3Llow rejects events with any opposite-charge
 same-flavour lepton combination of invariant mass 84 $ < m_{\ell\ell}<$
 98~\GeV. 
An additional \meff\ requirement is applied to maximise the expected significance of selected 
SUSY models in each signal region. 
This requirement on \meff\ is relaxed in the model-dependent limit-setting
procedure described in section~\ref{sec:interpretModelDep}.
The signal regions are all mutually exclusive. An SR3b veto, which
rejects events satisfying the SR3b selection, is included
in the definition of other signal regions that would otherwise have a small overlap with SR3b.

Each signal region is motivated by different SUSY scenarios and different SUSY
parameter settings.
The SR3b signal region targets gluino-mediated top squark scenarios resulting in 
signatures with four $b$-quarks.
This signal region does not require large values of  \met\ or \mt,
hence it is sensitive to compressed scenarios with small mass differences or to unstable LSPs.
The SR0b signal region is sensitive to gluino-mediated and directly produced squarks of
the first and second generations,
which do not enhance the production of $b$-quarks. 
Third-generation squark models resulting in 
signatures with two $b$-quarks, such as direct bottom squark or 
gluino-mediated top squark $\to c\ninoone$ production, are targeted by
SR1b.
The 3L signal regions have no requirement on the number of $b$-jets.
They target scenarios where squarks decay in multi-step cascades, such as
gluino-mediated (or direct) squark $\to q^{\prime} WZ\ninoone$ and
gluino-mediated (or direct) squark $\to$ sleptons (see figure \ref{fig:SUSYmodels}).
The signal region with low \met\ requirement, SR3Llow, targets compressed regions of the
phase space where SUSY decay cascades would produce off-shell $W$ and $Z$ bosons.
Backgrounds from $Z$ boson production in association with jets are
suppressed by a $Z$ boson veto.
Models with large \met\ and on-shell vector bosons are targeted by SR3Lhigh.
Hence no $Z$ boson veto is applied in this signal region, but $Z$ + jets backgrounds are 
suppressed by the larger \met\ requirement.

\section{Background estimation}
\label{sec:bkgd}

Searches in SS and 3L events are
characterised by low SM backgrounds. Three main classes of
backgrounds can be distinguished. They are, in decreasing order of importance
for this search: (1) prompt multi-leptons, (2) ``fake'' leptons, which denotes hadrons
mis-identified as leptons, leptons originating from heavy-flavour
decays, and electrons from photon conversions, and (3) charge
mis-measured leptons.

\subsection{Background estimation methods}
\label{sec:bkgmet}

\subsubsection{Prompt lepton background}
\label{sec:promptbkgd}
The background with prompt leptons arises
mainly from $W$ or $Z$ bosons, decaying
leptonically, produced in association with a top--antitop quark pair where at least one of the
top quarks decays leptonically, and from diboson processes ($WZ$,
$ZZ$, $W^\pm W^\pm$) in association with jets. The $\ttbar V$ and
diboson 
backgrounds are dominant for signal regions with and without $b$-jets,
respectively. The prompt multi-lepton backgrounds are estimated from
Monte Carlo samples normalised to NLO calculations as described in
section~\ref{sec:mc}.  The rarer processes $\ttbar H$, single top quark
plus a $Z$ boson, $\ttbar\ttbar$ and
$VVV$+jets, each of which constitutes at most 10\% of the background in
the signal regions, are also included. The production of $\ttbar WW$, 
$WH$ and $ZH$ (where the Higgs boson decay can produce isolated leptons
from $W$, $Z$ or $\tau$) were verified to give a negligible contribution
to the signal regions.

\subsubsection{Fake-lepton background}
\label{sec:fake}

The number of events with at least one fake lepton is
estimated using a data-driven method. 
A fake-enriched class of ``loose'' leptons is introduced,
composed of preselected leptons (defined in
section~\ref{sec:objectdefn}) with $\pt > 15 \gev$ failing the signal lepton
selection. 
If the ratio of the number of signal leptons to the number of loose leptons
is known separately for prompt and fake leptons, the number of events
with at least one fake lepton can be predicted. For illustration, when only pairs of
leptons are considered, the equation that relates the number
of events with signal ($S$) or loose ($L$) leptons to the number of
events with prompt ($P$) or fake ($F$) leptons: 

\begin{align}
\left(\begin{array}{c}
N_{SS} \\  N_{SL} \\ N_{LS} \\ N_{LL}
\end{array}\right) = 
\Lambda \cdot 
\left(\begin{array}{c}
N_{PP} \\  N_{PF} \\ N_{FP} \\ N_{FF}
\end{array}\right), 
\label{eq:mxm_start}
\end{align}
where the first and second indices refer to the leading and subleading
lepton of the pairs,
can be inverted to obtain the expected number of events with at least
one fake lepton.
The matrix $\Lambda$ is given by
 \begin{align}
\Lambda=
\left(\begin{array}{cccc}
\varepsilon_1\varepsilon_2 & \varepsilon_1\zeta_2 & \zeta_1\varepsilon_2 & \zeta_1\zeta_2\\
\varepsilon_1(1-\varepsilon_2) & \varepsilon_1(1-\zeta_2) & \zeta_1(1-\varepsilon_2) & \zeta_1(1-\zeta_2)\\
(1-\varepsilon_1)\varepsilon_2 & (1-\varepsilon_1)\zeta_2 & (1-\zeta_1)\varepsilon_2 & (1-\zeta_1)\zeta_2\\
(1-\varepsilon_1)(1-\varepsilon_2) & (1-\varepsilon_1)(1-\zeta_2) & (1-\zeta_1)(1-\varepsilon_2) & (1-\zeta_1)(1-\zeta_2)
\end{array}\right),
\label{eq:lambda}
\end{align}
where $\varepsilon_1$ and $\varepsilon_2$ ($\zeta_1$ and
$\zeta_2$) are
the ratios of the number of signal and loose leptons for the leading
and subleading prompt (fake) leptons, respectively.
This analysis employs a generalised matrix method where an arbitrary
number of loose leptons can be present in the event.  
For example, an event containing three 
leptons that pass, in decreasing order of $\pt$, the
signal--loose--signal selections is considered
a SS signal event if the first and third lepton have the same charge. In addition,
this event is included in the fake-lepton background calculation for
3L events since the second lepton passes only the loose
selections. In general, eqs.~\ref{eq:mxm_start}-\ref{eq:lambda} are
adapted by dynamically adjusting 
the size of the matrix $\Lambda$ according to the number of loose leptons
in the event under study. No upper limit on the number of loose leptons is
set. Each event is employed in all its possible incarnations (signal
and/or as part of the
background calculation) as illustrated in the example
above, but is only included in one of the signal regions, which are exclusive
by definition.

The efficiencies $\varepsilon$ and $\zeta$ are measured in data as a function
of the lepton $\pT$ 
and  $\eta$. The prompt lepton efficiencies are
determined from a data sample enriched with prompt leptons from
$Z\rightarrow \ell^+\ell^-$ decays, obtained by requiring $80 <
m_{\ell\ell}<100\GeV$. 
As the background is dominated by events with one real lepton and one
fake lepton, the fake-lepton efficiencies  are  measured from a data
set enriched 
with one prompt muon (by requiring it to pass the signal lepton
selection and $\pt>40\GeV$) and an 
additional fake lepton (by requiring it to pass the loose
selections). 
The fake electron background has contributions from heavy
flavour decays, as well as from conversions and fake pions.  The
fake-electron efficiency is therefore determined from two samples of SS $e\mu$
events to be sensitive to the different types of fake electrons,
one  with a $b$-jet veto and another with at least one  $b$-jet.
The fake-muon efficiency is determined from a
sample of same-sign dimuon events where at least two jets with $\pt > 25$ GeV are required. 
The event yields in these control regions are corrected for the
contamination of prompt SS using Monte Carlo 
simulation. The $e\mu$ SS control regions are also corrected for the presence of
charge mis-measured electrons using the likelihood fit method described
in Sec~\ref{sec:bkgflip}, but applied to loose electrons. The
contamination from signal events is verified to be
negligible in the same-sign $e\mu$ and $\mu\mu$ control regions. 
The size of the data sample is not sufficient to allow the extraction
of the fake-lepton efficiencies for muons
with $\pt > 40$ GeV or for events with at least three $b$-jets. For
these events the
fake-lepton efficiencies obtained from data in similar kinematic regions, i.e. muons
with $25 <\pt<40\GeV$ or events with at least one $b$-jet,
 are employed and corrected with extrapolation factors
obtained from the $\ttbar$ Monte Carlo samples.

\subsubsection{Background from lepton charge mis-measurement}
\label{sec:bkgflip}
Background from charge mis-measurement, commonly referred to as ``charge-flip'', consists of events with two
opposite-sign leptons for which the charge of a lepton is
mis-identified.  Such events 
constitute a background only for the SS signal regions.
The dominant mechanism of charge mis-identification is due to the
radiation of a hard photon from an electron followed by an asymmetric
conversion, for which the electron with the opposite charge has the
larger $\pt$
($e^{\pm} \rightarrow e^{\pm} \gamma \rightarrow e^{\mp} e^{\pm} e^{\pm}
$).  The probability of mis-identifying the charge of a muon is
determined in simulation to be negligible in the
kinematic range relevant to this analysis. The electron charge-flip background is
estimated using a fully data-driven technique. The charge-flip
probability is extracted in two $Z$ boson control samples, one with
same-sign electron pairs
and the other with opposite-sign electron pairs. The invariant mass of
these same-sign and opposite-sign electron pairs
is required\footnote{An asymmetric
  window around the $Z$ boson mass is chosen because charge-flip electrons lose more energy
  in the detector
  than electrons for which the charge is properly
  reconstructed.} to be between 75 GeV and 100 GeV. Background events are subtracted using the
invariant mass sidebands. A likelihood fit is employed which
takes as input the numbers of same-sign and opposite-sign electron pairs observed in the
sample. The charge-flip probability is a free parameter of the fit and
 is extracted as
a function of the electron $\pt$ and $\eta$.  The
probability of electron charge-flip varies from
approximately $10^{-4}$ to $10^{-2}$ in the range $0\le |\eta|\le 2.47$ and
$15<\pt< 200$~\GeV, increasing with electron $|\eta|$ and $\pt$. The
event yield of this background in the signal regions is 
obtained by applying the measured charge-flip probability to data regions with
the same kinematic requirements as 
the signal regions but with opposite-sign lepton pairs. The contamination
from fake leptons and signal events is found to be negligible in these
opposite-sign control regions.

\subsection{Systematic uncertainties on the background estimation}
\label{sec:bkgsys}

The systematic uncertainties on the sources of prompt SS and 3L events
arise from the Monte Carlo simulation and normalisation of these processes.
The cross sections used 
to normalise the Monte Carlo samples 
are varied according to the uncertainty on the theory calculation, i.e. $22\%$ for $t\bar{t}W$ \cite{Campbell:2012dh} 
and $t\bar{t}Z$ \cite{Garzelli:2012bn} and $7\%$ for
diboson production (computed with MCFM \cite{Campbell:2011bn}, considering
scales, parton distribution functions and $\alpha_s$
uncertainties). Normalisation uncertainties between 35\% and 100\% are
    applied to processes with smaller contributions.
Uncertainties caused by the limited accuracy of the $\ttbar V$+jets
and diboson+jets Monte Carlo generators are estimated by varying the renormalisation and factorisation
scales and the QCD initial- and final-state radiation used to generate these samples.
Additional Monte Carlo modelling uncertainties are included,
such as the
limited number of hard jets that can be produced from matrix element calculations
in the  {\sc MadGraph}+{\sc Pythia} and {\sc Sherpa}  samples, which
is the largest modelling uncertainty for the diboson+jets process, and
the difference between the predictions of various Monte Carlo
generators such as {\sc MadGraph} versus {\sc Alpgen},
which is the largest modelling uncertainty for the $\ttbar V$+jets process.

Monte Carlo simulation-based estimates also suffer from
detector simulation uncertainties. These are dominated by the uncertainties on the jet energy scale and the
$b$-tagging efficiency. The jet energy scale uncertainty is derived using a combination of simulations,
test beam data and {\it in situ} measurements \cite{Aad:2011he,Aad:2012vm}. Additional contributions
from the jet flavour composition, calorimeter response to different jet flavours,
pile-up and $b$-jet calibration uncertainties are taken into
account. The efficiency to tag real and fake $b$-jets is corrected in
Monte Carlo events by applying $b$-tagging scale factors, extracted in 
$\ttbar$ and dijet samples,
that compensate for the residual difference between data and simulation
\cite{ATLAS-CONF-2014-004,CON-2012-043, CONF-2012-040}.
The associated systematic uncertainty is computed by varying the scale factors 
within their uncertainty. 
Uncertainties in the jet energy resolution are
obtained with an {\it in situ} measurement of the jet response asymmetry 
in dijet events \cite{JER}. Other uncertainties on the lepton
reconstruction \cite{Aad:2011mk,ATLAS-CONF-2013-088}, 
calibration of calorimeter energy clusters not associated with physics objects 
in the $\met$ calculation \cite{Aad:2012re},  luminosity~\cite{Aad:2013ucp} and
simulation of pile-up events are included but have a negligible impact
on the final results.

The fake-lepton background uncertainty includes the
statistical uncertainty from the SS control regions, the dependence of
the fake-lepton efficiency
on the event
selections and the contamination of the SS control regions by real
leptons. Uncertainties on the extrapolation of the fake-lepton efficiency to
 poorly populated kinematic regions 
are estimated by comparing the prediction of different $\ttbar$ Monte
Carlo samples. For the charge-flip background
prediction, the main uncertainties originate from the statistical
uncertainty of the charge-flip probability measurements and the
background contamination of the sample used to extract the charge-flip
probability. 

\subsection{Cross-checks of the data-driven background estimates}
\label{sec:bkgdcrosscheck}

\begin{table}[tb]
\begin{center}
\resizebox{\textwidth}{!}{
\begin{tabular}{llccccc}
\hline\noalign{\smallskip}
\hline\noalign{\smallskip}
Background &  Method & SR3b &  SR0b &  SR1b & SR3Llow & SR3Lhigh \\
\hline\noalign{\smallskip}
\multirow{2}{*}{Charge-flip} & Nominal  & $0.2 \pm 0.1$ & $0.2 \pm 0.1$ & $0.5 \pm 0.1$ & $-$ & $-$ \\
            & Tag and probe  & $0.2 \pm 0.1$ &  $0.2 \pm 0.1$ & $0.5 \pm 0.2$ & $-$ & $-$ \\
\hline\noalign{\smallskip}
\multirow{2}{*}{Fake} & Nominal  & $0.7 \pm 0.6$ & $1.2^{+1.5}_{-1.2}$
& $0.8^{+1.2}_{-0.8}$ & $1.6 \pm 1.6$ &  $< 0.1$ \\
            & Monte Carlo based  & $2.0^{+1.4}_{-1.3}$  &  $5 \pm 5$ &
            $0.6^{+1.4}_{-0.6}$ &
            $1.0^{+0.8}_{-0.7}$ & $< 0.1$ \\
\hline\noalign{\smallskip}
\multirow{2}{*}{Total 3 $b$-jets} & Nominal  & $2.1 \pm 0.7$ & $-$ & $-$ &$-$ &$-$ \\
            & $b$-jets matrix method  & $2.9 \pm 0.9$ & $-$ & $-$ &$-$ &$-$ \\
\hline\noalign{\smallskip}
\end{tabular} }
\end{center}
\caption{Comparison of the predicted number of background events in the signal regions using the nominal
  and cross-check methods. Both the statistical
  and systematic uncertainties are included.}
\label{tab:cross-check}
\end{table}

Three alternative methods were developed to cross-check the
background estimates from data-driven methods.  The results are summarised in
table~\ref{tab:cross-check}, showing the background predictions for
the nominal methods, described in section~\ref{sec:bkgmet}, and the cross-check methods
described below. In each case consistent predictions are obtained, 
 but with generally larger uncertainties for the alternative methods.

For the electron charge-flip
background, a simpler ``tag and probe''  method is employed which selects
electron pairs with an invariant
mass consistent with a $Z$ boson decay.
One electron is required to have $|\eta| 
< 1.37$. Its charge is assumed to be measured correctly. The charge-flip
probability is extracted as a function of
$\pt$ and $\eta$ of the other electron, which is required to be in the pseudorapidity region
$1.52 < |\eta| 
< 2.47$, by computing the ratio of same-sign to opposite-sign pairs. The charge-flip
probability for central electrons is
extracted by requiring that both electrons are in
the same $\pt$ and $\eta$ region. This charge-flip probability is
applied in the same manner as the nominal charge-flip probability, as
described in section~\ref{sec:bkgmet}, to obtain a
prediction in the signal regions.

The fake-lepton background estimate were cross-checked with a
simulation-based technique. This method relies on kinematic extrapolation from
control regions, with low jet multiplicity and \met, to the
signal regions that require high jet multiplicity and \met.  
The separate control regions are characterised by the presence or the
absence of a $b$-jet, and by the
flavours of the two leading leptons.
Backgrounds with prompt leptons are obtained 
from Monte Carlo simulation as described in section~\ref{sec:promptbkgd}. 
Backgrounds with fake
leptons and charge-flip electrons are obtained from Monte Carlo
simulations normalised to match data in the control regions. The
normalisation is done using five multipliers. One
multiplier is used to correct the rate of electron charge
mis-identifications. The other four corrections are for processes producing either
fake electrons or muons that originate from $b$-jets or light
jets.

The background in the SR3b region is expected to be completely
dominated by events with at least 
one light or charm jet mis-tagged as a $b$-jet, i.e. a fake $b$-tag.
A cross-check of the background estimate in this signal region is performed by 
determining the number of events with at least one fake $b$-tag. 
A generalised matrix method applied to the estimation of fake $b$-tags
is used, similar to that described in section~\ref{sec:fake}, with the following differences. Loose leptons are replaced
by jets, signal leptons by $b$-tagged jets, and the different
tight/loose incarnations are combined in each event. 
The efficiency  for fake $b$-tags is estimated in a
$\ttbar$-enriched sample with at least one signal lepton, at least
four jets with 
$\pt > 20$ GeV, of which at least two must be $b$-tagged, and $100 
< \met < 200 \GeV$. The efficiency  for fake $b$-tags is calculated using the
additional $b$-jets
found in each event after subtracting contamination from events with
three or more real $b$-jets (such as $\ttbar \bbbar$). The efficiency to tag real
$b$-jets is determined independently of the efficiency  for fake $b$-tags,
as described in refs. \cite{ATLAS-CONF-2014-004,CON-2012-043}. 
The efficiencies for tagging real and fake $b$-jets are fed into the matrix
method to predict the background in SR3b. Small contributions from
processes with three real $b$-jets are estimated from simulation.

\subsection{Validation of background estimates}
\label{sec:bkgdval}

\begin{figure}
\addtolength{\subfigcapskip}{-10pt}
\begin{minipage}[b]{0.5\linewidth}
\centering 
\subfigure[\label{fig:valid:a}]{
\includegraphics[width=\textwidth]{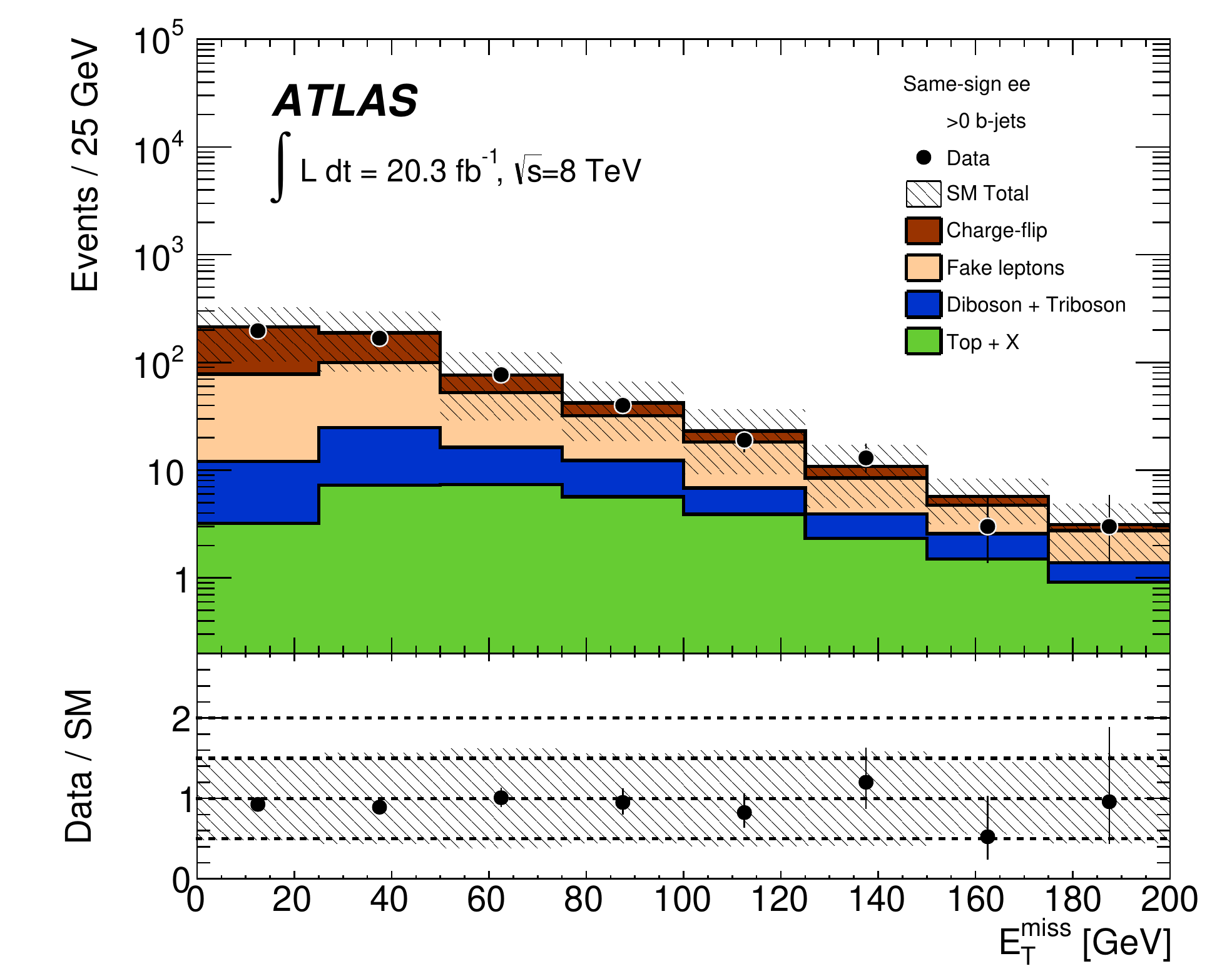} 
}
\end{minipage}
\begin{minipage}[b]{0.5\linewidth}
\centering 
\subfigure[\label{fig:valid:b}]{
\includegraphics[width=\textwidth]{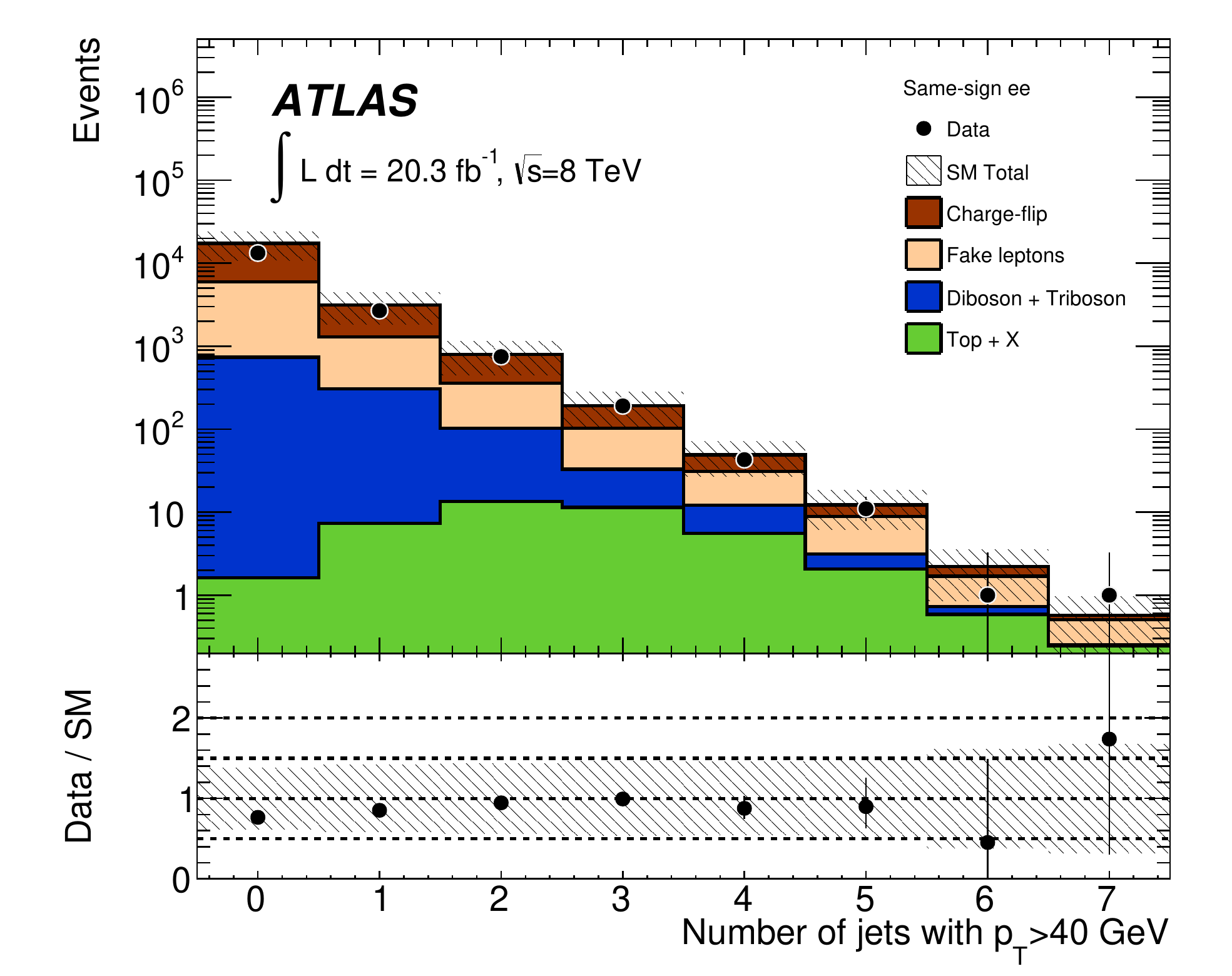} 
}
\end{minipage}
\begin{minipage}[b]{0.5\linewidth}
\centering 
\subfigure[\label{fig:valid:c}]{
\includegraphics[width=\textwidth]{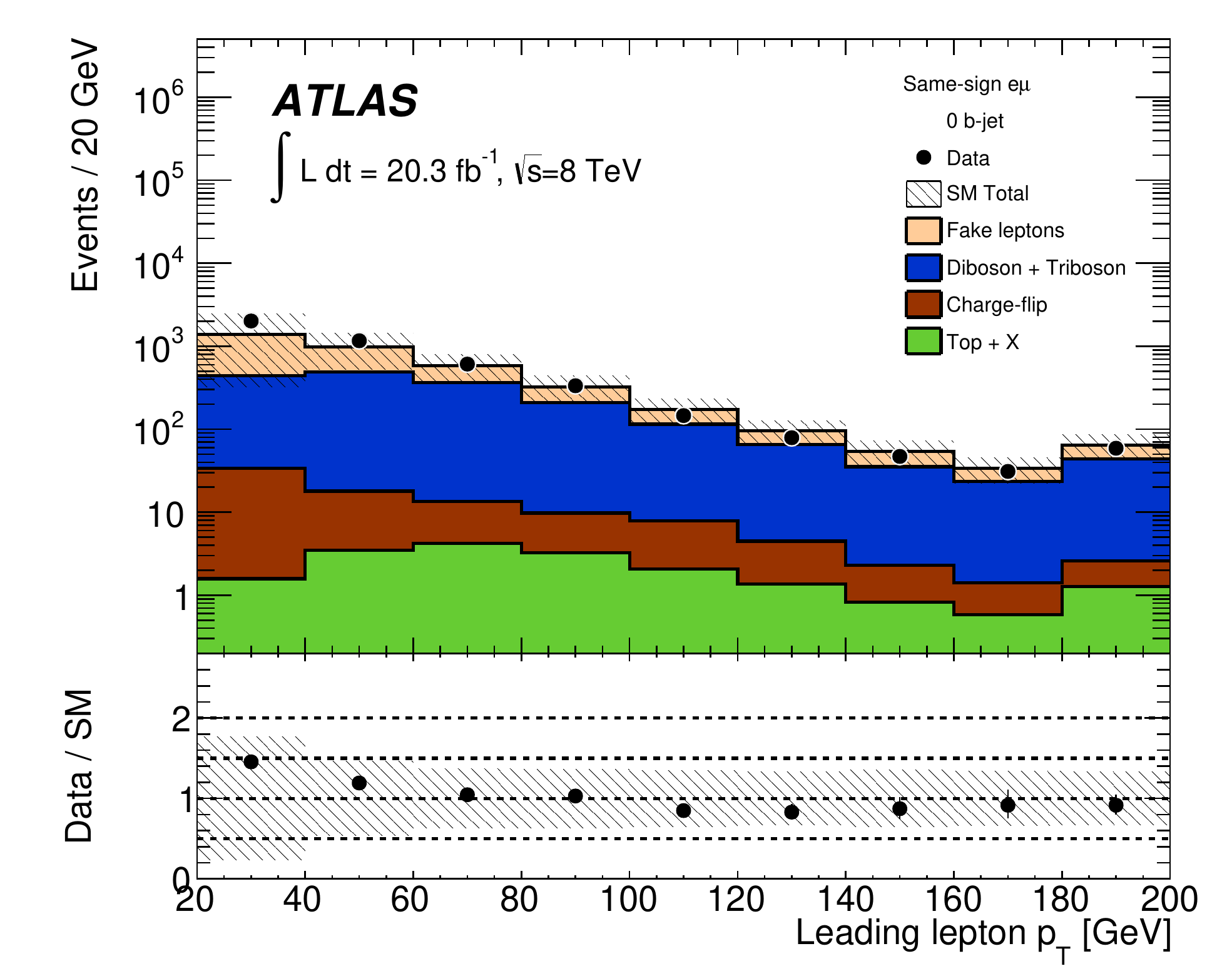}
}
\end{minipage}
\begin{minipage}[b]{0.5\linewidth}
\centering 
\subfigure[\label{fig:valid:d}]{
\includegraphics[width=\textwidth]{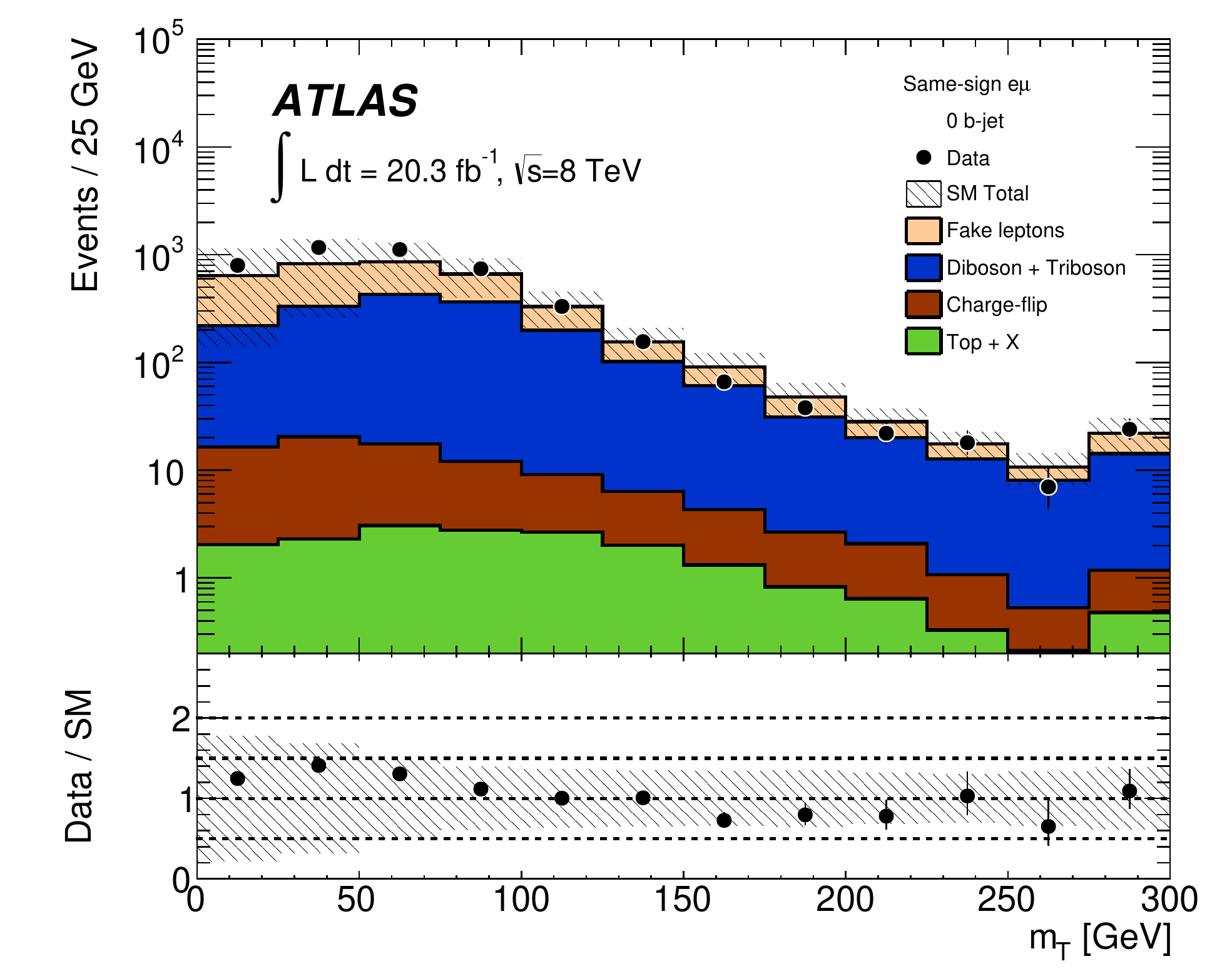}
}
\end{minipage}
\begin{minipage}[b]{0.5\linewidth}
\centering 
\subfigure[\label{fig:valid:e}]{
\includegraphics[width=\textwidth]{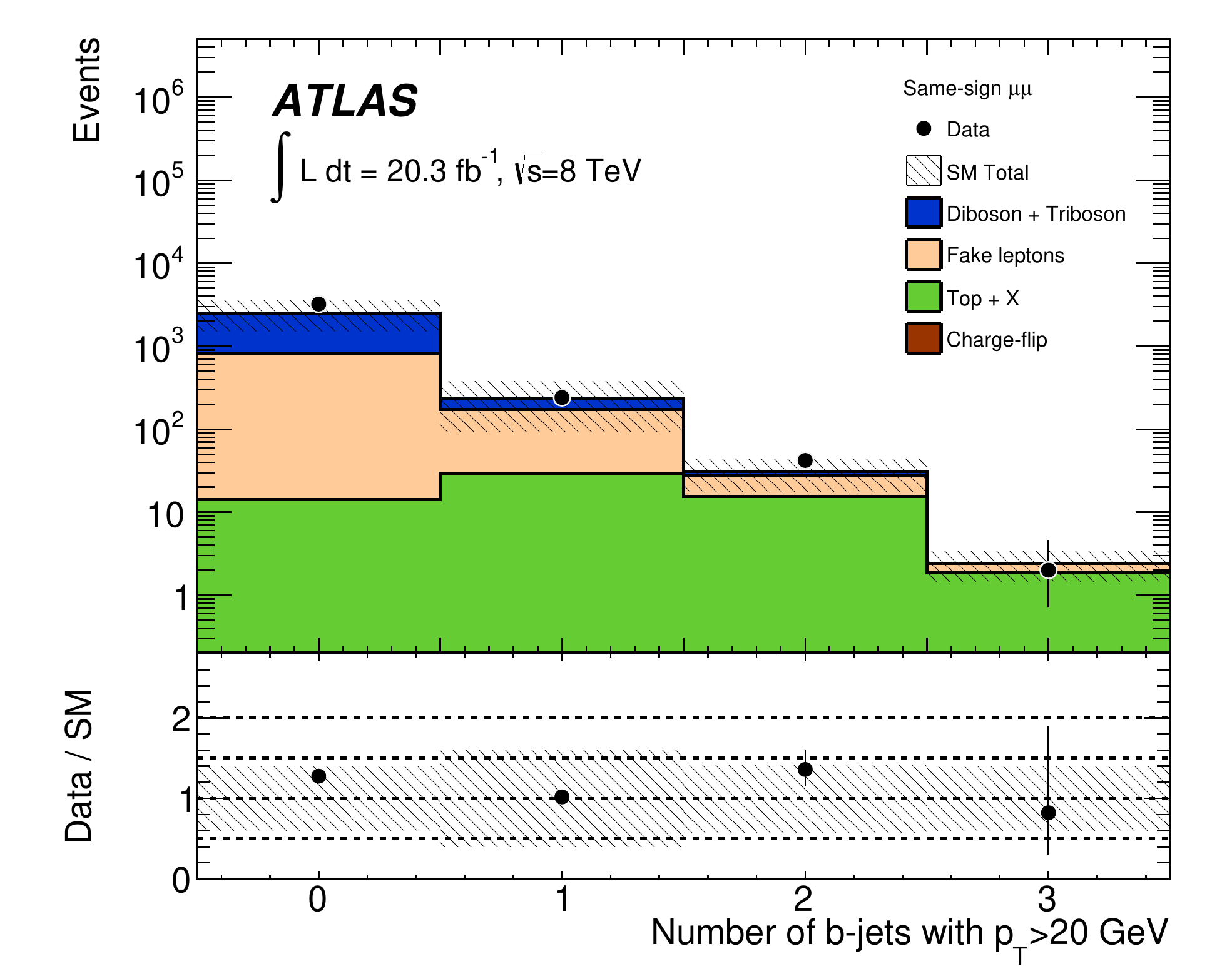} 
}
\end{minipage}
\begin{minipage}[b]{0.5\linewidth}
\centering 
\subfigure[\label{fig:valid:f}]{
\includegraphics[width=\textwidth]{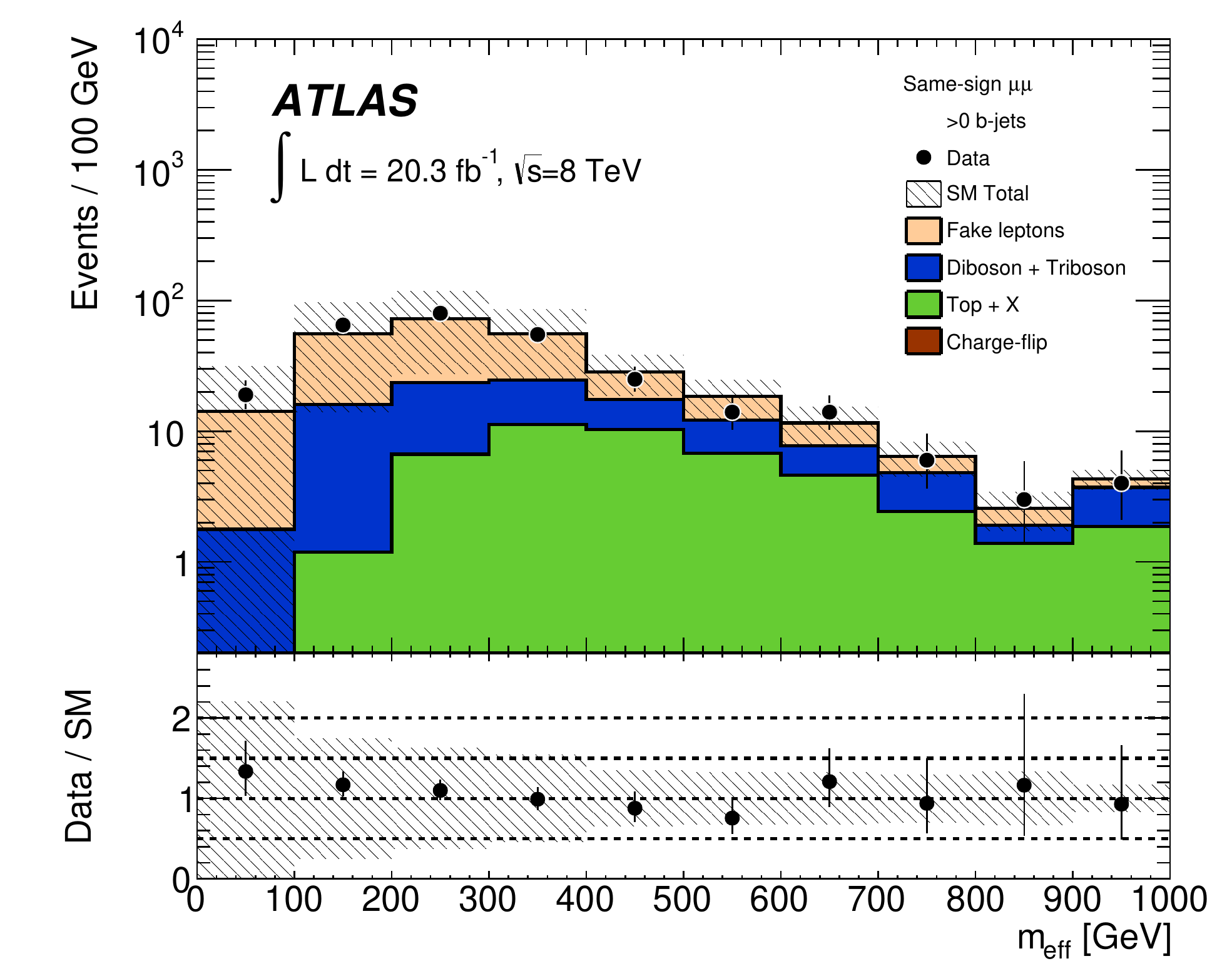} 
}
\end{minipage}
\caption{Distributions of kinematic variables in SS background
  validation regions: (a) $\met$ for events with at least one
  $b$-jet and (b) number of jets 
for the $ee$ channel, (c) leading lepton
$\pt$ for events with no
  $b$-jet and (d) transverse mass, $\mt$, for events with no
  $b$-jet
for the $e\mu$ channel, and (e) number of $b$-jets and (f) effective
mass, $\meff$, for events with at least one
  $b$-jet for the $\mu\mu$ channel. The statistical and
systematic uncertainties on the background prediction are included in the uncertainty band.  The last bin includes overflows.  The lower part of the figure shows
the ratio of data to
          the background prediction. }
\label{fig:valid}
\end{figure}

The data-driven background estimates are based on
control regions that employ less stringent requirements on the jet
and $b$-jet
multiplicities, total transverse energy and/or $\met$ than the signal regions. To ensure 
their validity in the signal regions, the background estimates are
validated in events with 
kinematic properties closer to the signal regions. This is first performed by individually probing
each of the kinematic variables used to define the signal
regions in events containing a same-sign lepton pair. The event is not
rejected if it contains more than two leptons.
Several relevant kinematic distributions are
studied for each lepton channel and for events with and without a $b$-jet.
No significant discrepancies are
observed. Some example distributions are shown in
figure~\ref{fig:valid}.

\begin{table}[tb]
\begin{center}
\resizebox{\textwidth}{!}{
\begin{tabular}{lcccccc}
\hline\noalign{\smallskip}
\hline\noalign{\smallskip}
 Background &  Leptons &  $N_{\rm jets}$  & $N_{b- \rm jets}$  & $\met$ (GeV) &
 $\mt$ (GeV) & Additional cuts \\
 Probed  & ($\pt > 20 \gev$) & & &  & & \\
\hline\noalign{\smallskip}
$\ttbar W$ & SS $\mu\mu$ & $\ge 1$ (30 GeV) & = 2 & 20
to 120  & $> 80$  & $-$ \\
$\ttbar Z$ & 3L & $\ge 2$ (40 GeV) & 1 or 2 & 20
to 120 & $-$ & $\meff >$ 300 GeV, \\
  & & & &  & & $Z$ boson mass \\
$WZ$+jets & SS $\mu\mu$ & $\ge 2$ (20 GeV) & Veto & 20
to 120 & $> 100$ & $-$ \\
\hline\noalign{\smallskip}
\end{tabular} }
\end{center}
\caption{Definition of the validation regions for rare SM backgrounds. 
The required jet $\pt$ threshold is indicated in parentheses under
the column $N_{jets}$. The $Z$ boson mass cut demands at least one
opposite-charge same-flavour lepton pair satisfying 84 $ < m_{\ell\ell} <$ 98 GeV.}
\label{tab:VR}
\end{table}
\begin{figure}
\addtolength{\subfigcapskip}{-10pt}
\begin{minipage}[b]{0.5\linewidth}
\centering 
\subfigure[\label{fig:resVR:a}]{
\includegraphics[width=\textwidth]{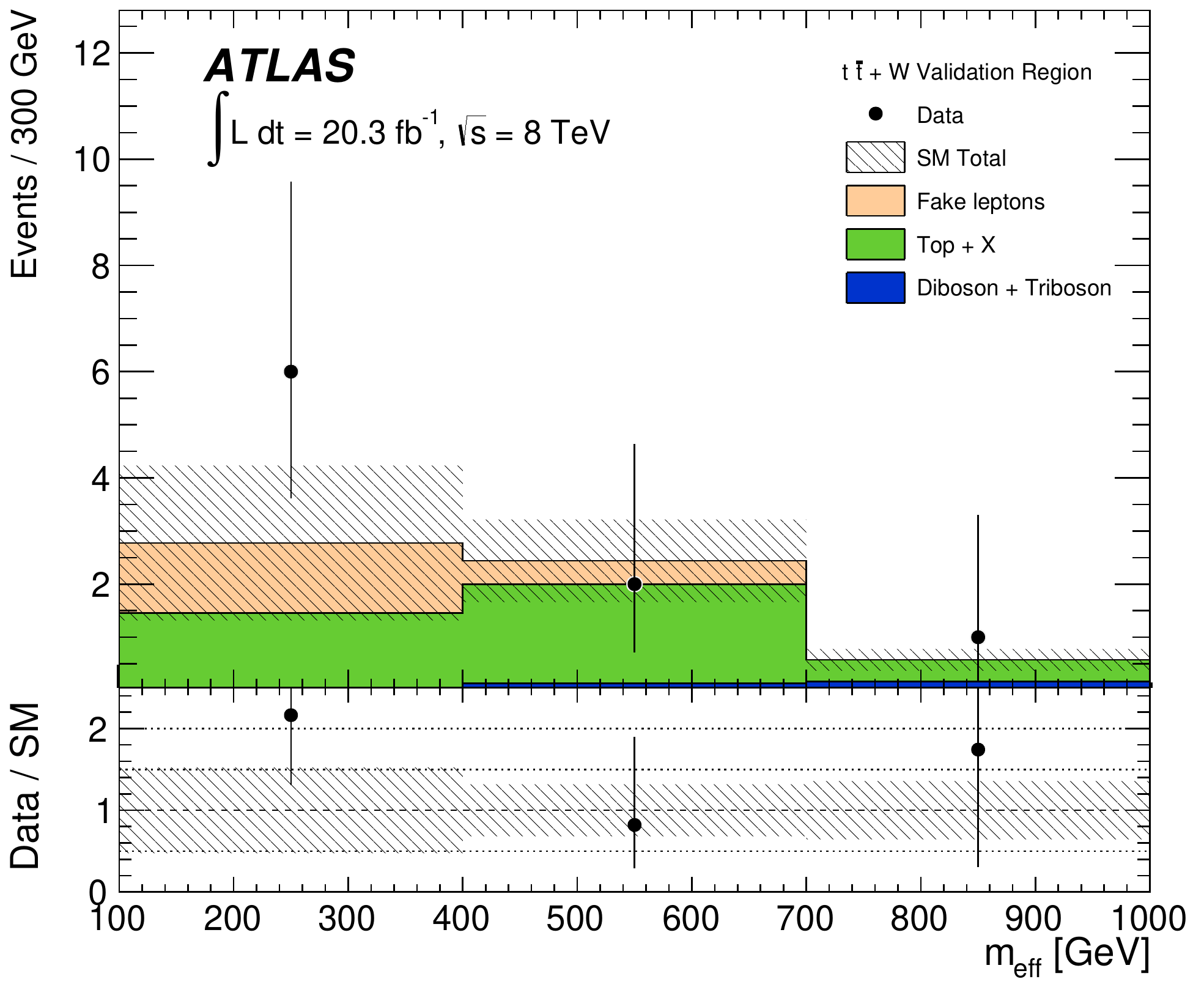}
}
\end{minipage}
\begin{minipage}[b]{0.5\linewidth}
\centering 
\subfigure[\label{fig:resVR:b}]{
\includegraphics[width=\textwidth]{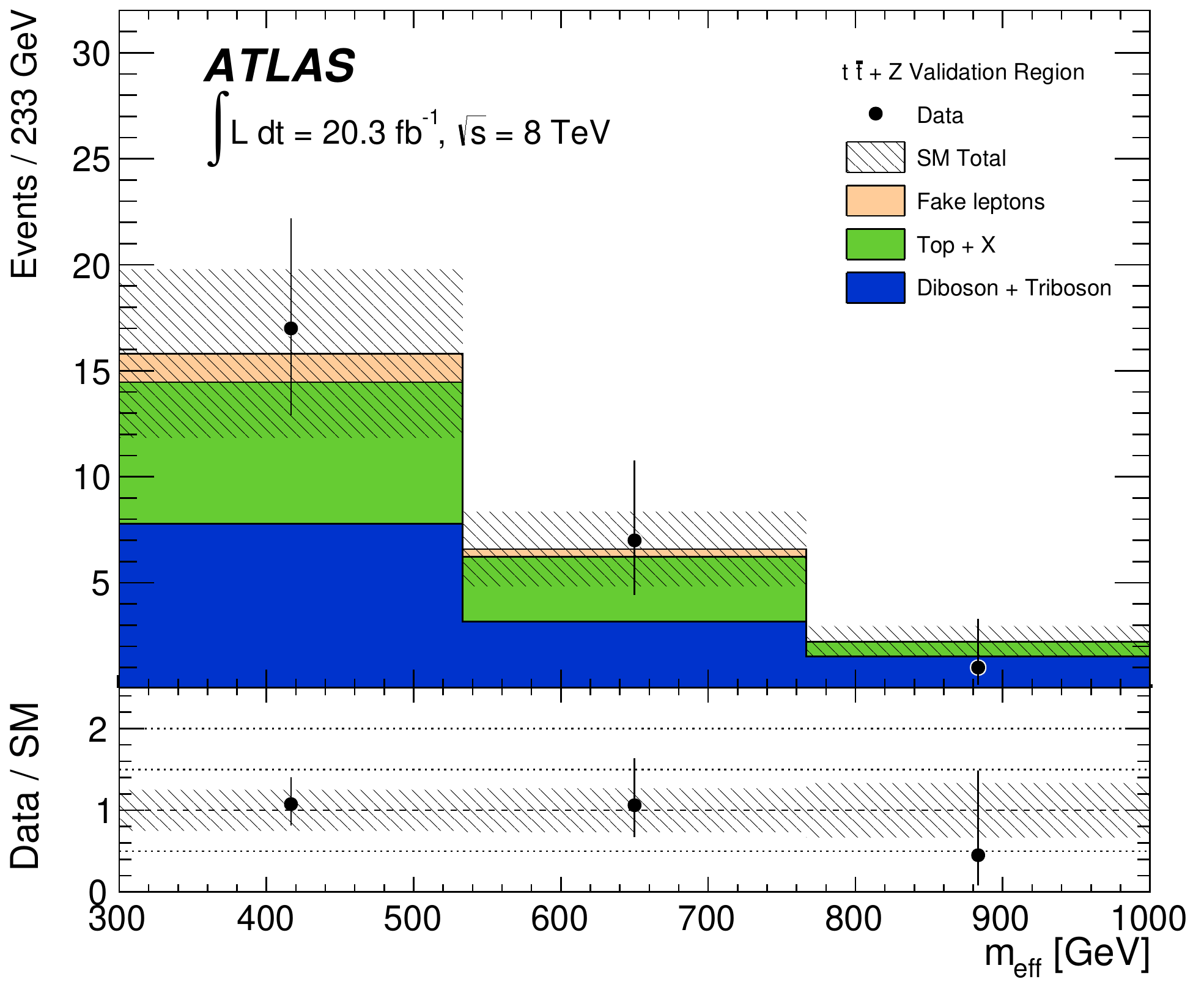}
}
\end{minipage}
\begin{minipage}[b]{0.5\linewidth}
\centering 
\subfigure[\label{fig:resVR:c}]{
\includegraphics[width=\textwidth]{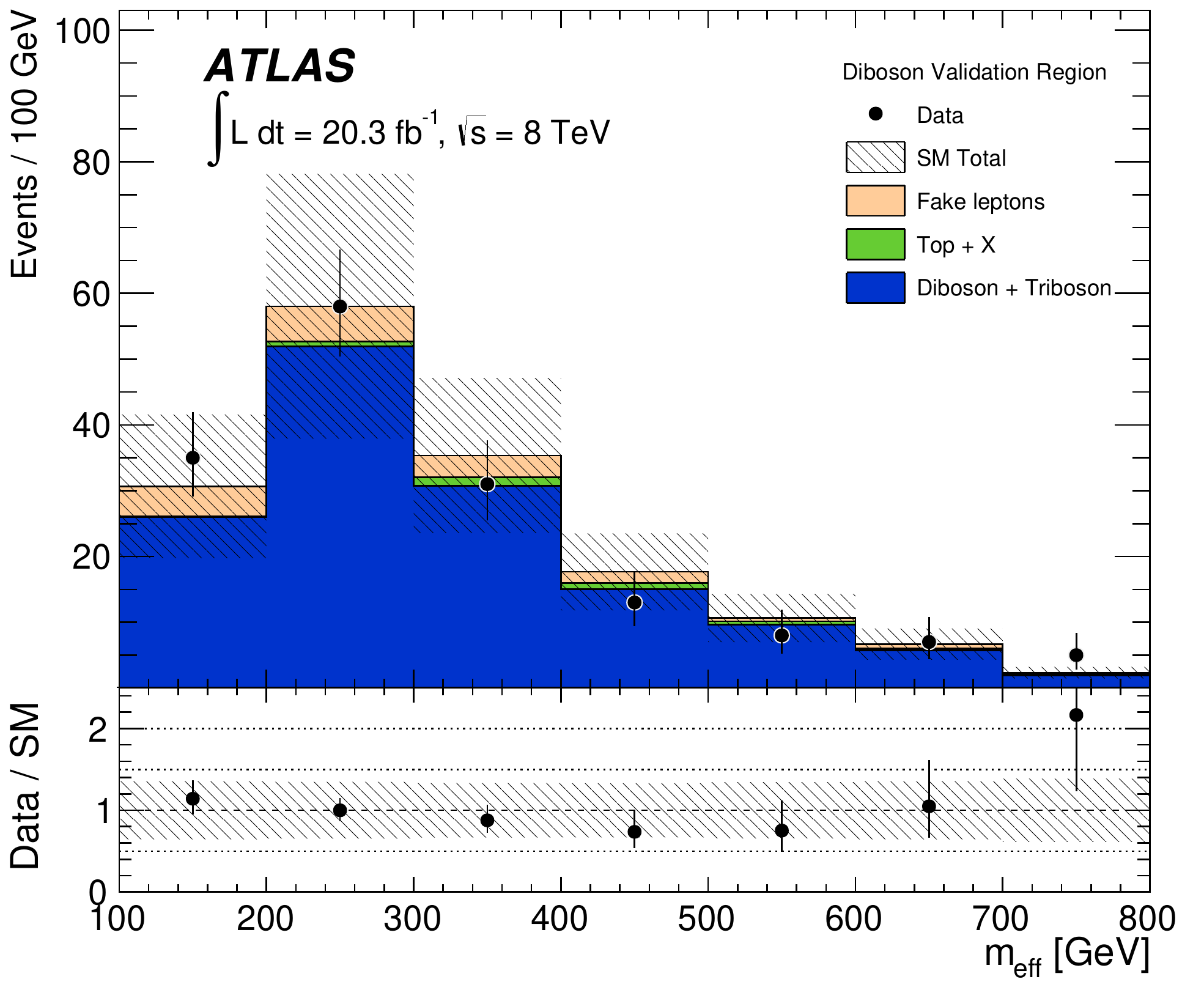}
}
\end{minipage}
\begin{minipage}[b]{0.5\linewidth}
\centering 
\subfigure[\label{fig:resVR:d}]{
\includegraphics[width=\textwidth]{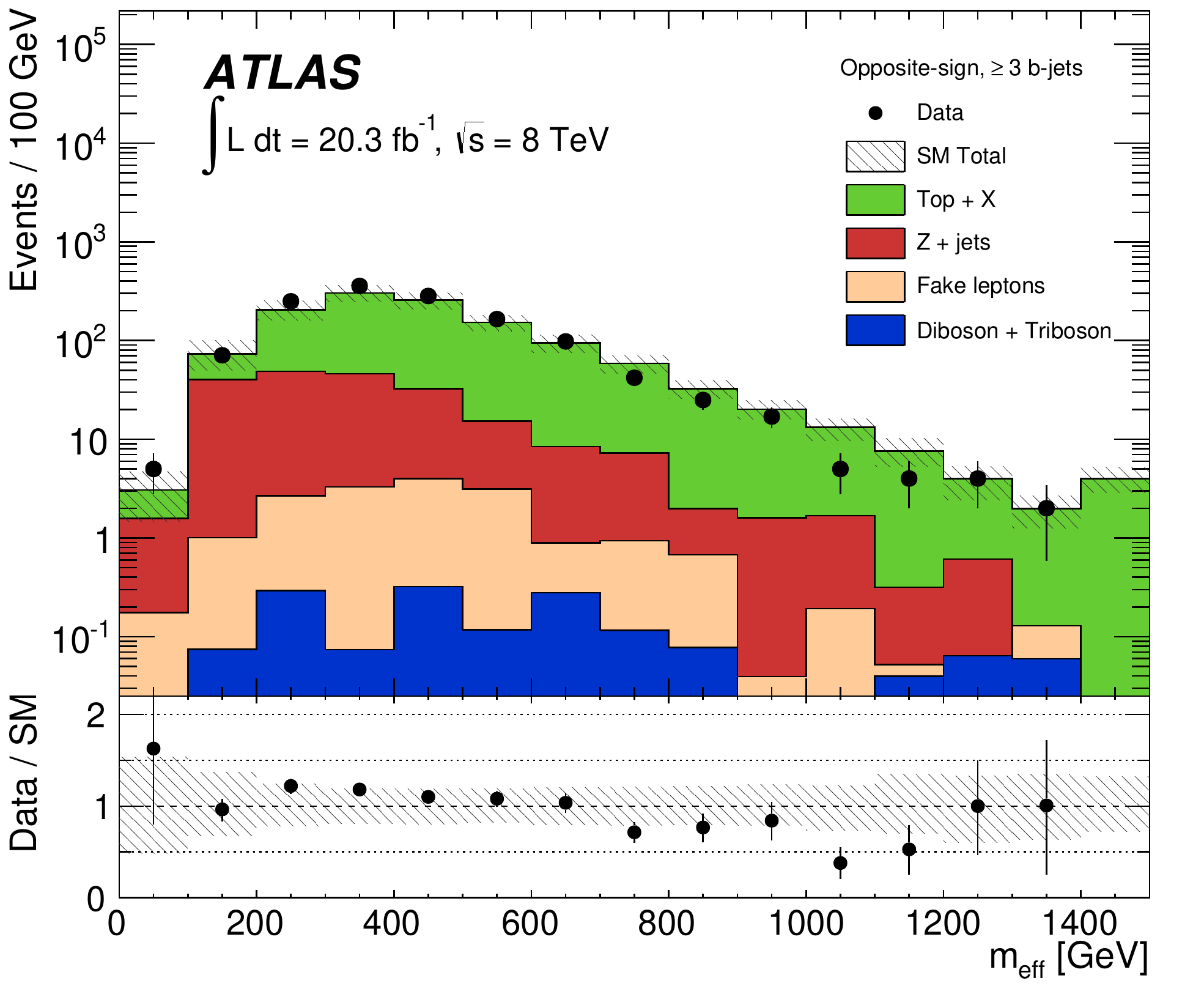}
}
\end{minipage}
\caption{Effective mass ($\meff$) distributions for the (a)
$\ttbar W$, (b) $\ttbar Z$, (c) $WZ$+jets
 and (d) OS plus three $b$-jets validation regions. The statistical and
systematic uncertainties on the background prediction are included in the uncertainty band.
The last bin includes overflows.  The lower part of the figure shows
the ratio of data to
          the background prediction.  }
\label{fig:resVR}
\end{figure}

Each of the background types (fake electron, fake
muon, charge-flip electron and prompt SS) is dominant, and thus validated directly, 
in particular regions of the kinematic phase space examined by
these SS validation regions. However, the prompt SS contributions are typically
dominated by inclusive $WZ$ production, while the prompt SS or 3L background in
the signal regions is expected to be dominated by $\ttbar V$ and
$WZ$ events produced in association with several hard
jets. The Monte Carlo
modelling of these rare processes is tested in a further set of dedicated
validation regions. The event selections are presented in
table~\ref{tab:VR}. They are based on the object
definitions described in section~\ref{sec:objectdefn}, and impose different
jet $\pt$ thresholds and require $\pt > 20 \gev$ for the
leptons to increase the rejection of fake-lepton events. 
The $\ttbar W$ and $WZ$+jets validation regions employ only SS
$\mu\mu$ events to avoid fake-electron events. The signal
contamination is verified to be negligible for the $\ttbar Z$ and $WZ$+jets
validation regions and at most $25\%$ for the $\ttbar W$
validation region for non-excluded SUSY models. The $\meff$
distributions of these validation regions are shown in
figures~\ref{fig:resVR:a}--\ref{fig:resVR:c}. The prediction 
 is observed to agree with the data, therefore validating the Monte Carlo modelling of these
rare SM processes.

The SR3b signal region receives a large contribution of $\ttbar V$
events where at
least one light or charm jet is mis-tagged as a $b$-jet. The Monte Carlo
modelling of this mis-tag
rate is validated in a large opposite-sign dilepton
sample where at least three $b$-tags are required. This sample
is dominated by dilepton $\ttbar$ events where the third $b$-jet is
mis-tagged. Figure~\ref{fig:resVR:d} shows the $\meff$
distribution in this sample, for which the Monte Carlo simulation prediction is
shown to describe the data.

\FloatBarrier
\section{Results and interpretation}
\label{sec:results}

\begin{figure}
\addtolength{\subfigcapskip}{-10pt}
\begin{minipage}{0.5\linewidth}
\centering
\subfigure[~SR3b]{
\includegraphics[width=\textwidth]{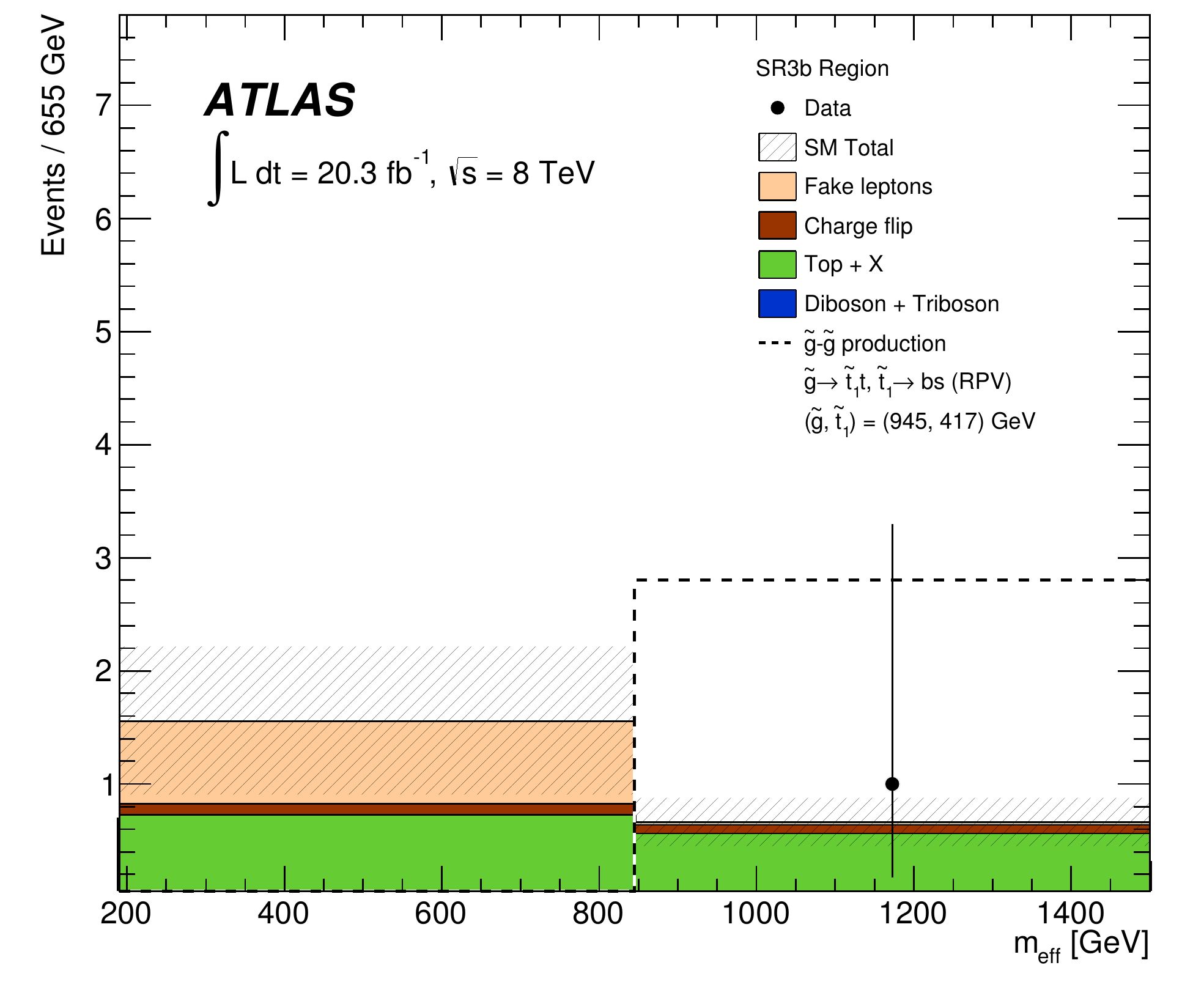}
}
\end{minipage}
\begin{minipage}{0.5\linewidth}
\centering
\subfigure[~SR0b]{
\includegraphics[width=\textwidth]{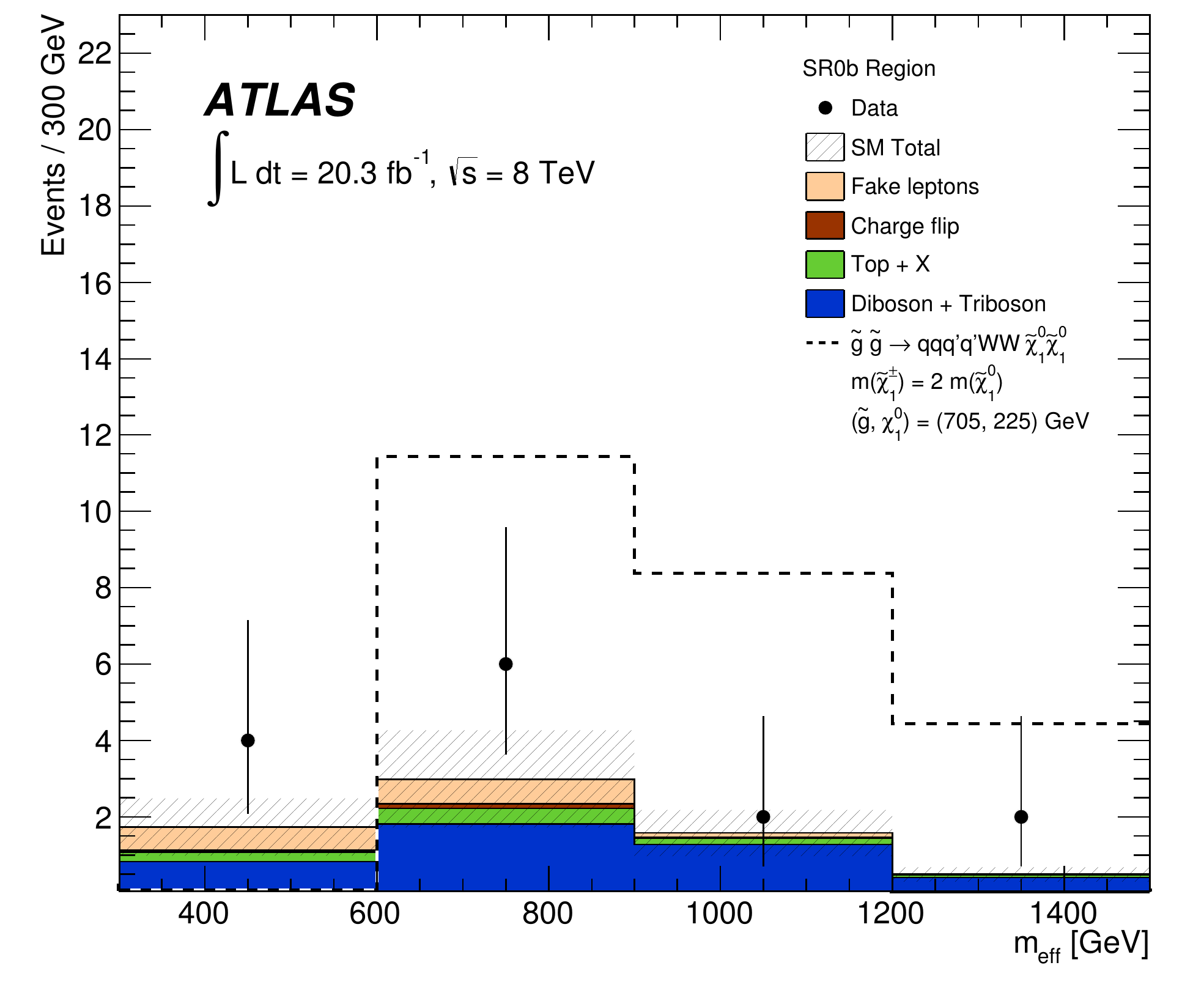}
}
\end{minipage}
\begin{minipage}{0.5\linewidth}
\centering
\subfigure[~SR1b]{
\includegraphics[width=\textwidth]{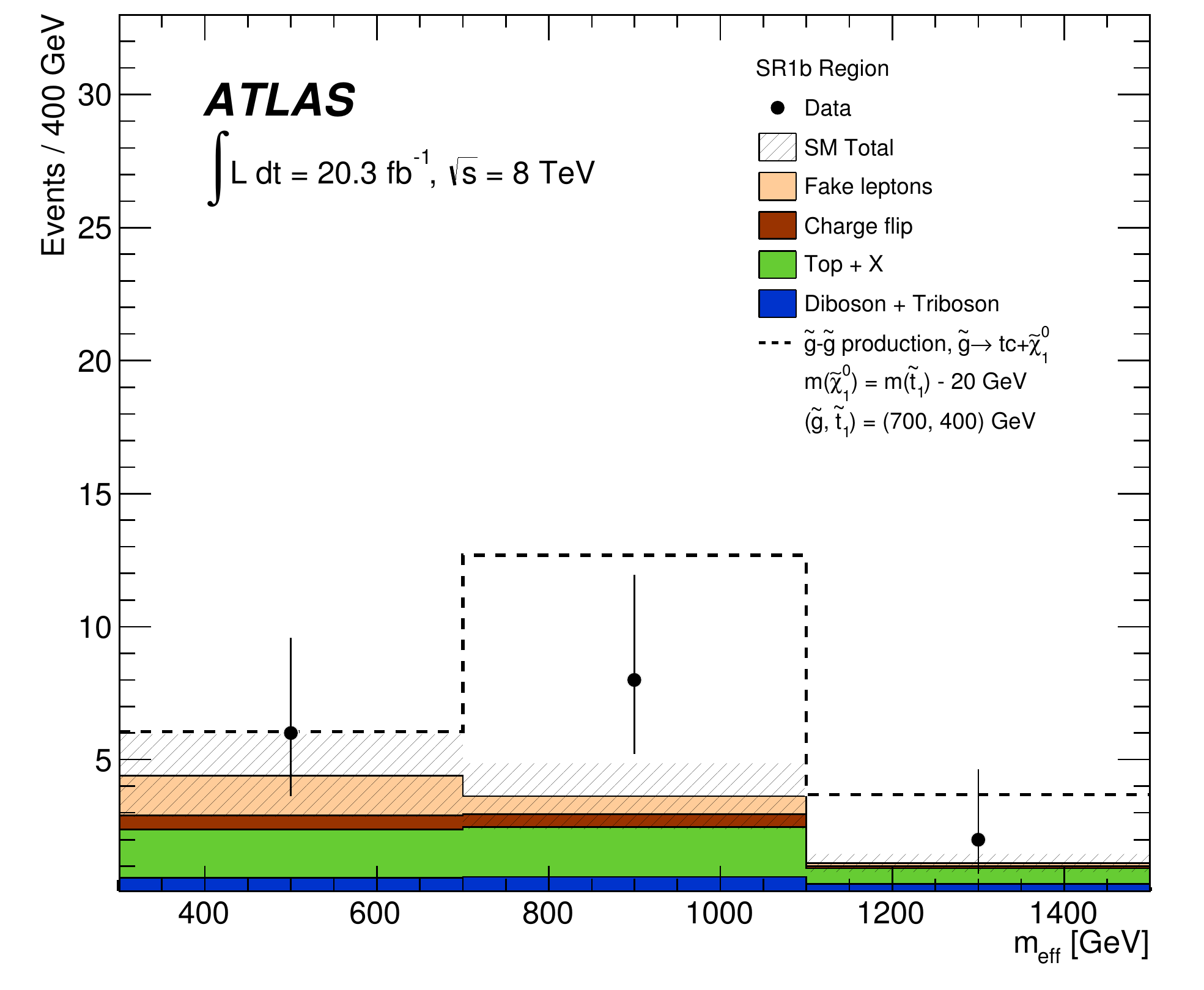}
}
\end{minipage}
\begin{minipage}{0.5\linewidth}
\centering
\subfigure[~SR3Llow]{
\includegraphics[width=\textwidth]{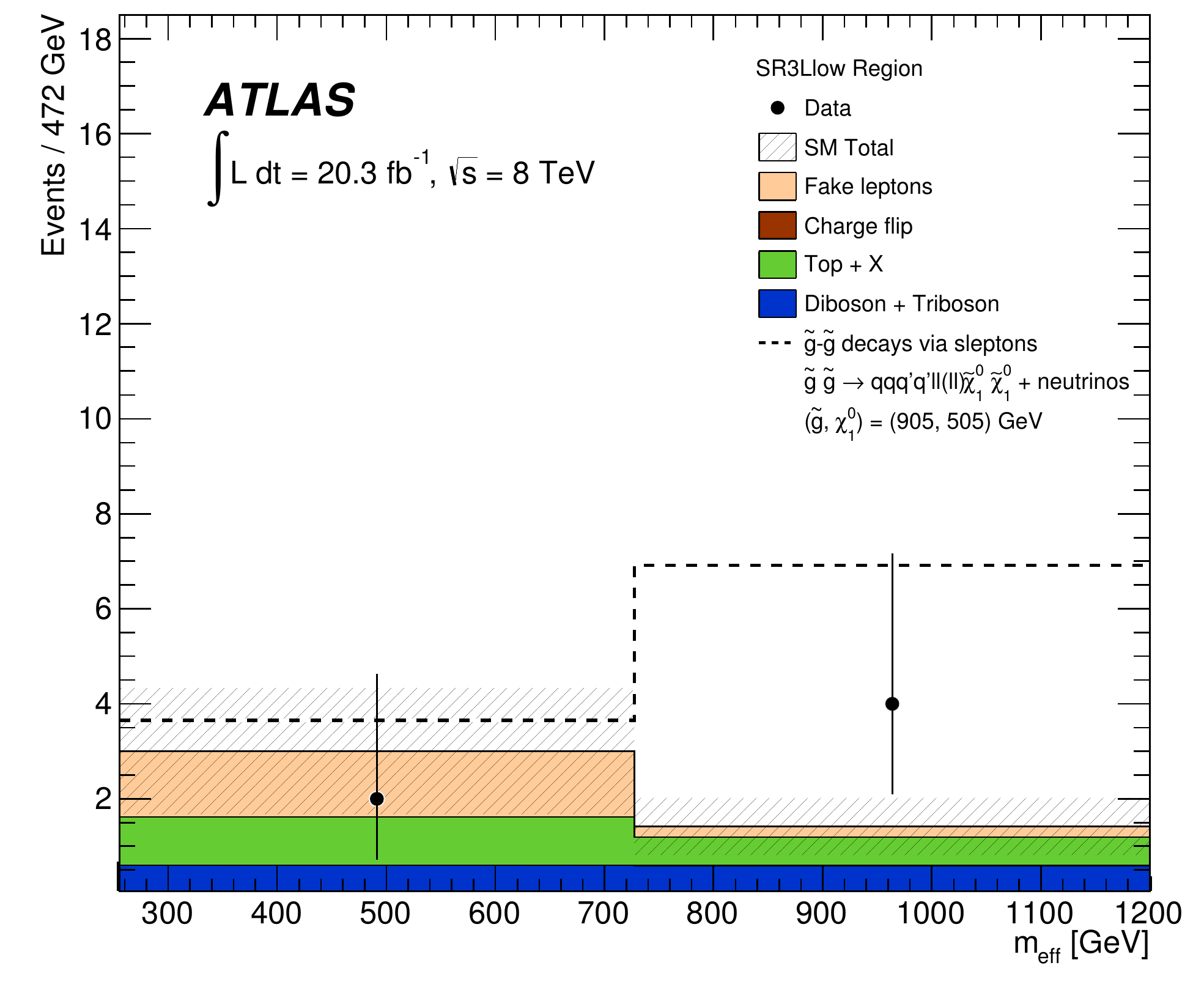}
}
\end{minipage}
\begin{minipage}{0.5\linewidth}
\centering
\subfigure[~SR3Lhigh]{
\includegraphics[width=\textwidth]{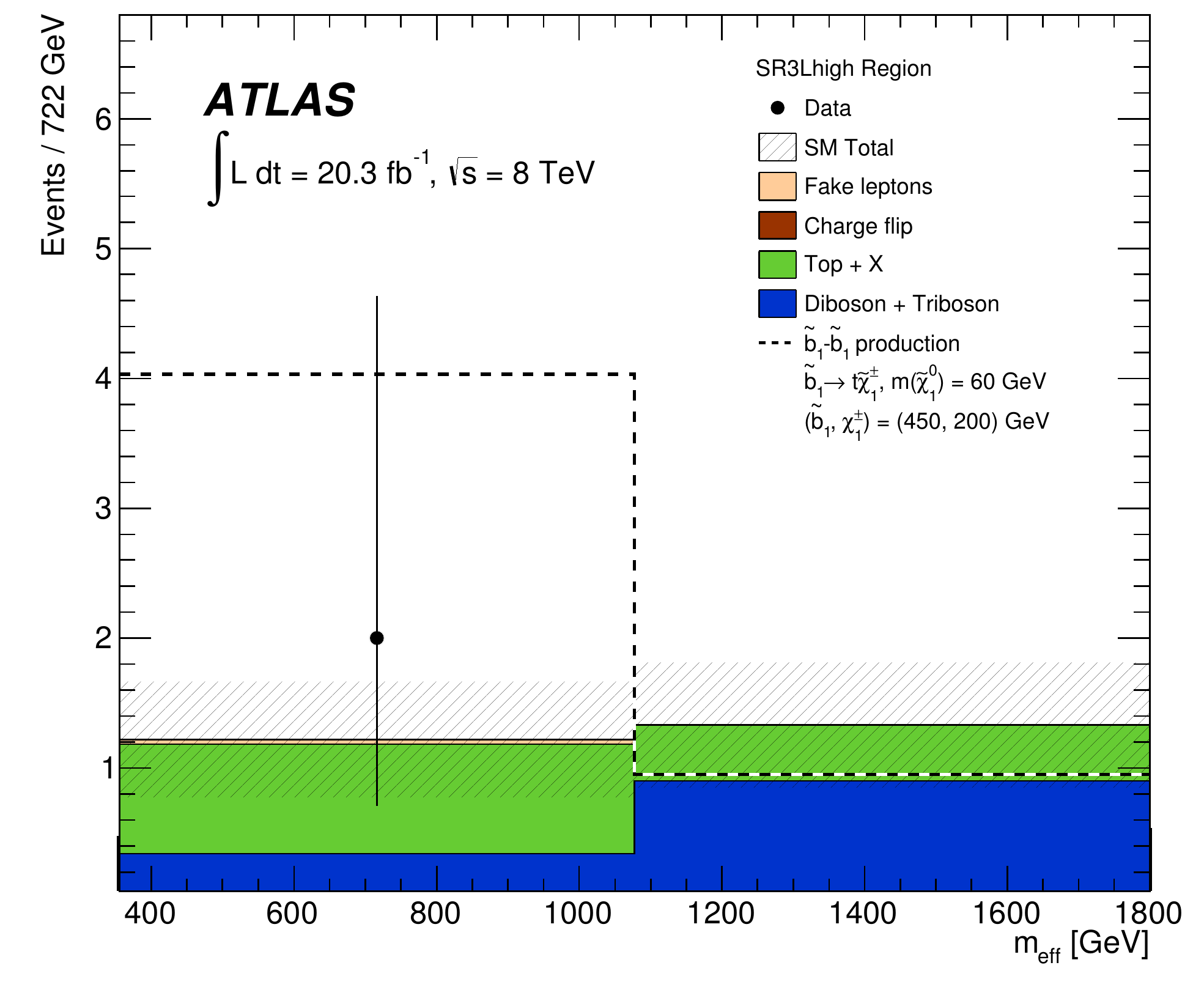}
}
\end{minipage}
\caption{Effective mass ($\meff$) distributions in the signal regions SR3b, SR0b,
  SR1b, SR3Llow and SR3Lhigh, used as input for the exclusion fits. 
  The statistical and systematic uncertainties on the background prediction are
  included in the uncertainty band.  The last bin includes
  overflows. 
 Signal expectations from SUSY models of particular sensitivity in
  each signal region
  are shown for illustration (see text).
}
\label{fig:resSR}
\end{figure}

\begin{table}
\begin{center}
\setlength{\tabcolsep}{0.0pc}
{\small
\begin{tabular*}{\textwidth}{@{\extracolsep{\fill}}lrrrrr}
                         & SR3b & SR0b            & SR1b            & SR3Llow            & SR3Lhigh             \\[-0.05cm]
\noalign{\smallskip}\hline\hline\noalign{\smallskip}
{\bf Observed events}           & $1$  & $14$              & $10$                         & $6$              & $2$                    \\
\noalign{\smallskip}\hline\noalign{\smallskip}
{\bf Total expected background events}  & $2.2 \pm 0.8$ & $6.5 \pm 2.3$          & $4.7 \pm 2.1$                  & $4.3 \pm 2.1$          & $2.5 \pm 0.9$              \\
$p(s=0)$  & $0.50$  & $0.03$          & $0.07$             &$0.29$          & $0.50$              \\
\noalign{\smallskip}\hline\noalign{\smallskip}
{\bf Expected signal events} & $3.4 \pm 0.7$  & $24.3 \pm 3.5$   & $16.4 \pm 3.0$     & $10.6 \pm 1.0$  & $5.0 \pm 0.8$      \\
{\bf for chosen benchmark models} &          &         &          &          &              \\

\noalign{\smallskip}\hline\noalign{\smallskip}
{\bf Components of the background}   \\
$\ttbar V$, $\ttbar H$, $tZ$ and $t\bar{t}t\bar{t}$   & $1.3 \pm 0.5$  & $0.9 \pm 0.4$     & $2.5 \pm 1.7$            & $1.6 \pm 1.0$          & $1.3 \pm 0.7$              \\
Dibosons and tribosons     & $<0.1$  & $4.2 \pm 1.7$          & $0.9 \pm 0.4$                 & $1.2 \pm 0.6$          & $1.2 \pm 0.6$              \\
Fake leptons  & $0.7 \pm 0.6$  & $1.2_{-1.2}^{+1.5}$    & $0.8_{-0.8}^{+1.2}$                & $1.6 \pm 1.6$          & $<0.1$    \\
Charge-flip electrons  & $0.2 \pm 0.1$   & $0.2 \pm 0.1$          & $0.5 \pm 0.1$               &  --          &  --              \\
\noalign{\smallskip}\hline\noalign{\smallskip}
{\bf Systematic uncertainties } \\
{\bf on expected background} \\
Fake-lepton background         & $\pm 0.6$  & $_{-1.2}^{+1.5}$          &$_{-0.8}^{+1.2}$           & $\pm 1.6$          & $<0.1$      \\
Theory unc. on dibosons       & $<0.1$     & $\pm 1.5$          & $\pm 0.3$                & $\pm 0.4$          & $\pm 0.4$       \\
Jet and \met\ scale and resolution         & $\pm 0.1$  & $\pm 0.7$          & $\pm 0.4$                  & $\pm 0.4$          & $\pm 0.3$       \\
Monte Carlo statistics        & $\pm 0.1$     & $\pm 0.5$          & $\pm 0.2$               & $\pm 0.4$          & $\pm 0.4$       \\
$b$-jet tagging          & $\pm 0.2$  & $\pm 0.5$          & $\pm 0.1$                & $<0.1$          & $\pm 0.1$       \\
Theory unc. on  $\ttbar V$, $\ttbar H$, $tZ$ and $t\bar{t}t\bar{t}$         & $\pm 0.4$  & $\pm 0.3$     & $\pm 1.7$         & $\pm 1.0$        & $\pm 0.6$       \\
Trigger, luminosity and pile-up         & $<0.1$      & $\pm 0.1$          & $\pm 0.1$             & $\pm 0.1$          & $\pm 0.1$       \\
Charge-flip background           & $\pm 0.1$       & $\pm 0.1$          & $\pm 0.1$          &  --          & --       \\
Lepton identification            & $<0.1$    & $\pm 0.1$          & $<0.1$            & $\pm 0.1$          & $\pm 0.1$       \\
\noalign{\smallskip}\hline\hline\noalign{\smallskip}
\end{tabular*}
}
\end{center}
\caption{Number of observed data events and expected backgrounds and
  summary of the systematic uncertainties on the background predictions for
 SR3b,  SR0b, SR1b, SR3Llow and SR3Lhigh. 
 The $p$-value of the observed events for the background-only
hypothesis is denoted by $p(s=0)$. By convention, the $p(s=0)$ value is truncated at
0.50 when the number of observed data events is smaller than the
expected backgrounds.
  The expected signal events correspond to the SUSY models considered
  for each signal region in figure~\ref{fig:resSR}
  with their experimental uncertainties. 
  The breakdown of the systematic uncertainties on the expected
  backgrounds, expressed in units of events, is also shown. The
  individual uncertainties are correlated and therefore do not
  necessarily add up in quadrature to
  the total systematic uncertainty.}
\label{table.results.yields.SRdisc}
\end{table}

\par Figure~\ref{fig:resSR} shows the effective mass distribution of
the observed data events and SM predictions for the five signal
regions, after all selections except the one on \meff. 
SUSY models of particular sensitivity to each signal region
are also shown for illustration purposes.
These models, illustrated in figure~\ref{fig:SUSYmodels} and described in section~\ref{sec:interpretModelDep}, are:
gluino-mediated top squark~$\to bs$ (RPV) 
with gluino mass of 945 \gev\ and top squark mass of 417 GeV for SR3b; 
gluino-mediated squark~$\to q^{\prime} W\ninoone$ 
with gluino mass of 705 \gev, 
\chinoonepm\ mass of 450 \gev\ 
and \ninoone\ mass of 225 \gev\ for SR0b;
gluino-mediated top squark~$\to c\ninoone$
with gluino mass of 700 \gev, top squark mass of 400 \gev\ and
\ninoone\ mass of 380 \gev\ for SR1b; 
gluino-mediated squark~$\to$ sleptons 
with gluino mass of 905 \gev, 
\ninotwo\ and \chinoonepm\ masses of 705 \GeV, 
slepton and sneutrino masses of 605 \GeV\ 
and \ninoone\ mass of 505 GeV for SR3Llow; 
and direct bottom squark~$\to t\tilde\chi^\pm_1$
with bottom squark mass of 450 \gev, 
\chinoonepm\ mass of 200 \gev\ 
and \ninoone\ mass of 60 GeV for SR3Lhigh.

The numbers of observed data events and expected background events in the
five signal regions, after the application of the additional requirements on \meff,
are presented in table~\ref{table.results.yields.SRdisc}.
Expected signal yields from the SUSY models appearing in figure~\ref{fig:resSR}
are also shown.
Diboson production in association with jets is a large source of
background for signal regions that do not require the presence of
$b$-jets, namely SR0b, SR3Llow and SR3Lhigh.
In SR1b and SR3b, which require one or more $b$-jets, the largest background
contribution arises from $\ttbar V$ events. 
The background from fake leptons is particularly significant in signal regions with
no or low requirements on \met, such as SR3b and SR3Llow. Background from
electron charge mis-identification is small in all SS signal regions,
and not applicable in the 3L signal regions. 

The level of agreement between the background prediction and data is
quantified by computing the $p$-value for the number of observed events to be consistent with the
background-only hypothesis, denoted by $p(s=0)$ in table~\ref{table.results.yields.SRdisc}.
To do so, the number of events in each signal region is described using a Poisson 
probability density function (pdf). The statistical and systematic uncertainties
on the expected background values are modelled with nuisance parameters
constrained by a Gaussian function with a width corresponding to the size of the uncertainty considered. 
The data and predicted background agree well for SR3b, SR3Llow and SR3Lhigh. 
No events with total electric charge of $\pm3$ are observed in the 3L signal regions. 
For SR0b and SR1b, small excesses are observed corresponding to
1.8 and 1.5 standard deviations, respectively. The significance is calculated  
using the uncertainty on the total expected background yields quoted in 
table~\ref{table.results.yields.SRdisc} and the Poissonian uncertainty of the 
total expected background value. If SR0b and
SR1b are combined, the significance of the excess becomes 2.1 standard
deviations. 

Table~\ref{table.results.yields.SRdisc} also presents the breakdown of 
uncertainties on the background predictions described in section~\ref{sec:bkgsys}.
For all signal regions the background uncertainty is dominated by the statistical
uncertainty on the expected number of background events. The largest systematic 
uncertainties arise 
from the estimation of the fake-lepton probability and from the theoretical
predictions for diboson+jets and $\ttbar V$+jets processes.
Uncertainties on the predicted background event yields are quoted as symmetric, 
except where the negative error reaches zero predicted events, 
in which case the negative error was truncated.

\FloatBarrier

\subsection{Model-independent upper limits}

\begin{table}
\begin{center}\setlength{\tabcolsep}{0.0pc}
\begin{tabular*}{\textwidth}{@{\extracolsep{\fill}}lccc}
\noalign{\smallskip}\hline\noalign{\smallskip}
{\bf Signal channel}                        & $\langle \sigma_{\rm vis}\rangle_{\rm obs}^{95}$[fb]  &  $S_{\rm obs}^{95}$  & $S_{\rm exp}^{95}$ \\
\noalign{\smallskip}\hline\noalign{\smallskip}
SR3b    &  $0.19$ &  $3.9$  & ${4.4}^{+1.7}_{-0.6}$ \\
SR0b    &  $0.80$ &  $16.3$ & ${8.9}^{+3.6}_{-2.0}$ \\
SR1b    &  $0.65$ &  $13.3$ & ${8.0}^{+3.3}_{-2.0}$ \\
SR3Llow &  $0.42$ &  $8.6$  & ${7.2}^{+2.9}_{-1.3}$ \\
SR3Lhigh&  $0.23$ &  $4.6$  & ${5.0}^{+1.6}_{-1.1}$ \\
\noalign{\smallskip}\hline\noalign{\smallskip}
\end{tabular*}
\end{center}
\caption[Breakdown of upper limits.]{
  The 95$\%$ CL upper limits on the visible cross section ($\langle \sigma_{\rm
  vis}\rangle_{\rm obs}^{95}$), defined as the product of acceptance,
  reconstruction efficiency and production cross section, and the observed and
  expected 95$\%$ CL upper limits on the number of BSM events ($S_{\rm
  obs}^{95}$ and $S_{\rm exp}^{95}$).  Results are obtained with
  pseudo-experiments.
\label{tab:SR1_modelIndependent_UPlim}}
\end{table}

No significant excess of events over the SM expectations is observed
in any signal region.
Upper limits at 95\% CL on the number of
beyond the SM (BSM) events for each signal region are derived
using the CL$_s$ prescription~\cite{Read:2002hq}. Normalising these by the
integrated luminosity of the data sample, they can be interpreted
as upper limits on the visible BSM cross section ($\sigma_{\rm vis}$), where
$\sigma_{\rm vis}$ is defined as the product of acceptance, reconstruction
efficiency and production cross section.
The results are given in table~\ref{tab:SR1_modelIndependent_UPlim},
where $\langle \sigma_{\rm vis}\rangle_{\rm obs}^{95}$ is the 95$\%$ CL upper limit on the visible cross section, and
$S_{\rm obs}^{95}$   and $S_{\rm exp}^{95}$ are the observed and expected 
95$\%$ CL upper limits on the number of BSM events, respectively.
The limits presented in table~\ref{tab:SR1_modelIndependent_UPlim} are calculated from pseudo-experiments.
For comparison, corresponding limits calculated with asymptotic formulae \cite{profilelh}
on the observed (expected) number of BSM events in SR3b, SR0b, SR1b, SR3Llow and SR3Lhigh are 
 3.8 (4.4), 15.9 (8.9), 12.6 (7.9), 8.4 (7.2), and 4.3 (5.0), respectively. 

\subsection{Model-dependent limits}
\label{sec:interpretModelDep}

The measurement is used to place exclusion limits on 14 SUSY models and one mUED model.
For each model, the limits are calculated from asymptotic formulae
with a simultaneous fit to all signal regions based on the profile likelihood method.
When doing so, the final \meff\ requirements are relaxed in each
signal region (i.e. the requirements in the rightmost column in
table~\ref{tab:SRDefn} are not applied)
and the fit inputs are the binned \meff\ distributions shown in figure~\ref{fig:resSR}. 
Most of the nuisance parameters are correlated between all bins, 
except for uncertainties of statistical nature, which are modelled with uncorrelated parameters.
The signal pdf is correlated in all bins and multiplied by an overall normalisation scale 
treated as a free parameter in the fit. 
This procedure increases the statistical power of the analysis for model-dependent exclusion.

The observed and expected limits resulting from the exclusion fits are displayed as 
solid red lines and dashed grey lines,
respectively, in figures~\ref{fig:expGTT}--\ref{fig:expPHYS}.
The $\pm1 \sigma^{\rm SUSY}_{\rm theory}$ lines around the observed limits are
obtained by changing the SUSY cross section by one standard deviation
($\pm1\sigma$), as described in section~\ref{sec:mc}. 
All mass limits on supersymmetric particles quoted later in this section are 
derived from the $-1 \sigma^{\rm SUSY}_{\rm theory}$ theory line. 
The yellow band around the expected limit shows the $\pm1\sigma$ uncertainty, 
including all statistical and systematic uncertainties 
except the theoretical uncertainties on the SUSY cross section. The
uncertainties on the SUSY signal include the
 detector simulation uncertainties described in section~\ref{sec:bkgsys}.
For simplified models, 95\% CL upper limits on cross sections obtained 
using the signal efficiency and acceptance specific to each model are available
in the HepData database~\cite{HepData}.
When available, exclusion limits set by previous ATLAS searches 
\cite{Aad:2012ms,Atlas:0lep7TeV,Atlas:1Tau7TeV,Atlas:3b7TeV,ATLAS:1568342} 
are also shown for comparison.

Three categories of simplified models are used to design the signal regions and interpret the results: 
gluino-mediated top squark, gluino-mediated (or direct) first- and second-generation squark, and direct bottom squark production,
as illustrated in figure~\ref{fig:SUSYmodels}. 
In addition, three complete SUSY models and one mUED model are used for interpretation only.

\FloatBarrier
\subsubsection{Gluino-mediated top squarks}

\begin{figure}[h]
\addtolength{\subfigcapskip}{-10pt}
\begin{minipage}[b]{0.5\linewidth}
\centering
\subfigure[\label{fig:expGTT:a}]{
\includegraphics[scale=0.38]{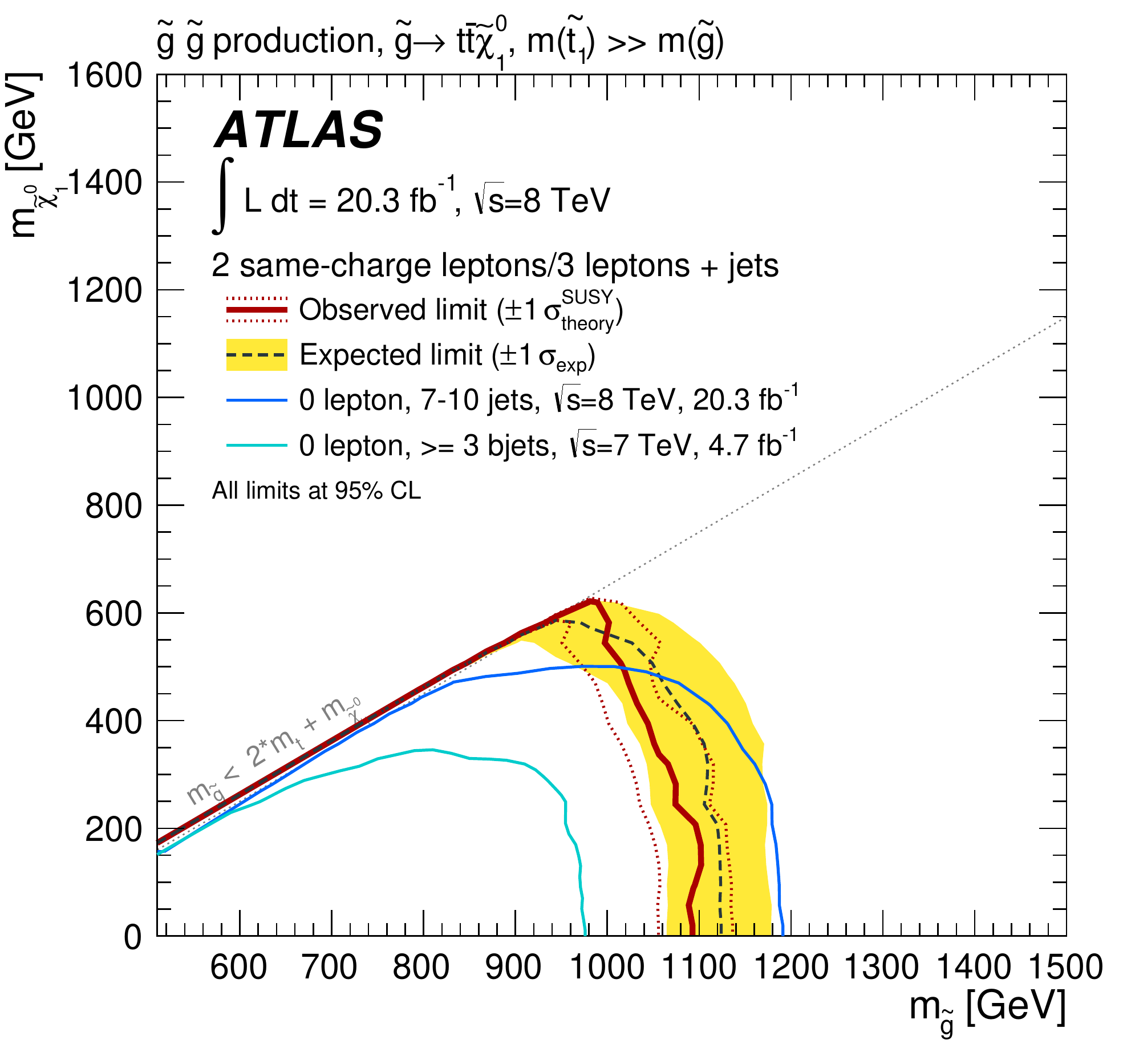}
}
\end{minipage}
\begin{minipage}[b]{0.5\linewidth}
\centering
\subfigure[\label{fig:expGTT:b}]{
\includegraphics[scale=0.38]{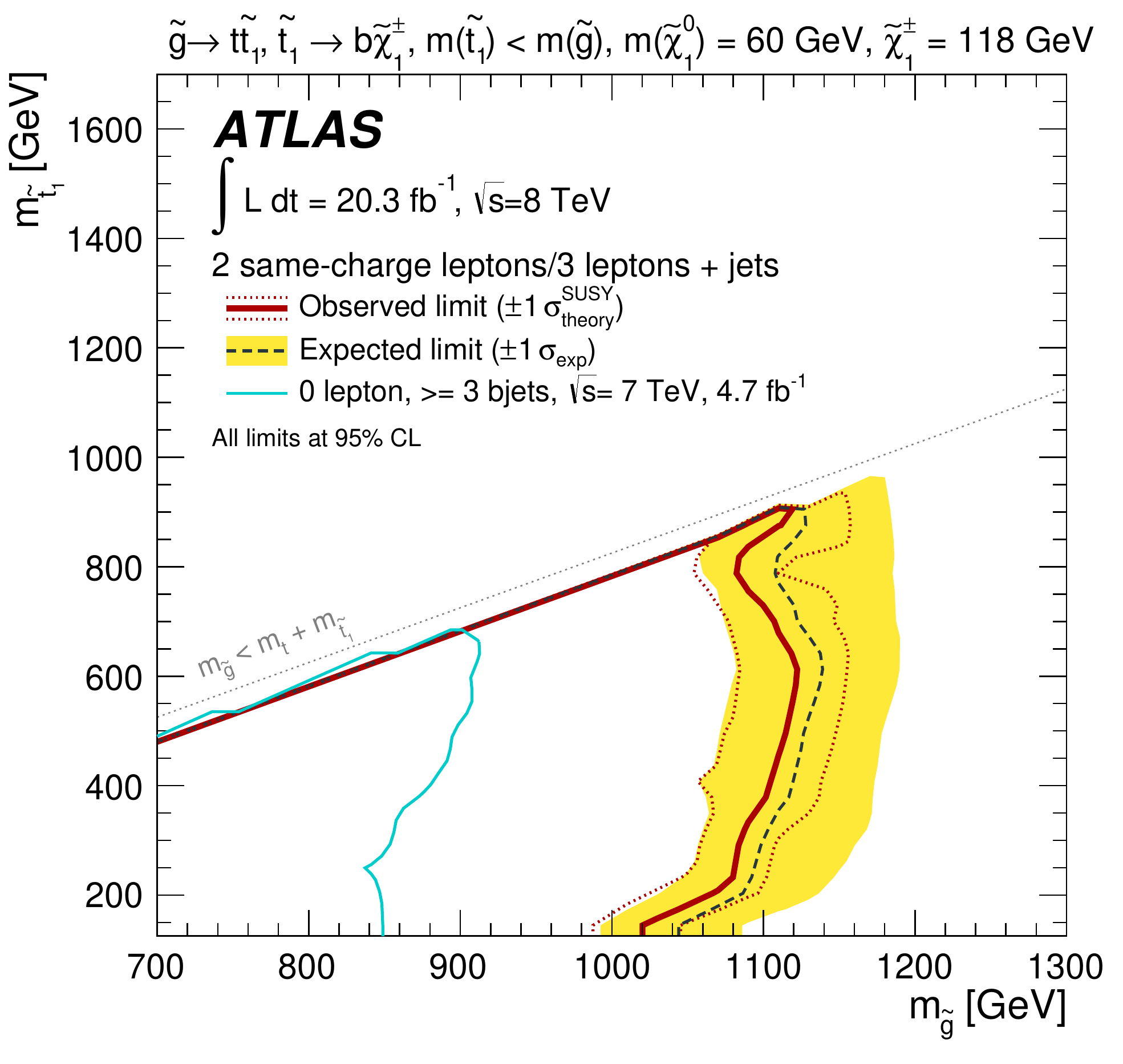} }
\end{minipage}
\begin{minipage}[b]{0.5\linewidth}
\centering
\subfigure[\label{fig:expGTT:c}]{
 \includegraphics[scale=0.38]{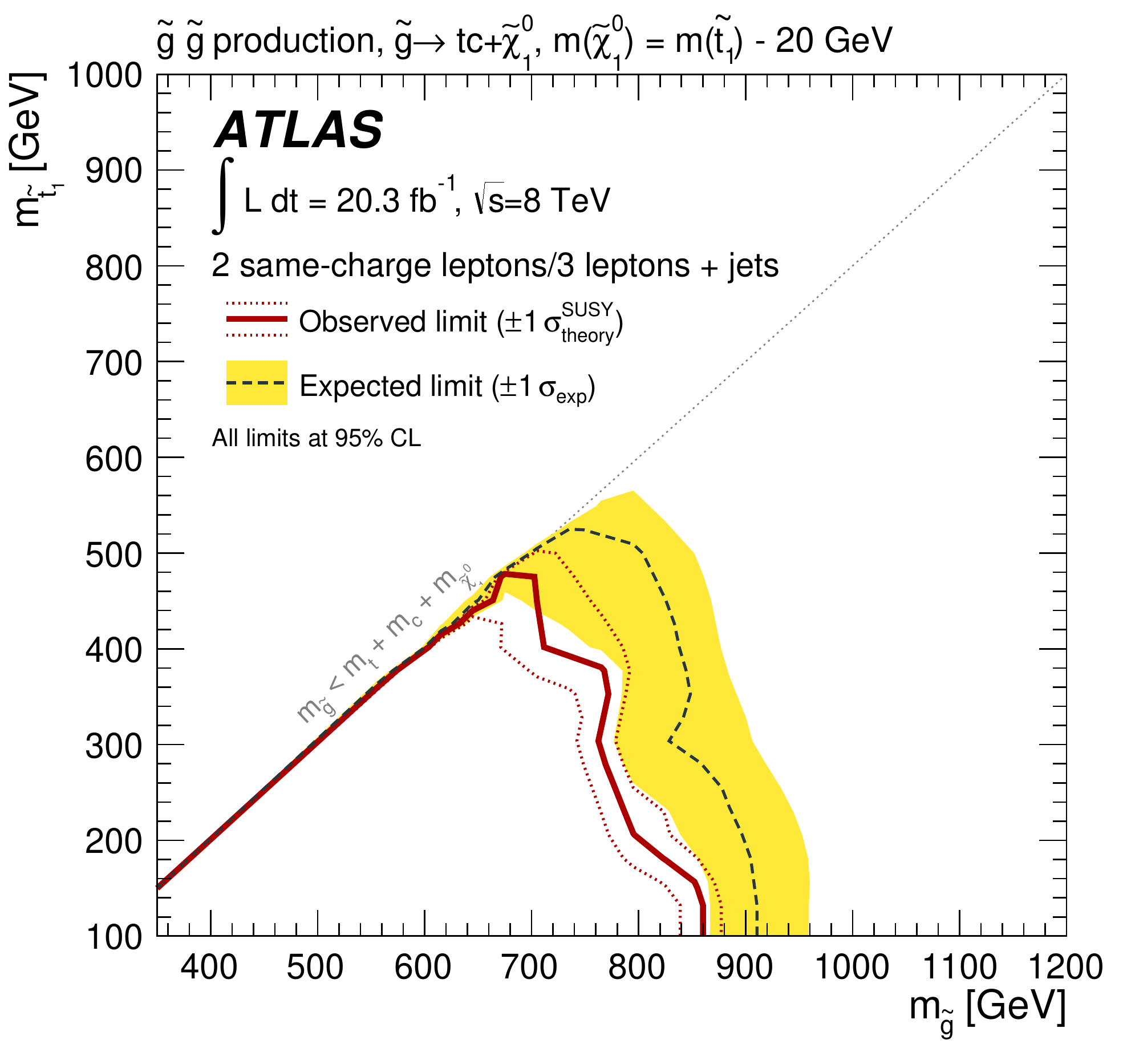}
}
\end{minipage}
\begin{minipage}[b]{0.5\linewidth}
\centering
\subfigure[\label{fig:expGTT:d}]{
\includegraphics[scale=0.38]{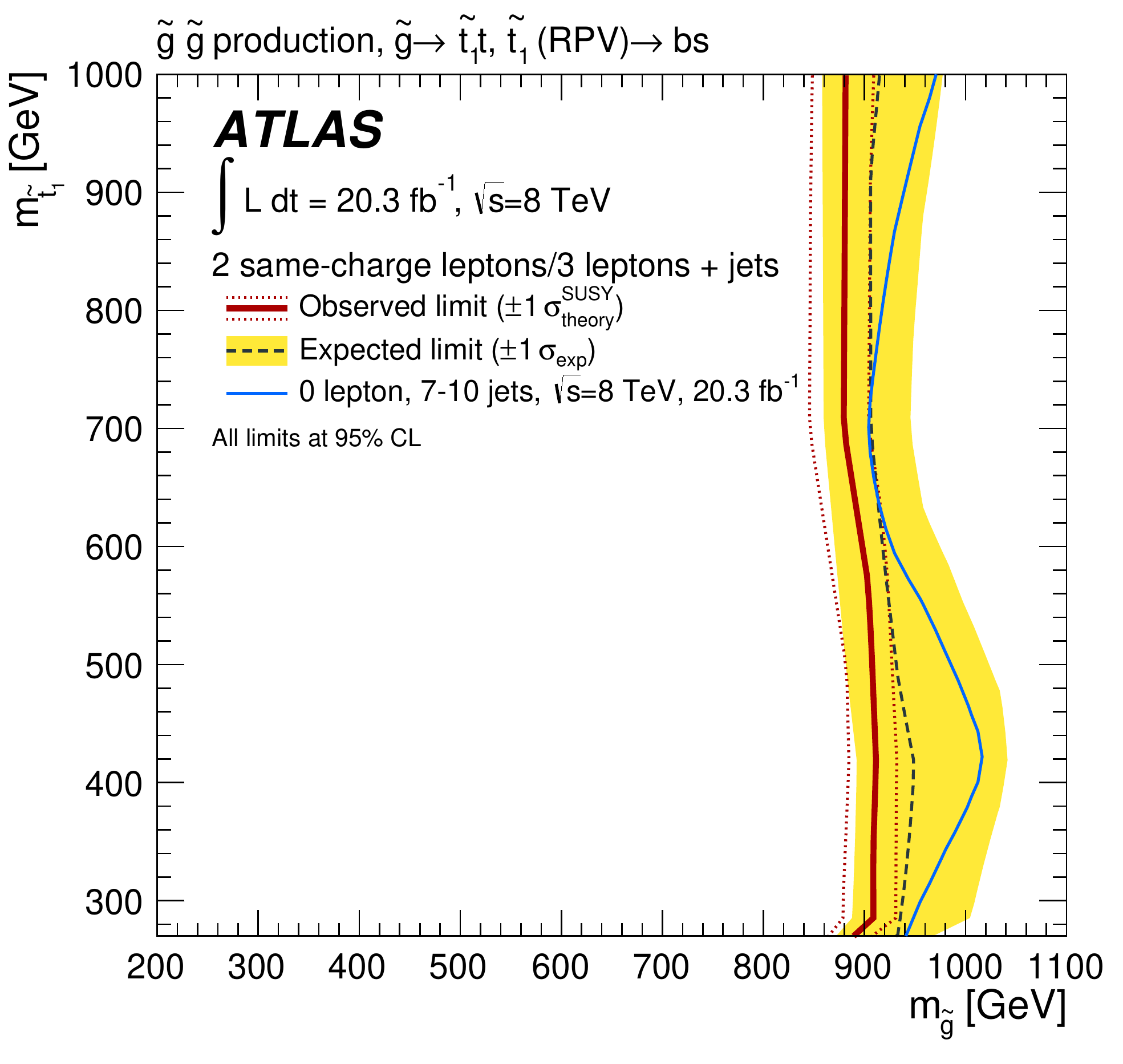} 
}
\end{minipage}
\caption{Observed and expected exclusion limits on gluino-mediated top squark production, 
obtained with 20.3\,fb$^{-1}$ of $pp$ collisions at $\sqrt{s}$=8~\TeV, for four different
top squark decay modes (see text). When available, results are compared with
the limits obtained by previous ATLAS
searches~\cite{Atlas:3b7TeV,ATLAS:1568342}. }
\label{fig:expGTT}
\end{figure}

\par Results for four simplified models of gluino-mediated top squark production are presented 
in figure~\ref{fig:expGTT}.
In each case, gluinos are produced in pairs, the top squark $\tilde{t}_1$ is assumed to be 
the lightest squark, and the 
$\gluino \to t\stop_1^{(*)}$ branching fraction is set to 100\%.
The top squark, however, decays to a different channel in each model: 
$\stopone\rightarrow t\ninoone$, $\stopone\rightarrow b\tilde\chi^\pm_1$, 
$\stopone\rightarrow c\ninoone$ or $\stopone\rightarrow bs$,
with a 100\% branching fraction.

In the gluino-mediated top squark $\to  t\ninoone$ model, the mass of the 
top squark is set to $m_{\tilde{t}_1}=2.5$~TeV and the masses of all
other squarks are much higher (they are assumed to be decoupled).
Gluinos decay through mediation by an off-shell top squark to a pair
of top quarks and a stable neutralino, 
$\gluino \to t\stop_1^* \to t\tbar~\ninoone$.  The final state is therefore
$\gluino\gluino\to bbbb~WWWW~\ninoone\ninoone$, with the constraint that
$m_{\gluino} > 2m_{t} + m_{\ninoone}$.
Results are interpreted in the parameter space of the gluino and
\ninoone\ masses (see figure~\ref{fig:expGTT:a}).
Gluino masses below 950~GeV are excluded at 95\% CL, for any $\ninoone$ mass.
The sensitivity is dominated by SR3b. 

In the gluino-mediated top squark $\to b\tilde\chi^\pm_1$ model, the
top squark is on-shell,
the \chinoonepm\ mass is set to $118$ \GeV, the \ninoone\ mass set to
$60$ \GeV\ and the \ninoone\ is stable.
Hence the chargino decays through an off-shell $W$ boson, and
the final state is $\gluino\gluino\to bbbb~WWW^*W^*~\ninoone\ninoone$,
with the constraint that $m_{\gluino}> m_{t} + m_{\stopone}$.
Results are interpreted in the parameter space of the gluino and top
squark masses (see figure~\ref{fig:expGTT:b}).
Gluino masses below 1 \TeV\  are excluded at 95\% CL for top squark masses above 200 \GeV.
The sensitivity is dominated by SR3b.

In the gluino-mediated top squark $\to c\ninoone$ model, the on-shell
top squark and stable neutralino have close-by masses, $\Delta
m(\stop,\ninoone)=20$ \GeV,
which forbids the top squark decay to a top quark but allows the decay to a charm quark.
The final state is therefore  $\gluino\gluino\to bb~cc~WW~\ninoone\ninoone$,
with the constraint that $m_{\gluino}> m_{t} + m_c + m_{\ninoone}$.
Results are interpreted in the parameter space of the gluino and top
squark masses (see figure~\ref{fig:expGTT:c}).
Gluino masses below 640 \GeV\  are excluded at 95\% CL, 
for any $\stopone$ and $\ninoone$ masses.
The sensitivity is dominated by SR1b. 

In the gluino-mediated top squark $\to bs$ (RPV) model, top squarks are assumed to decay with 
an $R$-parity-violating and  baryon-number-violating coupling
$\lambda''_{323}=1$, as proposed in ref.~\cite{Allanach:2012vj}.
The final state is therefore  $\gluino\gluino\to bbbb~ss~WW$,
characterised by the presence of four $b$-quarks but only moderate missing transverse momentum. 
Results are interpreted in the parameter space of the gluino and top
squark masses (see figure~\ref{fig:expGTT:d}).
Gluino masses below 850~\GeV\  are excluded at 95\% CL, almost independently of the top squark mass.
The sensitivity is dominated by SR3b. 

Stringent limits are hence placed on gluino-mediated top squark scenarios favoured 
by naturalness arguments. 
The SR3b signal region is sensitive to almost any scenario with SS or
$\geq$3 leptons and $\geq$3 $b$-quarks.
This is demonstrated in the gluino-mediated top squark $\to bs$ (RPV) model, 
where $m_{\gluino}<850 \GeV$ is excluded by SR3b alone in the absence of a large \met\ signature.
In $R$-parity-conserving scenarios, the sensitivity is further
increased by including the SR3Lhigh and SR1b
signal regions. Results on gluino-mediated $\stopone\rightarrow t\ninoone$ and 
$\stopone\rightarrow b\tilde\chi^\pm_1$ show that $m_{\gluino}\lesssim 950 \GeV$ is excluded for 
on-shell or off-shell top squarks, largely independently of the top squark mass, 
as long as the top squark decay involves $b$-quarks.
As shown for the gluino-mediated top squark $\to  t\ninoone$ model, this conclusion holds
for $\Delta m(\gluino,\ninoone) \simeq 2 m_{t}$ as well. 
In the especially difficult gluino-mediated top squark $\to c\ninoone$ case,
where only two $b$-quarks and two $W$ bosons are produced,
gluino masses can still be excluded up to 840 \GeV.

\FloatBarrier
\subsubsection{Gluino-mediated (or direct) first- and second-generation squarks} 

\begin{figure}[h]
\addtolength{\subfigcapskip}{-10pt}
\begin{minipage}[b]{0.5\linewidth}
\centering
\subfigure[\label{fig:expGluinoSquark:a}]{
\includegraphics[scale=0.34]{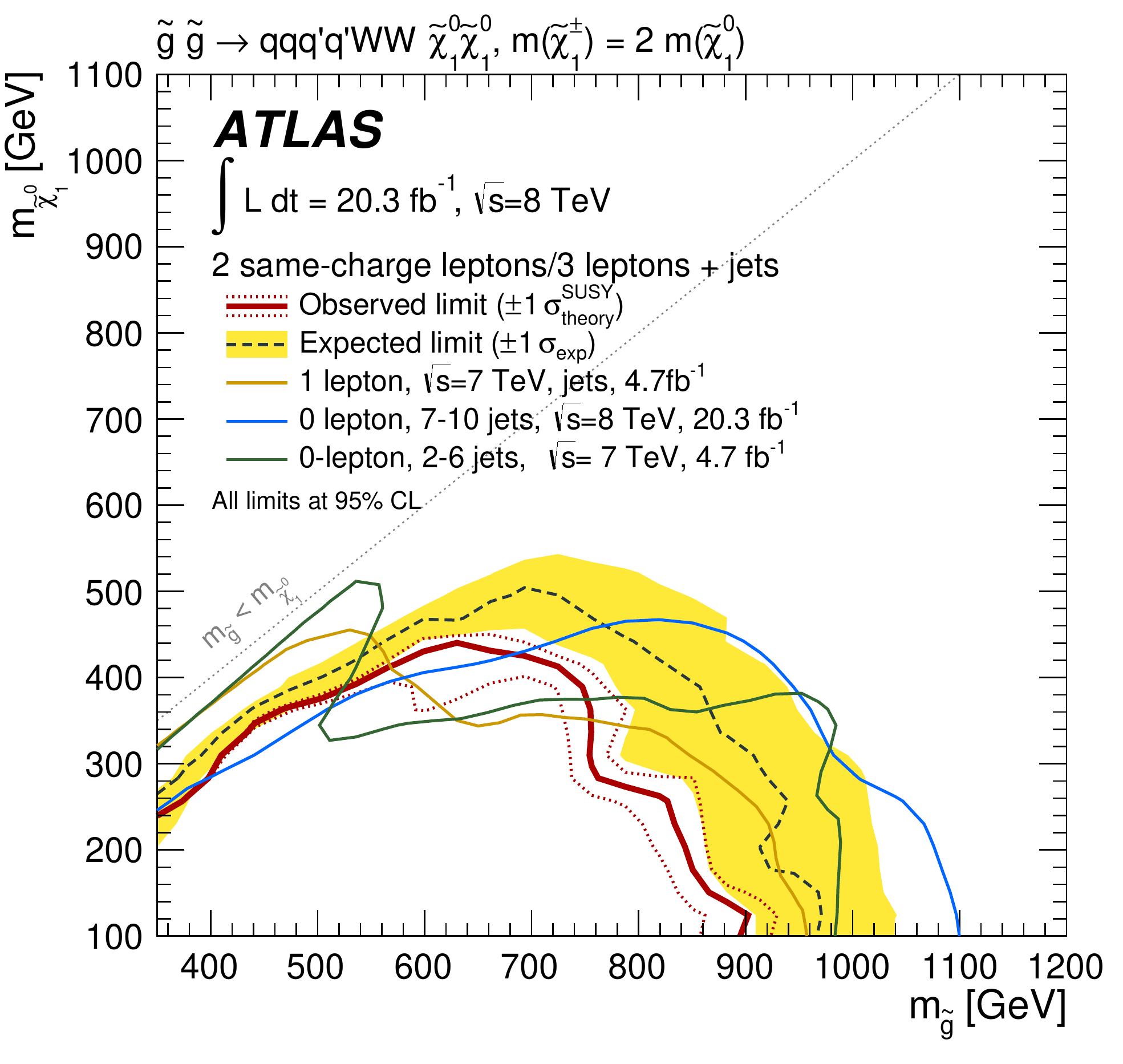}}
\end{minipage} \\
\begin{minipage}[b]{0.5\linewidth}
\centering
\subfigure[\label{fig:expGluinoSquark:b}]{
 \includegraphics[scale=0.34]{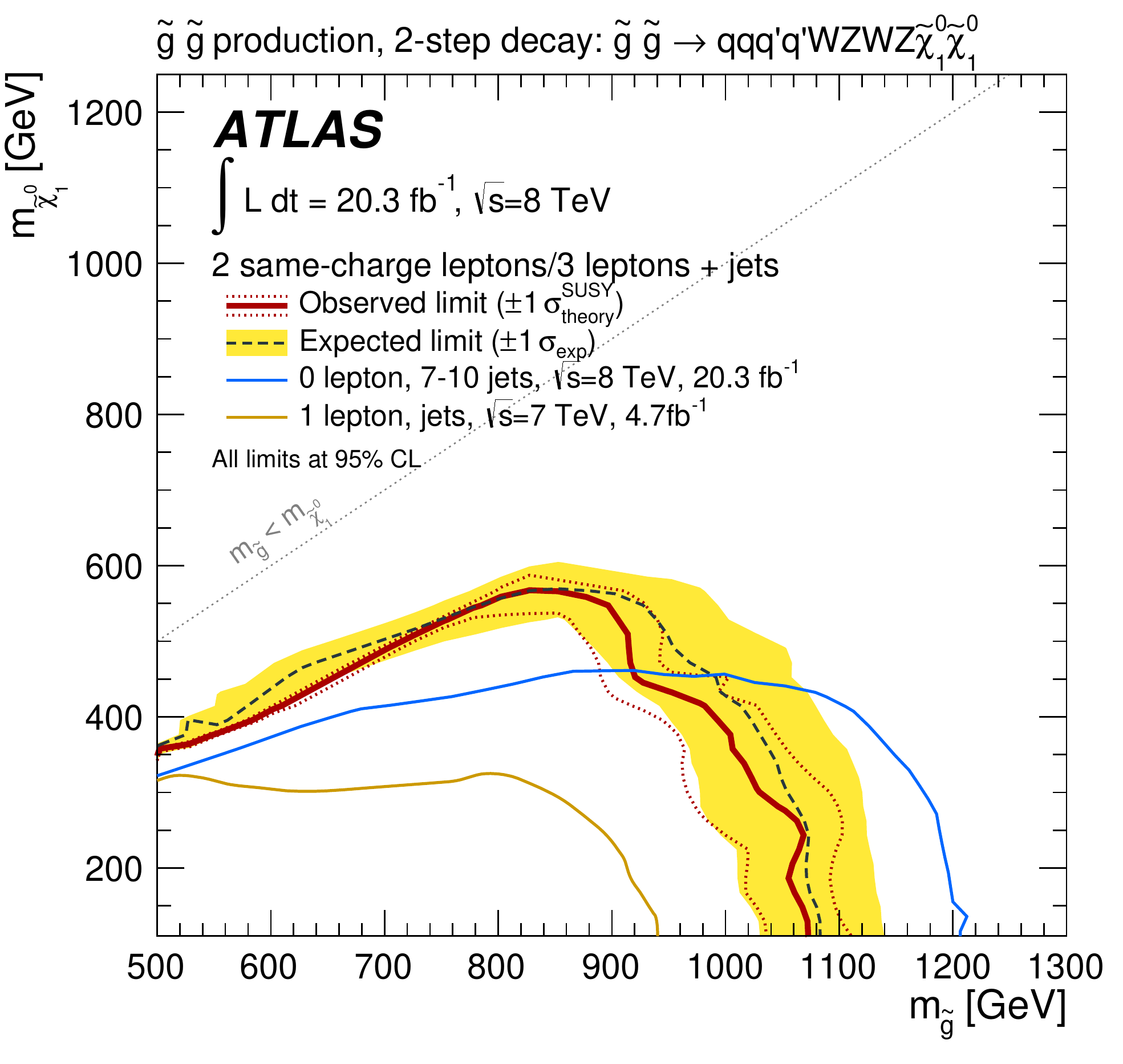}}
\end{minipage}
\begin{minipage}[b]{0.5\linewidth}
\centering
\subfigure[\label{fig:expGluinoSquark:c}]{
 \includegraphics[scale=0.34]{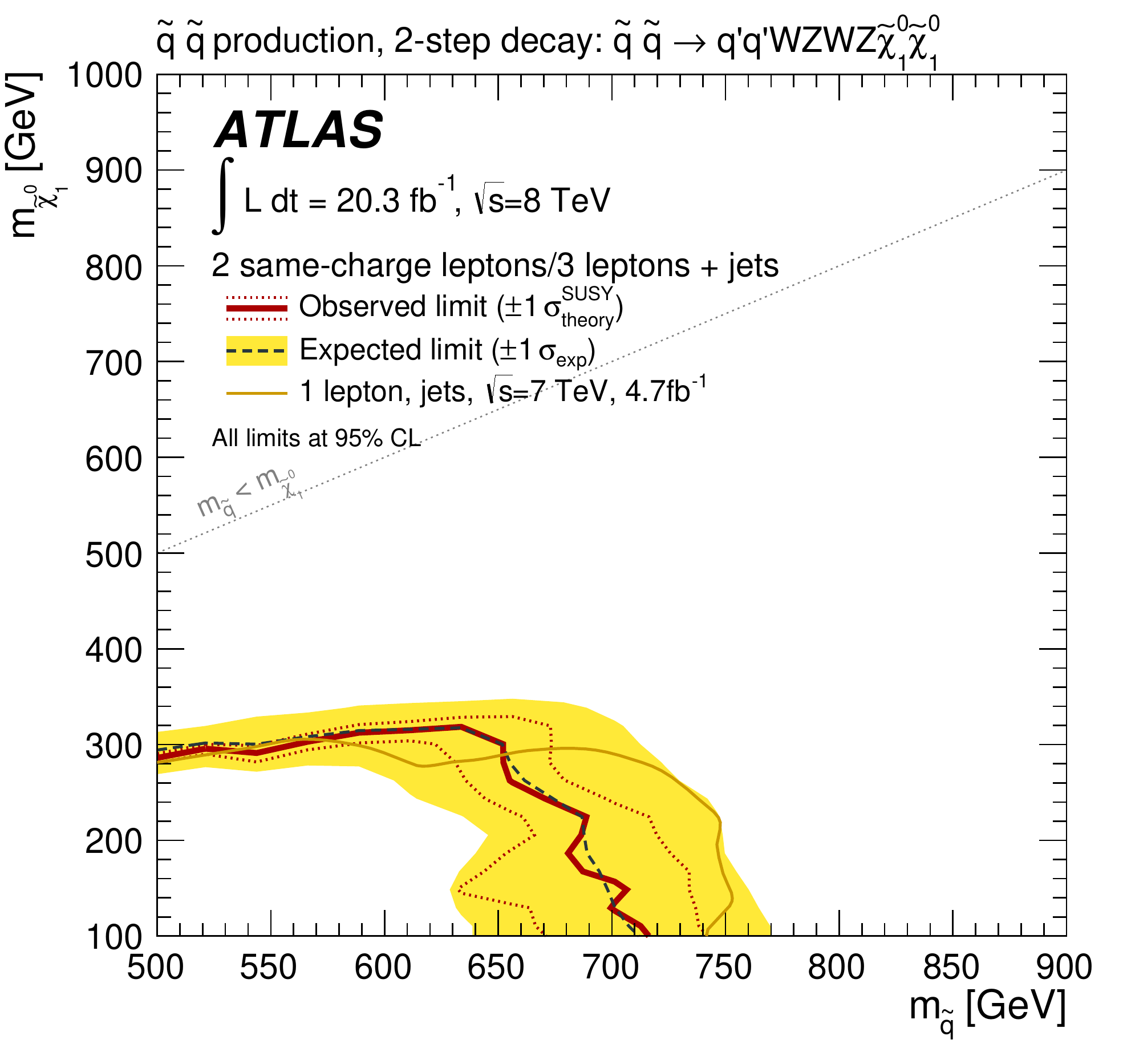}}
\end{minipage}
\begin{minipage}[b]{0.5\linewidth}
\centering
\subfigure[\label{fig:expGluinoSquark:d}]{
 \includegraphics[scale=0.34]{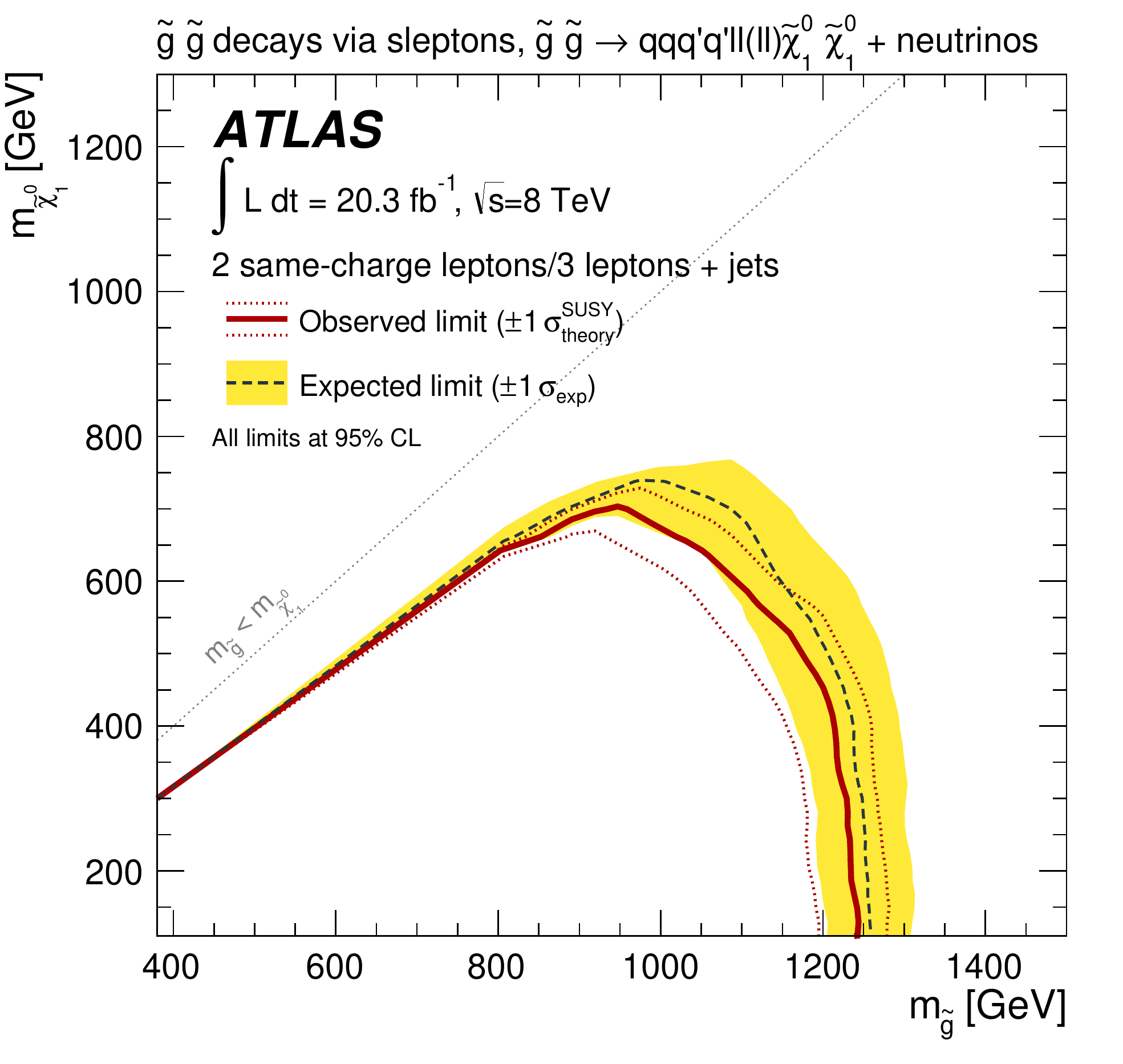} }
\end{minipage}
\begin{minipage}[b]{0.5\linewidth}
\centering
\subfigure[\label{fig:expGluinoSquark:e}]{
 \includegraphics[scale=0.34]{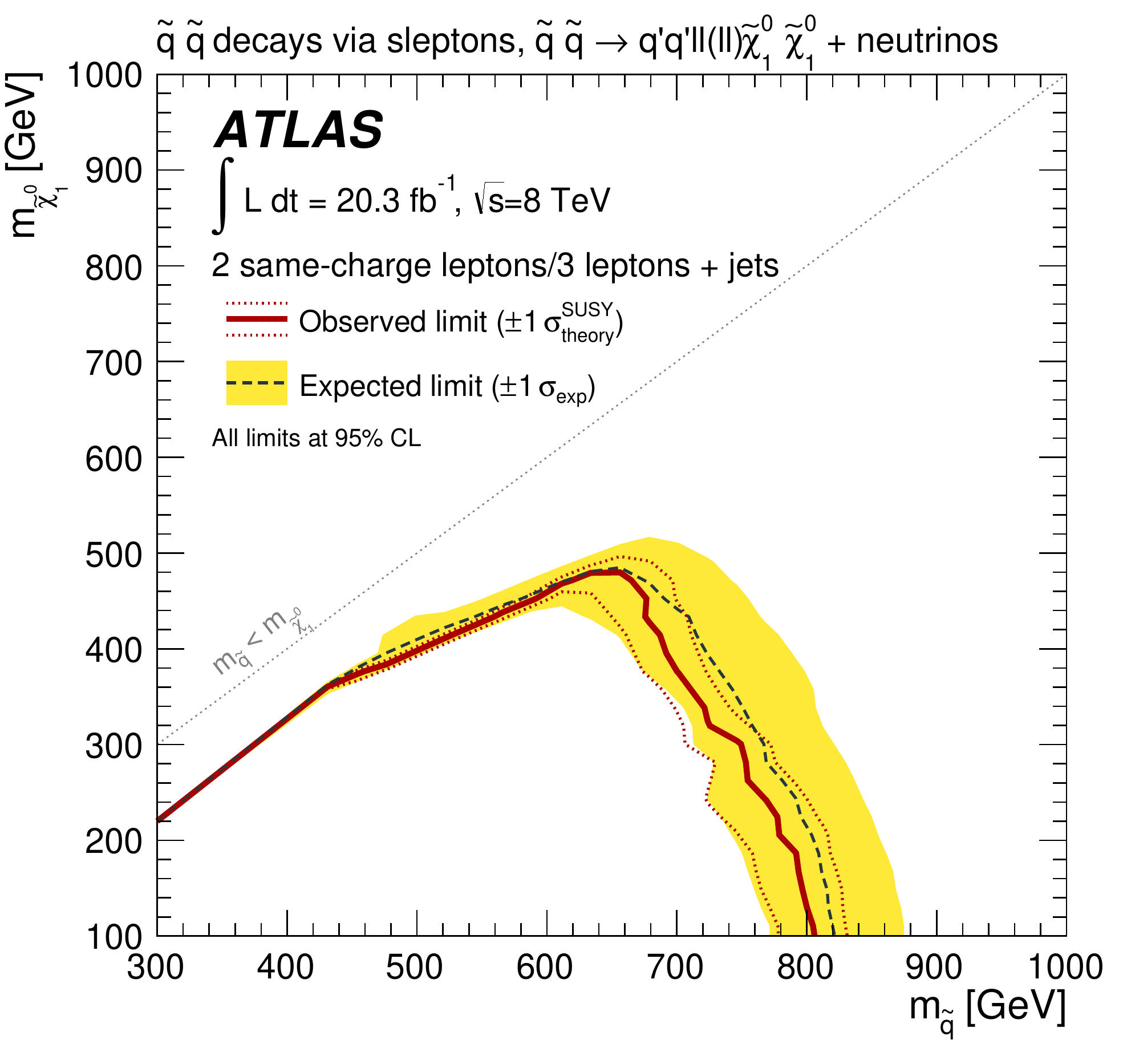} }
\end{minipage}
\caption{
Observed and expected exclusion limits on gluino-mediated production of
first- and second-generation squarks (left)
and direct production of first- and second-generation squarks (right), 
obtained with 20.3\,fb$^{-1}$ of $pp$ collisions at $\sqrt{s}$=8~\TeV,
for three different squark decay cascades (see text). 
When available, results are compared with
the limits obtained by previous ATLAS
searches~\cite{Aad:2012ms,Atlas:0lep7TeV,ATLAS:1568342}. }
\label{fig:expGluinoSquark}
\end{figure}

\par Results for five simplified models of direct and gluino-mediated
first- and second-generation squark 
production are presented in figure~\ref{fig:expGluinoSquark}. In all models, the four 
squarks of first and second generations, collectively referred to as ``squarks'' ($\squark$), 
are assumed to be left-handed and degenerate in mass.
These squarks are pair-produced, either directly ($\squark\squark$) 
or via gluinos ($\gluino\gluino\to qq\squark\squark$), and the
\ninoone\ is assumed to be stable. Different assumptions on the decay of
the squarks are considered:
$\squark \to q^{\prime} W\ninoone$, $\squark \to q^{\prime} WZ\ninoone$ and $\squark\to$~sleptons.
The masses of the resulting supersymmetric particles 
are set according to commonly used conventions in order to cover a variety of scenarios. 

\medskip
In the gluino-mediated or direct squark $\to q^{\prime} W\ninoone$ model, 
the \chinoonepm\ and \ninoone\ masses are related by $m_{\chinoonepm}=2m_{\ninoone}$. 
For gluino-mediated and direct squark production, the final states are therefore
\begin{align*}
\gluino\gluino & \to qqq^{\prime} q^{\prime}~W^{(*)}W^{(*)}~\ninoone\ninoone  \text{,}\\
\squark\squark & \to q^{\prime} q^{\prime}~W^{\pm(*)}W^{\mp(*)}~\ninoone\ninoone \text{.}
\end{align*}
The $\gluino\gluino$ model is the simplest strong-production scenario from which prompt same-sign leptons can arise, due to the Majorana nature of gluinos.
However, the $\squark\squark$ model can only produce opposite-sign leptons, for which this search has no sensitivity.
Results are interpreted in the parameter space of the gluino and
$\ninoone$ masses (see figure~\ref{fig:expGluinoSquark:a}).
This scenario is excluded at 95\% CL for gluino masses up to 860 \GeV\ 
and $\ninoone$ masses up to 400 \GeV.
The sensitivity is dominated by SR0b. 

\medskip
In the gluino-mediated or direct squark $\to q^{\prime} WZ\ninoone$ model, squarks decay as
\[\tilde{q}\to q^{\prime}~\tilde{\chi}^{\pm}_{1}\to q^{\prime} W\ninotwo\to q^{\prime} W Z\ninoone\text{.}\]
The intermediate particle masses are set to 
\begin{align*}
m_{\tilde{\chi}^{\pm}_{1}} & = (m_{\tilde{g}}+m_{\ninoone})/2 \text{,}\\
m_{\ninotwo} & = (m_{\tilde{\chi}^{\pm}_{1}}+m_{\ninoone})/2\text{.}
\end{align*}
The final states are therefore
\begin{align*}
\gluino\gluino\to qqq^{\prime}
q^{\prime}~W^{(*)}W^{(*)}Z^{(*)}Z^{(*)}~\ninoone\ninoone \text{,}\\
\squark\squark\to q^{\prime}
q^{\prime}~W^{\pm(*)}W^{\mp(*)}Z^{(*)}Z^{(*)}~\ninoone\ninoone\text{.}
\end{align*}

The $W$ and $Z$ bosons are on-shell (off-shell) at large (small) mass differences $\Delta
m(\gluino,\ninoone)$ and $\Delta
m(\squark,\ninoone)$. 
Results are interpreted in the parameter space of the gluino (squark)
and $\ninoone$ masses (see figures~\ref{fig:expGluinoSquark:b} and \ref{fig:expGluinoSquark:c}).
These scenarios are excluded at 95\% CL for gluino (squark) masses up to 1040 (670) \GeV\
and $\ninoone$ masses up to 520 (300) \GeV.
The sensitivity is dominated by SR3Lhigh at large $\Delta
m(\gluino,\ninoone)$ and $\Delta m(\squark,\ninoone)$ and 
by SR3Llow at small $\Delta m(\gluino,\ninoone)$ and $\Delta m(\squark,\ninoone)$.  

\medskip
In the gluino-mediated or direct squark $\to$~sleptons model, squarks
are assumed to decay as
$\squark\to q^{\prime}~\chinoonepm$ or 
$\squark\to q~\ninotwo$ with equal probability. 
The \chinoonepm~ decays  with equal probability 
as $\chinoonepm \to \slepton \nu$ or $\chinoonepm \to \ell \tilde\nu$. The
$\ninotwo$ decays with equal probability as $\ninotwo \to \ell
\slepton$ or $\ninotwo \to \nu \tilde\nu$. 
Finally the slepton decays as $\slepton \to \ell \ninoone$, 
and the sneutrino decays as $\tilde \nu \to \nu \ninoone$. 
All three flavours of sleptons are considered and are degenerate in mass.
The masses of the 
\chinoonepm~ and \ninotwo~ are assumed to be equal and set to
$m_{\tilde{\chi}^{\pm}_{1}}=m_{\ninotwo}=(m_{\tilde{g}/\tilde{q}} +m_{\ninoone})/2$.
The masses of the slepton and sneutrino are assumed to be equal and set to
$m_{\tilde{\ell}}=m_{\tilde{\nu}}=(m_{\ninotwo} + m_{\ninoone})/2$.
   The resulting decay chains are
\begin{eqnarray*}
 \squark &\to& q^{\prime}~\chinoonepm \to q^{\prime} \slepton \nu \to q^{\prime} \ell\nu\ninoone,\\
 \squark&\to& q^{\prime}~\chinoonepm \to q^{\prime} \ell \tilde\nu\to q^{\prime} \ell\nu\ninoone,\\
 \squark&\to& q~\ninotwo \to q \ell \slepton\to q \ell\ell \ninoone,\\
 \squark &\to& q~\ninotwo \to q \nu \tilde\nu \to q \nu\nu \ninoone.
  \end{eqnarray*}
Pair production of squarks or gluinos therefore leads to final states with missing transverse momentum,
two or four light jets, and up to four charged leptons.
Results are interpreted in the parameter space of the gluino (squark)
and $\ninoone$ masses (see figures~\ref{fig:expGluinoSquark:d} and \ref{fig:expGluinoSquark:e}).
These scenarios are excluded at 95\% CL for gluino (squark) masses up to 1200 (780) \GeV\
and $\ninoone$ masses up to 660 (460) \GeV.
The sensitivity is dominated by SR3Lhigh.

The signal regions SR0b, SR3Lhigh and SR3Llow are thus sensitive to
first- and second-generation squark production in $R$-parity-conserving scenarios.
The reach in gluino and \ninoone\ masses varies by more than 300 \GeV\ between
the most difficult case (gluino-mediated squark $\to q^{\prime} W\ninoone$, with lowest leptonic 
branching fraction) and the most favourable case
(gluino-mediated squark decaying via sleptons, with largest leptonic branching fraction).
In an intermediate case, the gluino-mediated squark $\to q^{\prime} WZ\ninoone$ model demonstrates the sensitivity 
of the SR3Lhigh signal region to signals involving on-shell $Z$ bosons,
which is an improvement compared to ref.~\cite{ATLAS:2012ai}.
Similarly, the limits on direct squark production are most stringent for long decay cascades 
involving sleptons, and less stringent for decays involving $W$ and $Z$ bosons because of
the smaller leptonic branching fractions. 
However, none of the signal regions are sensitive to compressed first- and second-generation 
squark scenarios with $\Delta m(\gluino,\ninoone)$ or $\Delta m(\squark,\ninoone)$ smaller than $\sim$100 \GeV.

\FloatBarrier
\subsubsection{Direct bottom squarks}
\label{sec:DBOTTOM}

\begin{figure}[h]
\addtolength{\subfigcapskip}{-10pt}
\begin{minipage}[b]{0.5\linewidth}
\centering
\subfigure[\label{fig:expDBOTTOM:a}]{
\includegraphics[scale=0.38]{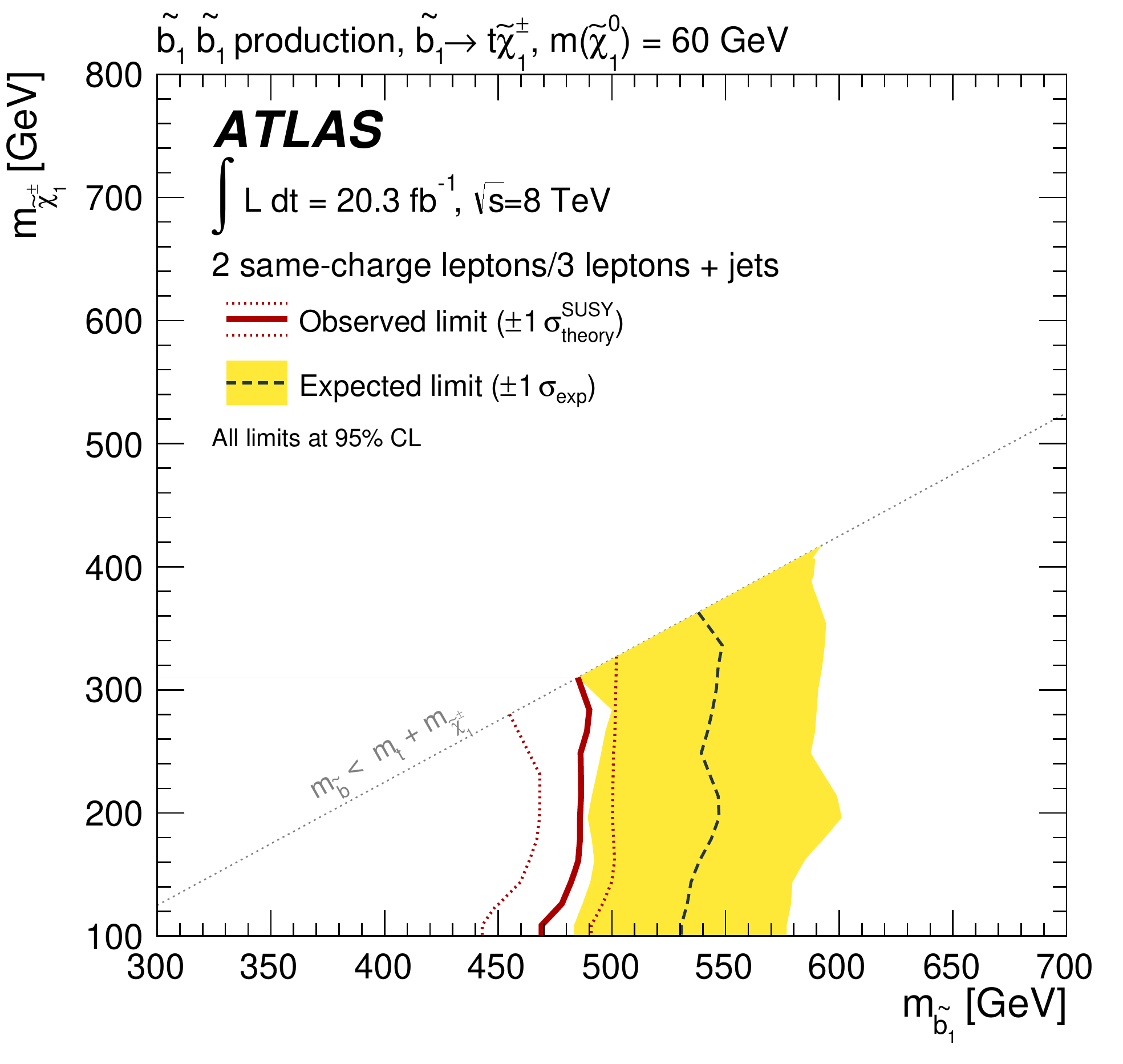}  }
\end{minipage}
\begin{minipage}[b]{0.5\linewidth}
\subfigure[\label{fig:expDBOTTOM:b}]{
\centering\includegraphics[scale=0.38]{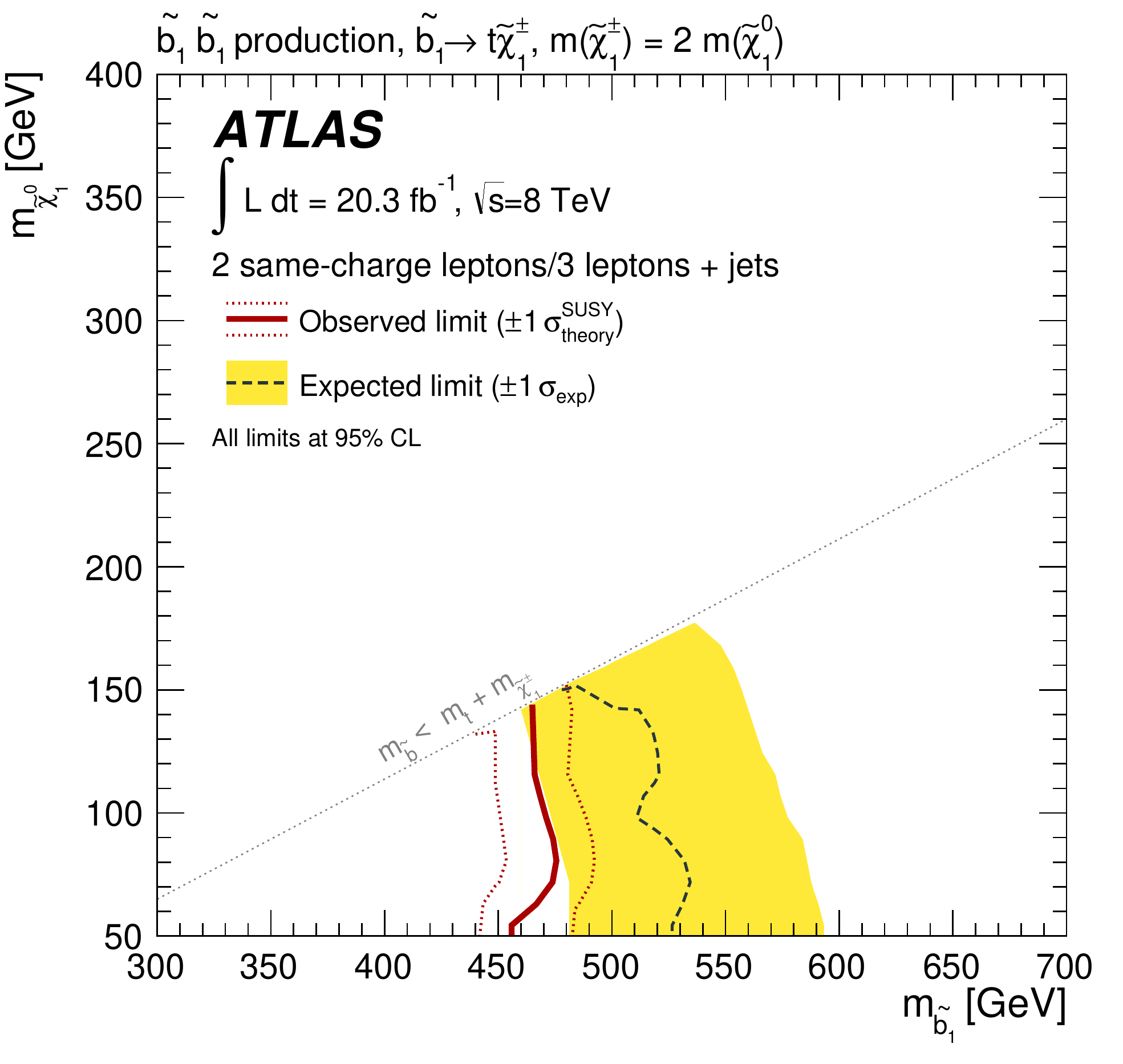}  }
\end{minipage}
\caption{Observed and expected exclusion limits on direct bottom squark production,
obtained with 20.3\,fb$^{-1}$ of $pp$ collisions at $\sqrt{s}$=8~\TeV,
for $\sbottom\to t \chinoonepm$ decays
with $m_{\ninoone}=60$~\GeV\ (left) and
$m_{\ninoone}=m_{\chinoonepm}/2$  (right). }
\label{fig:expDBOTTOM}
\end{figure}

\medskip
Results for direct bottom squark production are shown in figure~\ref{fig:expDBOTTOM} for two simplified models.  
Both models assume bottom squark pair production, decaying as
$\sbottom_1\to t \chinoonepm$,
followed by the chargino  decay
$\chinoonepm\rightarrow W^{(*)\pm} \ninoone$,
with branching fractions of 100\%.
In one model, the neutralino mass is fixed to $m_{\ninoone}=60$ \GeV.
In the other model, the \chinoonepm\ and \ninoone\ masses are related by 
$m_{\chinoonepm}=2m_{\ninoone}$. The \ninoone\ is stable in both models.
The final state is therefore
$\sbottom_1\sbottom_1\to bb~WWW^{(*)}W^{(*)}~\ninoone\ninoone$,
with the constraint that $m_{\sbottom_1}> m_{t} + m_{\chinoonepm}$.
In the fixed $m_{\ninoone}$ model, results are interpreted in the
parameter space of the bottom squark and 
$\chinoonepm$ masses (see figure~\ref{fig:expDBOTTOM:a}). 
In the varied $m_{\ninoone}$ model, results are interpreted in the
parameter space of the bottom squark and 
$\ninoone$ masses (see figure~\ref{fig:expDBOTTOM:b}). 
Bottom squark masses are excluded below 440 \GeV\ at 95\% CL,
for any chargino (neutralino) mass in the fixed (varied) $m_{\ninoone}$ model. 
The sensitivity is dominated by SR3Lhigh, SR3Llow and SR1b in both models.
These limits on $\sbottom_1\to t \chinoonepm$ complement the study of
$\sbottom_1\to b \ninoone$ 
processes performed in ref.~\cite{Aad:2013ija}.

\FloatBarrier
\subsubsection{MSUGRA/CMSSM, bRPV, GMSB and mUED}

\begin{figure}[h]
\addtolength{\subfigcapskip}{-10pt}
\begin{minipage}[b]{0.5\linewidth}
\subfigure[\label{fig:expPHYSICS:a}]{
\includegraphics[scale=0.38]{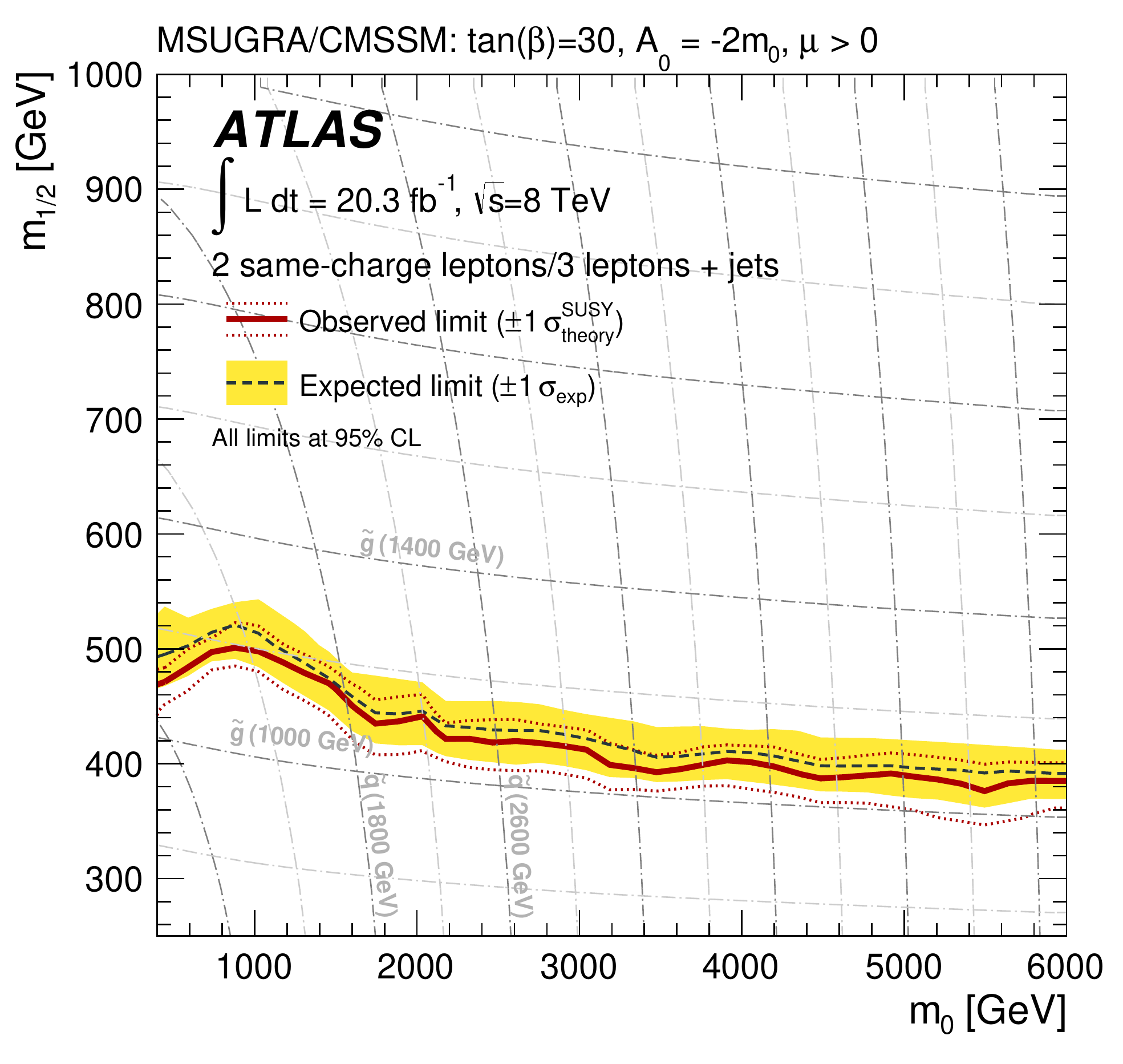}}
\end{minipage}
\begin{minipage}[b]{0.5\linewidth}
\subfigure[\label{fig:expPHYSICS:b}]{
 \includegraphics[scale=0.38]{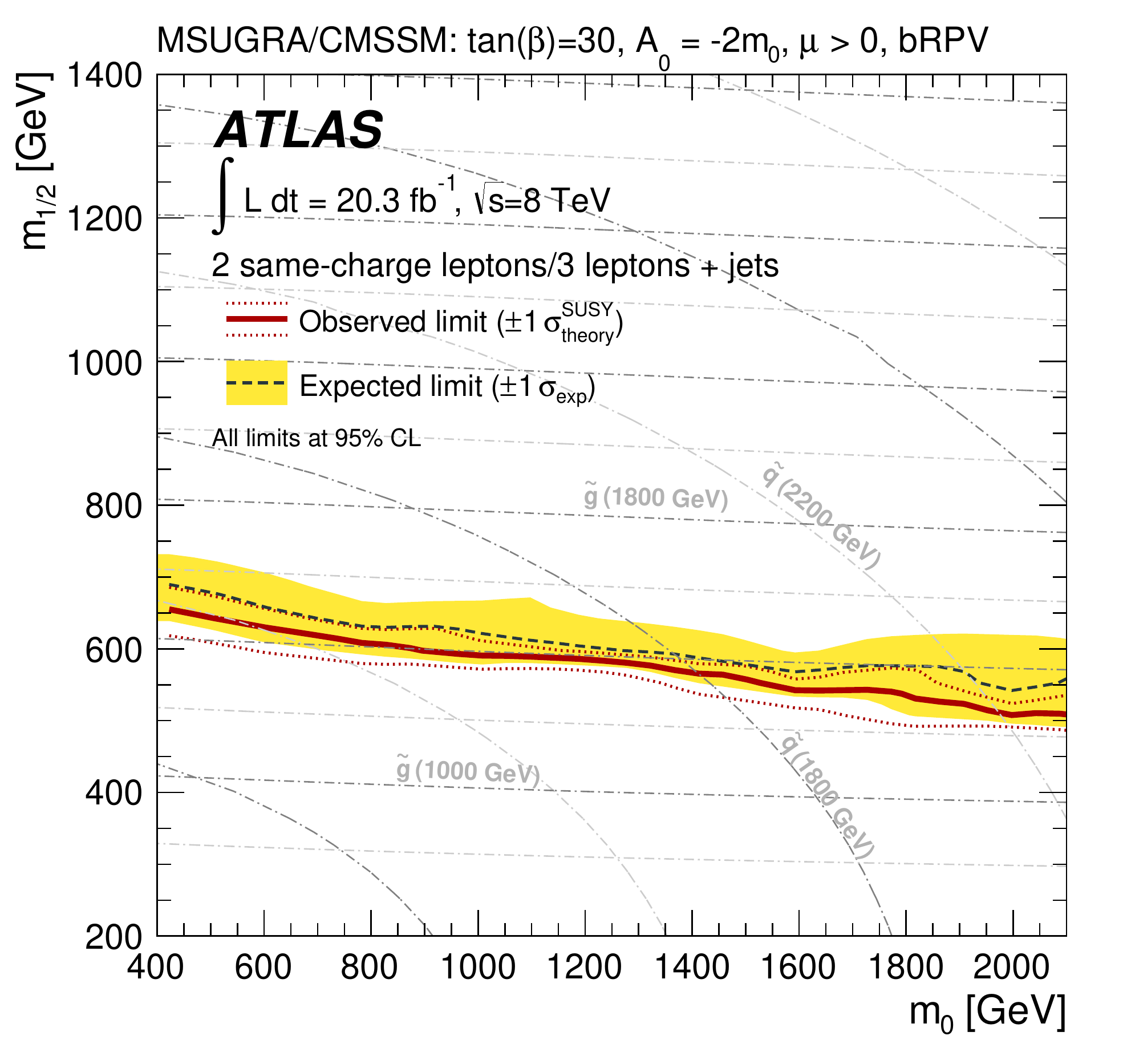}}
\end{minipage}
\begin{minipage}[b]{0.5\linewidth}
\subfigure[\label{fig:expPHYSICS:c}]{
\includegraphics[scale=0.38]{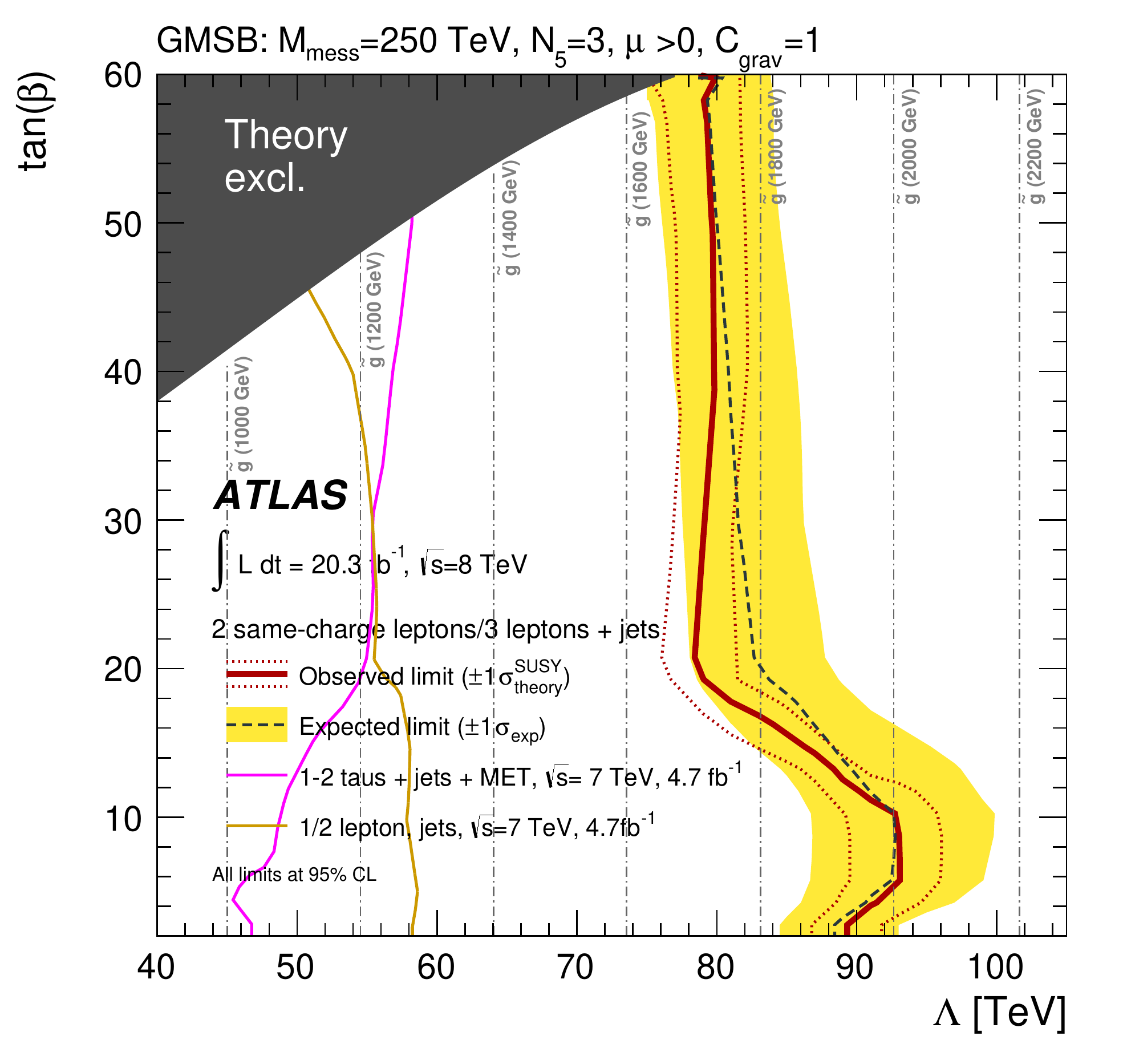}}
\end{minipage}
\begin{minipage}[b]{0.5\linewidth}
\subfigure[\label{fig:expPHYSICS:d}]{
\includegraphics[scale=0.38]{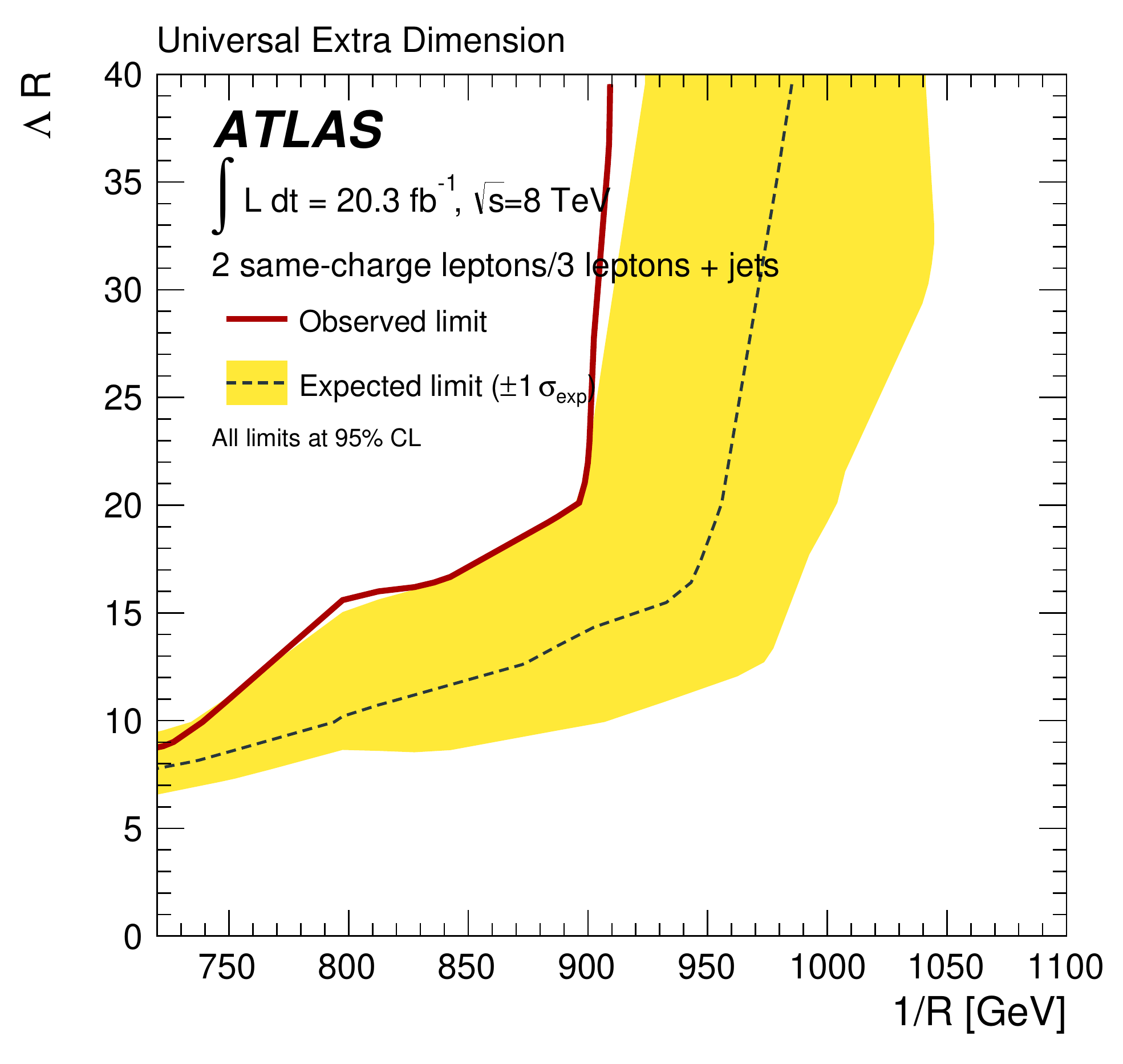}}
\end{minipage}
\caption{Observed and expected exclusion limits for the MSUGRA/CMSSM, bRPV, GMSB and mUED models,
obtained with 20.3\,fb$^{-1}$ of $pp$ collisions at
$\sqrt{s}$=8~\TeV. When available, results are compared with
the limits obtained by previous ATLAS searches~\cite{Aad:2012ms,Atlas:1Tau7TeV,ATLAS:1568342}.}
\label{fig:expPHYS}
\end{figure}

This analysis is designed and optimised to cover the SUSY processes
included in the simplified models presented above. 
To demonstrate the versatility of the search,
results are also interpreted in the context of complete models that include
a mixture of different processes.
Results are presented in figure~\ref{fig:expPHYS} for four standard benchmark models: 
MSUGRA/\allowbreak CMSSM \cite{Chamseddine:1982jx,Barbieri:1982eh,Ibanez:1982ee,Hall:1983iz,Ohta:1982wn,Kane:1993td}, bilinear $R$-parity violation (bRPV) \cite{Roy:1996bua}, 
minimal Gauge-Mediated Supersymmetry Breaking (GMSB)
\cite{Dine:1981gu,AlvarezGaume:1981wy,Nappi:1982hm,Dine:1993yw,Dine:1994vc,Dine:1995ag} 
and mUED. 

The MSUGRA/\allowbreak CMSSM model 
is characterised by five parameters: the universal scalar
and gaugino mass parameters $m_0$ and $m_{1/2}$, the universal
trilinear coupling parameter $A_0$, the ratio of the vacuum
expectation values of the two Higgs doublets $\tan(\beta)$, and
the sign of the Higgsino mass parameter $\mu$. 
Three of the parameters are fixed: $\tan(\beta) = 30$, $A_0=-2m_0$ and $\mu>0$, 
which accommodates a lightest Higgs boson mass between 122 and 128 \GeV\ for 
$m_{1/2} >$ 400 \GeV\ and $m_0 >$ 1400 \GeV.
Results are expressed as a function of $m_0$ and $m_{1/2}$ (see figure~\ref{fig:expPHYSICS:a}).
Values of $m_{1/2}$ are excluded below 360 \GeV\ at 95\% CL, for $m_0$ values below 6 TeV. 
The sensitivity is dominated by SR3b, SR3Lhigh and SR1b.

The bRPV model 
allows for bilinear $R$-parity-violating terms in the superpotential,
resulting in an unstable LSP. The $R$-parity-violating couplings are
embedded in an MSUGRA/\allowbreak CMSSM SUSY production model as described above.
For a chosen set of MSUGRA/\allowbreak CMSSM parameters, the bilinear $R$-parity violating
parameters are determined under the tree-level-dominance
scenario \cite{Grossman:2003gq} by fitting them to the neutrino
oscillation data as described in ref.~\cite{Carvalho:2002bq}. The
neutralino LSP decays within the detector through
decay modes that include neutrinos \cite{Porod:2000hv}.
Results are expressed as a function of $m_0$ and $m_{1/2}$ (see figure~\ref{fig:expPHYSICS:b}).
Values of $m_{1/2}$ are excluded between 200~GeV and 490~GeV at 95\%
CL for $m_0$ values below 2.1 TeV. Signal models with
$m_{1/2} < 200 \gev$ are not considered in this analysis because the lepton acceptance is
significantly reduced due to the increased LSP lifetime in that region.
The sensitivity is dominated by SR3b in the entire plane.

The GMSB
model is described by six parameters: the SUSY-breaking mass
scale felt by the low-energy sector ($\Lambda$), the messenger mass
(\Mmess), the number of SU(5) messenger fields (\Nfive), the ratio of the
vacuum expectation values of the two Higgs doublets ($\tan(\beta)$), the
sign of Higgs sector mixing parameter $\mu$ and the scale factor for the
gravitino mass (\Cgrav). Four parameters are fixed to
the values previously used in refs.~\cite{Aad:2012rt,ATLAS:2012ag,Aad:2012ms}:
$\Mmess=250$ TeV, $\Nfive=3$, $\mu>0$ and $\Cgrav=1$. 
Results are expressed as a function of $\Lambda$ and $\tan(\beta)$
(see figure~\ref{fig:expPHYSICS:c}).
The region of small $\Lambda$ and large $\tan(\beta)$ shown as a grey
area in  figure~\ref{fig:expPHYSICS:c} is excluded theoretically since it leads to tachyonic states.
Values of $\Lambda$ below 75~TeV are excluded at 95\% CL, for $\tan(\beta)$ below 60. 
The sensitivity is dominated by SR3Lhigh. 

The mUED model is the minimal extension of the SM with one additional universal spatial dimension.
In this non-SUSY model, the Kaluza--Klein (KK) quark decay 
chain to the lightest KK particle, the KK photon, gives a signature very similar to the supersymmetric 
decay chain of a squark to the lightest neutralino. 
The properties of the model depend on two parameters: 
the compactification radius $R$ and the cut-off scale $\Lambda$. 
The Higgs boson mass is fixed to 125 \GeV.
Results are expressed as a function of $1/R$ and $\Lambda R$ (see figure~\ref{fig:expPHYSICS:d}).
Uncertainties on the signal cross section are not considered for this model.
Values of $1/R$ below 850 \GeV\ are excluded at 95\% CL,
for $\Lambda R$ above 18. The sensitivity drops with decreasing
$\Lambda R$ because of the reduced mass splitting between the KK states. 
The sensitivity is dominated by SR3Lhigh and SR0b.

\section{Conclusion}
\label{sec:conclusion}

A search for supersymmetry in multi-jets events with exactly two same-sign leptons or at least three leptons 
is presented.
Proton--proton collision data from the full 2012 data-taking period were analysed, corresponding to 
an integrated luminosity of 20.3 \ifb\ collected at $\rts$=8 TeV by
the ATLAS detector at the LHC.
The search also utilises $b$-jets, missing
transverse momentum and other observables to extend its sensitivity.
Five signal regions were 
determined with a quantitative optimisation procedure
using a large number of simplified models.
Standard Model backgrounds were estimated using Monte Carlo simulations and data-driven techniques, and were tested in validation regions. 
Observations are in agreement with SM expectations in each signal region
and constraints are set on the visible cross section for new physics processes.
Exclusion limits are placed on 14 SUSY models and one mUED model,
using a binned fit performed simultaneously in the five signal regions.
Gluino-mediated top squark scenarios, favoured by naturalness arguments, are excluded
for $m_{\gluino}<[600\mbox{--}1000] \GeV$, largely independently of the top
squark mass and decay mode. 
Similar limits are placed on gluino-mediated production of
first- and second-generation quarks for $m_{\ninoone}<[300\mbox{--}600] \GeV$. Limits are
also placed on pair-production  of bottom squarks and squarks of the first
and second generations decaying in long cascades.
These results put new or
significantly improved limits in SUSY parameter regions where the lightest squark can be of the first,
second or third generation, where the mass differences between the supersymmetric particles can be large
or compressed, and where $R$-parity can be conserved or violated.


\section*{Acknowledgements}

We thank CERN for the very successful operation of the LHC, as well as the
support staff from our institutions without whom ATLAS could not be
operated efficiently.

We acknowledge the support of ANPCyT, Argentina; YerPhI, Armenia; ARC,
Australia; BMWF and FWF, Austria; ANAS, Azerbaijan; SSTC, Belarus; CNPq and FAPESP,
Brazil; NSERC, NRC and CFI, Canada; CERN; CONICYT, Chile; CAS, MOST and NSFC,
China; COLCIENCIAS, Colombia; MSMT CR, MPO CR and VSC CR, Czech Republic;
DNRF, DNSRC and Lundbeck Foundation, Denmark; EPLANET, ERC and NSRF, European Union;
IN2P3-CNRS, CEA-DSM/IRFU, France; GNSF, Georgia; BMBF, DFG, HGF, MPG and AvH
Foundation, Germany; GSRT and NSRF, Greece; ISF, MINERVA, GIF, I-CORE and Benoziyo Center,
Israel; INFN, Italy; MEXT and JSPS, Japan; CNRST, Morocco; FOM and NWO,
Netherlands; BRF and RCN, Norway; MNiSW and NCN, Poland; GRICES and FCT, Portugal; MNE/IFA, Romania; MES of Russia and ROSATOM, Russian Federation; JINR; MSTD,
Serbia; MSSR, Slovakia; ARRS and MIZ\v{S}, Slovenia; DST/NRF, South Africa;
MINECO, Spain; SRC and Wallenberg Foundation, Sweden; SER, SNSF and Cantons of
Bern and Geneva, Switzerland; NSC, Taiwan; TAEK, Turkey; STFC, the Royal
Society and Leverhulme Trust, United Kingdom; DOE and NSF, United States of
America.

The crucial computing support from all WLCG partners is acknowledged
gratefully, in particular from CERN and the ATLAS Tier-1 facilities at
TRIUMF (Canada), NDGF (Denmark, Norway, Sweden), CC-IN2P3 (France),
KIT/GridKA (Germany), INFN-CNAF (Italy), NL-T1 (Netherlands), PIC (Spain),
ASGC (Taiwan), RAL (UK) and BNL (USA) and in the Tier-2 facilities
worldwide.

\clearpage
\addcontentsline{toc}{section}{Bibliography}
\bibliographystyle{JHEP}
\bibliography{jhep_ss3L_susy}

\onecolumn 
\clearpage 
\begin{flushleft}
{\Large The ATLAS Collaboration}

\bigskip

G.~Aad$^{\rm 84}$,
B.~Abbott$^{\rm 112}$,
J.~Abdallah$^{\rm 152}$,
S.~Abdel~Khalek$^{\rm 116}$,
O.~Abdinov$^{\rm 11}$,
R.~Aben$^{\rm 106}$,
B.~Abi$^{\rm 113}$,
M.~Abolins$^{\rm 89}$,
O.S.~AbouZeid$^{\rm 159}$,
H.~Abramowicz$^{\rm 154}$
% ...
\bigskip
\\
% ...
$^{11}$ Institute of Physics, Azerbaijan Academy of Sciences, Baku, Azerbaijan\\
% ...
$^{84}$ CPPM, Aix-Marseille Universit{\'e} and CNRS/IN2P3, Marseille, France\\
% ...
$^{89}$ Department of Physics and Astronomy, Michigan State University, East Lansing MI, United States of America\\
% ...
$^{106}$ Nikhef National Institute for Subatomic Physics and University of Amsterdam, Amsterdam, Netherlands\\
% ...
$^{112}$ Homer L. Dodge Department of Physics and Astronomy, University of Oklahoma, Norman OK, United States of America\\
$^{113}$ Department of Physics, Oklahoma State University, Stillwater OK, United States of America\\
% ...
$^{116}$ LAL, Universit{\'e} Paris-Sud and CNRS/IN2P3, Orsay, France\\
% ...
$^{152}$ Institute of Physics, Academia Sinica, Taipei, Taiwan\\
% ...
$^{154}$ Raymond and Beverly Sackler School of Physics and Astronomy, Tel Aviv University, Tel Aviv, Israel\\
% ...
$^{159}$ Department of Physics, University of Toronto, Toronto ON, Canada\\
% ...
\end{flushleft}


\end{document}